\documentclass{jfm}
\usepackage{graphicx,psfrag}
\usepackage{caption} 
\usepackage{subcaption}
\usepackage{natbib}
\usepackage{float}
\usepackage{rotating}
\usepackage{mathrsfs,amsmath}
\usepackage{upmath}
\usepackage{upgreek}
\usepackage{latexsym}
\usepackage{mathrsfs}
\usepackage{layout}
\usepackage{bm}
\usepackage{url}
\usepackage{wrapfig}
\usepackage{tabularx}
\usepackage{fancyhdr}
\usepackage{color}
\usepackage[normalem]{ulem}
\usepackage[section]{placeins}

\ifCUPmtlplainloaded \else
  \checkfont{eurm10}
  \iffontfound
    \IfFileExists{upmath.sty}
      {\typeout{^^JFound AMS Euler Roman fonts on the system,
                   using the 'upmath' package.^^J}%
       \usepackage{upmath}}
      {\typeout{^^JFound AMS Euler Roman fonts on the system, but you
                   dont seem to have the}%
       \typeout{'upmath' package installed. JFM.cls can take advantage
                 of these fonts,^^Jif you use 'upmath' package.^^J}%
       \providecommand\upi{\pi}%
      }
  \else
    \providecommand\upi{\pi}%
  \fi
\fi

\ifCUPmtlplainloaded \else
  \checkfont{msam10}
  \iffontfound
    \IfFileExists{amssymb.sty}
      {\typeout{^^JFound AMS Symbol fonts on the system, using the
                'amssymb' package.^^J}%
       \usepackage{amssymb}%
         \let\leq=\leqslant
         \let\geq=\geqslant
      }{}
  \fi
\fi

\ifCUPmtlplainloaded \else
  \IfFileExists{amsbsy.sty}
    {\typeout{^^JFound the 'amsbsy' package on the system, using it.^^J}%
     \usepackage{amsbsy}}
    {\providecommand\boldsymbol[1]{\mbox{\boldmath $##1$}}}
\fi

\DeclareMathAlphabet\mathsfbi{OT1}{cmss}{m}{sl}

\title[Settling bidisperse particle dynamics in turbulence]{Small-scale dynamics of settling, bidisperse particles in turbulence}

\author[Rohit Dhariwal and Andrew D. Bragg]%
{R\ls O\ls H\ls I\ls T\ns D\ls H\ls A\ls R\ls I\ls W\ls A\ls L$^1$ 
   \and %
    ~A\ls N\ls D\ls R\ls E\ls W\ls \ns D.\ns B\ls R\ls A\ls G\ls G$^1$%
      \thanks{Email address for correspondence: andrew.bragg@duke.edu}
}

\affiliation{$^1$ 
Department of Civil and Environmental Engineering, Duke University, Durham, \\ North Carolina 27708, USA}

\pubyear{2010}
\volume{650}
\pagerange{119--126}
\date{?; revised ?; accepted ?. - To be entered by editorial office}

\begin{document}

\maketitle

\begin{abstract}
	We use Direct Numerical Simulations (DNS) to investigate the dynamics of settling, bidisperse particles in isotropic turbulence. In agreement with previous studies, we find that without gravity (i.e. $Fr=\infty$, where $Fr$ is the Froude number), bidispersity leads to an enhancement of the relative velocities, and a suppression of their spatial clustering. For  $Fr<1$, the relative velocities in the direction of gravity are enhanced by the differential settling velocities of the bidisperse particles, as expected. However, we also find that gravity can strongly enhance the relative velocities in the ``horizontal" directions (i.e. in the plane normal to gravity). This non-trivial behavior occurs because fast settling particles experience rapid fluctuations in the fluid velocity field along their trajectory, leading to enhanced particle accelerations and relative velocities. Indeed, the results show that even when $Fr\ll1$, turbulence can still play an important role, not only on the horizontal motion, but also on the vertical motion of the particles, with significant implications for understanding the mixing rates of settling bidisperse particles in turbulence. We also find that gravity drastically reduces the clustering of bidisperse particles. These results are strikingly different to the monodisperse case, for which recent results have shown that when $Fr<1$, gravity strongly suppresses the relative velocities in all directions, and can enhance clustering. We then consider the implications of these results for the collision rates of settling, bidisperse particles in turbulence. We find that for $Fr=0.052$, the collision kernel is almost perfectly predicted by the collision kernel for bidisperse particles settling in quiescent flow, such that the effect of turbulence may be ignored. However, for $Fr=0.3$, turbulence plays an important role, and the collisions are only dominated by gravitational settling when the difference in the particle Stokes numbers is $\geq O(1)$. 
\end{abstract}

\begin{keywords}
	Authors should not enter keywords on the manuscript, as these must be chosen by the author during the online submission process and will then be added during the typesetting process (see http://journals.cambridge.org/data/\linebreak[3]relatedlink/jfm-\linebreak[3]keywords.pdf for the full list)
\end{keywords}

\pagebreak

\section{Introduction}\label{sec:intro}

Understanding the dynamics of small particles in turbulent flows is relevant to a broad range of problems, such as aerosol manufacturing \citep{moody03}, drug delivery \citep{li96}, spray combustion \citep{faeth96}, microorganism motion in aquatic environments \citep{delillo14}, and planetesimal formation in the early universe \citep{johansen07}. A great deal of research on the motion of particles at the small-scales of turbulent flows has been motivated by the need to better understand the microphysical mechanisms governing atmospheric cloud formation, for which standard models can fail in important regimes \citep{shaw03,grabowski13}. In particular, standard models of droplet growth in clouds capture two main mechanisms: (1) condensation, which is effective for droplets with radii smaller than 30$\mu$m, and (2) coalescence due to differential sedimentation, which is effective for drops with radii larger than $80\mu$m.  In the intermediate size range of $30$-$80\mu$m, neither mechanism can account for the rapid droplet growth observed in clouds, and hence this size range is commonly referred to as the ``condensation-coalescence bottleneck'' or the ``size-gap'' \citep{prupp97}.  It is believed that turbulence-induced collisions lead to rapid growth in the size-gap, though many open questions concerning this remain \citep{devenish12,grabowski13}. Turbulence is understood to give rise to enhanced particle collision rates by generating spatial clustering of the particles, and by leading to enhanced collision velocities \citep{pumir16b}. Turbulence-induced collisions are also thought to be important for understanding the formation of protoplanetary disks, and the atmospheres of planets and dwarf stars \citep{cuzzi96,johansen07,pan11}. 

In this paper we are concerned with the motion of heavy, inertial particles, namely, particles whose density is much greater than that of the fluid in which they are suspended. This regime is relevant in both the cloud and astrophysical contexts, and gas-solid flows in general. The small-scale motion of such particles in turbulence is controlled by a range of physical mechanisms, whose effect depends upon the particle Stokes number, $St\equiv \tau_p/\tau_\eta$, where $\tau_p$ is the particle response time, and $\tau_\eta$ is the Kolmogorov timescale. One of the important effects is preferential sampling, which describes the fact that inertial particles tend to avoid regions of strong rotation due to the centrifuge effect, and favor strain dominated regions of the flow \citep{maxey87,squires91a,eaton94}. In the regime $St\ll1$, preferential sampling is the sole mechanism by which inertia affects the particle motion, and it gives rise to spatial clustering of the particles \citep{chun05}. When $St\geq O(1)$, the particles are strongly affected by the finite memory they possess of their interaction with the turbulence along their path-history \citep{bragg14b,bragg14c}. This path-history effect gives rise to ``caustics'' \citep{wilkinson05}, associated with inertial particles having very large  relative velocities at small-separations. The path-history effect also leads to strong spatial clustering of the particles through the non-local clustering mechanism \citep{bragg14d,bragg14e}. When $St\geq O(1)$, the filtering effect is also important \citep{bec06a,sathya08a}, describing the lack of response of the inertial particles to flow fluctuations having time scales less than $\tau_p$.

These physical mechanisms have been primarily studied for the case of monodisperse particles, however, in almost all real systems, the particles are polydisperse. In clouds and many other problems, the particle collisions are usually binary due to the dilute particle loading \citep{grabowski13}, and furthermore, mixing and dispersion of the particles in the flow is typically analyzed by considering the relative motion of particle-pairs \citep{salazar09}. In this case, the problem of polydispersity reduces to understanding the effect of bidispersity on the small-scale, particle-pair motion. The polydisperse statistics can then be obtained through suitable integrals of the bidisperse statistics, weighted according to the size distribution of the particles. Several studies have considered how bidispersity affects the spatial clustering \citep{zww01,chun05,pan11,chen17}, relative velocities \citep{pan14,pan14b,chen17} and collisions \citep{zww01,pan14c} of particles in turbulence. It was found that bidispersity reduces the clustering of inertial particles in turbulent flows \citep{zww01,chun05,pan11,chen17}, whereas it enhances their relative velocities \citep{zww01,chen17,kruis97}. Bidispersity was found to increase collision rates, since the enhancement of the collision velocities due to bidispersity outweighs the reduction in the clustering \citep{pan14c}. We defer discussion of the physical mechanisms responsible for these effects of bidispersity to \S\ref{theory}.

In the context of cloud microphysics and many other applications, another effect that is crucial to account for is the effect of gravity on the particle motion. The relative importance of gravity to the turbulent accelerations can be quantified by the Froude number $Fr \equiv \epsilon^{3/4}/(\nu^{1/4}\mathfrak{g})$, where $\epsilon$ is the mean fluid kinetic energy dissipation rate, $\nu$ is the kinematic viscosity (such that $\epsilon^{3/4}/\nu^{1/4}$ is the Kolmogorov acceleration scale), and $\mathfrak{g}$ is the magnitude of the gravitational acceleration vector. The impact of varying $Fr$ has recently been investigated for the case of monodisperse particles \citep{bec14b,gustavsson14,ireland16b}. These studies found that reducing $Fr$ always reduces the relative velocities of the inertial particles, but has a non-trivial effect on the clustering, such that in some regimes of $St$ and $Fr$, gravity reduces clustering, and in other regimes it enhances clustering. \citet{parishani15} and \citet{ireland16b} also showed that gravity can significantly enhance inertial particle accelerations, though the explanations they gave for this behavior were very different (see \S\ref{theory}).

Despite being crucial for understanding droplet motion in clouds and many other problems, it appears that the effect of gravity on the small-scale dynamics of bidisperse particles in turbulence has scarcely been addressed. Indeed, to the best of our knowledge, the only studies that have attempted to address this are those by \citet{woittiez09} and \citet{parishani15}. The DNS study of \citet{woittiez09} showed that gravity leads to a strong suppression of the clustering of bidisperse particles, but a strong enhancement of their collision velocities, due to the contribution from the differential settling of the particles. Furthermore, they found that the collision kernel based on bidisperse particles settling in quiescent flow significantly underpredicted the collision rates measured in their DNS. However, their study did not comprehensively explore the effects of varying the bidispersity and $Fr$, and also focused exclusively on particle collisions, rather than more general small-scale processes. The DNS study of \citet{parishani15} presented a preliminary study on the influence of gravity and bidispersity on the relative velocities of particle-pairs in turbulence. Their approach was to decompose the relative velocity of the particle-pair into various contributions that emerge from the particle equation of motion, and then analyze some of the statistics of those contributions. While interesting, these results do not give clear insight into how gravity and bidispersity affect particle-pair dynamics in turbulence, since the statistics of the relative velocities depend not only on the individual behavior of these contributions, but also on various correlations between these contributions, and this was not examined in \citet{parishani15}. Also, they only considered the relative velocities at the contact separation, and therefore how gravity and bidispersity affects the relative motion of the particle-pair at different scales/separations is unknown. There therefore remain many open questions with respect to how gravity affects bidisperse particle motion at the small-scales of turbulence, and these knowledge gaps motivate the present work. 

The organization of the paper is as follows. In \S\ref{theory}, we consider theoretically the effect of bidispersity and gravity on the relative motion of particle-pairs at the small-scales of turbulence. In \S\ref{CompD}, we explain the numerical
methods and parameters for our simulations. In \S\ref{RD}, we present the results of our simulations, exploring how bidispersity and $Fr$ impact the particle relative velocities, clustering and collisions. Finally, in \S\ref{conc}, we draw conclusions and highlight open issues that remain to be explored.

\section{Theoretical Considerations}\label{theory}
For investigating the motion of droplets in clouds, it is usually assumed (e.g. \cite{shaw03,grabowski13}) that the particle loading is sufficiently small to ignore the effect of the particles on the turbulence (i.e. the system is ``one-way coupled''), and that the particles are well described as being small (i.e., $d/\eta \ll 1$, where $d$ is the particle diameter and $\eta$ is the Kolmogorov length scale) and heavy ($\rho_p/\rho \gg 1$, 
where $\rho_p$ is the particle density and $\rho$ is the fluid density). Under these conditions, each particle is treated as a point particle whose 
motion obeys a simplified version of the equation by \cite{maxey83}
\begin{align}
\dot{\bm{x}}^p(t) & = \bm{v}^p(t), \label{eq:part1}\\
\ddot{\bm{x}}^p(t) & = \dot{\bm{v}}^p(t) = \frac{\bm{u}(\bm{x}^p(t),t)-\bm{v}^p(t)}{\tau_p} + \bm{\mathfrak{g}}, \label{eq:part2}
\end{align}
where $\bm{x}^p(t)$ and $\bm{v}^p(t)$ are the particle position and velocity,
 respectively, $\tau_p = \rho_p d^2/18 \mu$ is the particle response time ($\mu$ is the fluid dynamic viscosity) and 
$\bm{\mathfrak{g}}$ is the gravitational acceleration. As discussed in \citet{ireland16b}, the use of a linear rather than nonlinear drag force for the particles is likely sufficient for $St\leq O(1)$, which is the range we restrict attention to in this study. Furthermore, using a linear drag force facilitates comparison with previous studies on the effect of gravity on the small-scale motion of monodisperse particles in turbulence that were also based on a linear drag force \citep{bec14b,gustavsson14,ireland16b}.

We are interested in understanding how settling, bidisperse particle-pairs move relative to each other, especially at the small-scales of turbulence. We therefore consider the motion of a ``satellite'' particle relative to a ``primary'' particle, for which the equation of relative motion is obtained by subtracting \eqref{eq:part2} for the primary particle from that for the satellite particle
\begin{equation}
\ddot{\bm{r}}^p(t) = \dot{\bm{w}}^p(t) = 
\frac{\Delta\bm{ u}(\bm{x}^p(t),\bm{r}^p(t),t) - \bm{w}^p(t)}{\tau_p} + 
\frac{\theta \left( \bm{a}^{p}(t) - \bm{\mathfrak{g}} \right)}{\tau_p},
\label{eq:eom_pair} 
\end{equation}
where $\bm{x}^p(t)$ and $\bm{x}^p(t)+\bm{r}^p(t)$ are the positions of the primary and satellite particles, respectively, $\bm{w}^p(t)$ is their relative 
velocity, \[\Delta\bm{u}(\bm{x}^p(t),\bm{r}^p(t),t)\equiv \bm{u}(\bm{x}^p(t)+\bm{r}^p(t),t)-\bm{u}(\bm{x}^p(t),t)\]is the difference in the fluid velocity at 
the two particle positions, $\tau_p$ is now the response time of the satellite particle, $\theta \equiv \Delta \tau_p$ is the response time of the satellite particle subtracted from that of the primary particle, and 
$\bm{a}^{p}(t)=(\theta+\tau_p)^{-1}(\bm{u}(\bm{x}^p(t),t)-\bm{v}^p(t)) + \bm{\mathfrak{g}}$ is the primary particle acceleration. Since we are considering homogeneous turbulence, we will drop the $\bm{x}^p(t)$ argument from statistical quantities. 

If we non-dimensionalize the variables in (\ref{eq:eom_pair}) using appropriate combinations of the Kolmogorov acceleration $a_\eta$, velocity $u_\eta$, length $\eta$, and time $\tau_\eta$ scales, we obtain
\begin{equation}
\ddot{\widetilde{\bm{r}^p}}(\tilde{t}) = \dot{\widetilde{\bm{w}^p}}(\tilde{t}) = 
\frac{\Delta \widetilde{\bm{u}}(\widetilde{\bm{x}^p}(\tilde{t}),\widetilde{\bm{r}^p}(\tilde{t}),\tilde{t}) - \widetilde{\bm{w}^p}(\tilde{t})}{St_2} + 
\frac{\Delta St \left(\widetilde{\bm{a}^p}(\tilde{t}) - \bm{e}_{\mathfrak{g}} Fr^{-1} \right)}{St_2},
\label{eq:eom_pairND} 
\end{equation}
where $\widetilde{\cdot}$ denotes a non-dimensional quantity, $St_2\equiv\tau_p/\tau_\eta$ is the Stokes number of the satellite particle, $\Delta St\equiv \theta/\tau_\eta\equiv St_1-St_2$, where $St_1$ is the Stokes number of the primary particle, and $\bm{e}_{\mathfrak{g}}$ is the unit vector in the direction of gravity. Another important parameter that we shall make use of is the scale-dependent Stokes number $St_r\equiv\tau_p/\tau_r$, where $\tau_r$ is the eddy turnover time at scale $r$.
\subsection{Relative Velocities and Particle Accelerations}
To gain insight into the effect of bidispersity and gravity on the particle-pair relative velocities, it is informative to consider the formal solution of (\ref{eq:eom_pairND}), namely (in the rest of this section we drop $\widetilde{\cdot}$ for notational ease)
\begin{equation}
\bm{w}^p(t) =\frac{1}{St_2} \int_0^t {\rm e}^{-(t-s)/St_2}\Delta \bm{u}^p(s) \,ds -\Delta St Fr^{-1}\bm{e}_{\mathfrak{g}} +\frac{\Delta St}{St_2} \int_0^t {\rm e}^{-(t-s)/St_2}\bm{a}^{p}(s)\,ds,
\label{eq:sol_w}
\end{equation}
where $\Delta \bm{u}^p(s)\equiv \Delta\bm{ u}(\bm{x}^p(s),\bm{r}^p(s),s) $. In (\ref{eq:sol_w}), and throughout, we assume $t\gg St_2$ such that initial conditions/effects can be ignored (valid since we are interested in the statistically stationary state of the system).

In the monodisperse case, $\Delta St=0$, and only the first integral in  (\ref{eq:sol_w}) remains. This case has been studied in great detail using theoretical and numerical approaches \citep{bec10a,gustavsson11,salazar12a,gustavsson14,bragg14c,ireland16a,ireland16b}. Here we summarize the findings and refer the reader to those papers for detailed discussions.

When $St_r$ is finite, the inertial particles preferentially sample the fluid velocity field, showing a tendency to avoid regions of strong rotation due to the centrifuge effect, and accumulate in regions of strain \citep{maxey87}. At scales outside of the dissipation range, the same effect occurs and can be understood in terms of the particle interaction with the coarse-grained strain and rotation fields at those scales \citep{bragg14e}. When $St_r\geq O(1)$, the dynamics becomes temporally non-local; $\bm{w}^p(t)$ depends upon $\Delta\bm{u}^p$ along the path-history of the particle pair over the time-span $t-s\leq O(\tau_p)$. In the dissipation and inertial ranges where the statistics of $\Delta\bm{u}$ depend upon separation, this can have a dramatic effect. In particular, if the particle separation changes substantially over a time span $O(\tau_p)$, then the particle-pairs will remember their interaction with scales of motion of the turbulence that have very different properties from those at their current separation. This ``path-history effect'' gives rise to ``caustics'' \citep{wilkinson05}, where $\|\bm{w}^p(t)\|\gg\|\Delta \bm{u}^p(t)\|$ at small separations. Finally, when $St_r\geq O(1)$ there also exists the ``filtering effect'', which describes the fact that due to their inertia, the particles cannot fully respond to all of the turbulent fluctuations and so behave sluggishly.

In the dissipation range, the path-history effect dominates the behavior of $\bm{w}^p(t)$ when $St_r\geq O(1)$. As the separation increases, the path-history effect remains important when $St_r\geq O(1)$, but its effect weakens because the separation dependence of $\Delta\bm{u}$ weakens. In this case, the preferential sampling and filtering effects play important roles. Finally, at the large scales of homogeneous turbulence, the path-history effect vanishes (since the statistics of $\Delta\bm{u}$ no longer depend upon separation), the preferential sampling plays a weak role, and the filtering effect dominates.

When $\Delta St=0$, the effect of gravity on $\bm{w}^p(t)$ is only implicit; gravity affects $\Delta\bm{u}^p$ by changing the way the particles interact with the turbulent flow, and most importantly, by reducing the correlation timescale of $\Delta\bm{u}^p$. This leads to a \emph{reduction} of the preferential sampling and path-history effects (and therefore a reduction of caustics), and to an \emph{enhancement} of the filtering effect (see \cite{ireland16b} for a detailed discussion of these effects).

When $|\Delta St|>0$ and $Fr=\infty$ (i.e. no gravity), the main effect of the bidispersity is through the ``acceleration term'' in (\ref{eq:sol_w}), with an implicit effect upon the first integral in (\ref{eq:sol_w}) through the modification to the properties of $\Delta\bm{u}^p$. An essential difference between the two integrals in (\ref{eq:sol_w}) is that the first depends upon the particle separation, whereas the second does not. Since $\Delta \bm{u}^p$ decreases, on average, with decreasing separation, then provided $|\Delta St|>0$, there will be a separation below which the second integral is $\gg$ than the first. In this regime, the statistics of $\bm{w}^p(t)$ will not depend upon the particle separation. This acceleration contribution causes the relative velocities of bidisperse particles to exceed those of monodisperse particles at small-separations, as observed in a number of previous theoretical and numerical studies \citep{zww01,zaichik06a,zaichik09b,pan10,pan14}. In homogeneous turbulence, there is no path-history effect associated with the second integral in (\ref{eq:sol_w}); the statistical properties of $\bm{a}^p(s)$ are equivalent along all primary particle trajectories $\bm{x}^p$. The accelerations are, however, strongly affected by both the preferential sampling and filtering effects \citep{bec06a,sathya08a,salazar12b,ireland16b}, with filtering causing the fluctuations of $\bm{a}^p$ to decrease rapidly with increasing $St$.

When $|\Delta St|>0$ and $Fr<\infty$, gravity can lead to a number of effects on the relative motion of bidisperse particles. First, we note that the role of gravity in \eqref{eq:eom_pairND} is a little subtle since $\bm{a}^{p}(t) -\bm{e}_{\mathfrak{g}} Fr^{-1}=St_1^{-1}(\bm{u}(\bm{x}^p(t),t)-\bm{v}^p(t))$, i.e. the explicit role of gravity appears to be absent. However, the direct effect of gravity is hidden in $\bm{v}^p(t)$, which may be brought out using the formal solution for $\bm{v}^p(t)$ to obtain
\begin{align}
\begin{split}
\bm{a}^{p}(t) &=\frac{1}{St_1}\Big(\bm{u}(\bm{x}^p(t),t)-\bm{v}^p(t)\Big)+\bm{e}_{\mathfrak{g}} Fr^{-1}\\
&=\frac{1}{St_1}\Big(\bm{u}(\bm{x}^p(t),t)-\frac{1}{St_1}\int_0^t {\rm e}^{-(t-t')/St_1}\bm{u}^p(t') \,dt' -St_1\Big[1-{\rm e}^{-t/St_1}\Big]\bm{e}_{\mathfrak{g}} Fr^{-1}\Big)+\bm{e}_{\mathfrak{g}} Fr^{-1}\\
&=\frac{1}{St_1^2}\int_0^t {\rm e}^{-(t-t')/St_1}\Big(\bm{u}^p(t) -\bm{u}^p(t')\Big) \,dt',
\end{split}
\label{eq:sol_ag}
\end{align}
where we have again assumed $t\gg St_1$. The important point then is that $\bm{a}^{p}$ is in fact only implicitly affected by gravity, despite the fact that gravity appears explicitly in its dynamical equation.

The effect of gravity on the first term on the rhs of (\ref{eq:sol_w}) is qualitatively the same as in the monodisperse case, described earlier. The effect of gravity on the second term on the rhs of (\ref{eq:sol_w}) is clear; it increases as $Fr$ decreases, and becomes increasingly important as $|\Delta St|$ increases. However, this term only acts in the direction of gravity, and plays no role in the plane normal to this. Therefore, in the plane normal to $\bm{e}_{\mathfrak{g}}$ the effect of gravity is only implicit. The third term on the rhs of (\ref{eq:sol_w}) involves $\bm{a}^p$, and in \cite{ireland16b} we showed that gravity can enhance $\bm{a}^p$, the effect being dramatic when $Fr\ll1$. This enhancement of inertial particle accelerations due to gravity was also observed by \citet{parishani15}, who proposed that the enhancement is due to the preferential sweeping effect \citep{wang93}. However, in \cite{ireland16b} we argued that this cannot be the explanation since the acceleration enhancement due to gravity is strongest in a regime of $St$ and $Fr$ where the preferential sampling of the fluid velocity gradient field is absent. We argued in contrast, that the acceleration enhancement occurs because when $St$ is finite, the inertial particles have large settling velocities when $Fr$ is small, causing the
fluid velocity seen by the particles to change rapidly in time. That is, fast settling leads to large values of $\bm{u}^p(t) -\bm{u}^p(t')$ even when $t-t'\ll St$, which in turn through (\ref{eq:sol_ag}) implies large fluctuations in $\bm{a}^p(t)$. Further, we showed in \cite{ireland16b} that fluctuations in $\bm{a}^p(t)$ are weaker in the direction of gravity than in the plane normal to gravity, which we explained in terms of the difference in the longitudinal and transverse correlation lengthscales of $\bm{u}$ in isotropic turbulence.

In the plane normal to gravity, the third term, which always dominates over the first term at sufficiently small separations, can increase as $Fr$ is decreased due to the enhanced accelerations. In \cite{ireland16b} we derived a theoretical model for the effect of strong gravity on the particle accelerations. For simplicity, let us define the gravitational force to act in the $x_3$ direction, so that $\bm{e}_{\mathfrak{g}}=(0,0,1)$. This will be referred to as the ``vertical'' direction, whereas, $x_1$ and $x_2$ will be referred to as the ``horizontal'' directions. (Note also that since we are considering isotropic turbulence, the statistics of the particle motion are axisymmetric about $\bm{e}_{\mathfrak{g}}$ when $Fr<\infty$). Using this notation, the result derived in \cite{ireland16b} for the inertial particle accelerations in the hortizontal direction is
\begin{align}
\frac{\langle a^p_1(t)a^p_1(t)\rangle}{a_\eta^2}=Fr^{-1}\Bigg(\frac{u'}{u_\eta}\Bigg)^2\Bigg(\frac{1}{St_1^2 Fr^{-1}+\ell\eta^{-1}}\Bigg),\quad\text{for}\,\,St_1 Fr^{-1}\gg u'u_\eta^{-1},\label{aperp}
\end{align}
where $u'$ is the fluid r.m.s. velocity, and $\ell$ is the integral lengthscale of the flow. Comparisons of \eqref{aperp} with DNS in \cite{ireland16b} showed excellent agreement in the regime $St_1 Fr^{-1}\gg u'u_\eta^{-1}$. For fixed $R_\lambda$, \eqref{aperp} increases as $Fr$ decreases, resulting in strong accelerations in the plane normal to gravity. Since $w^p_1$ depends upon $a^p_1$, gravity can therefore lead to an enhancement of the relative velocities in the plane normal to its action, and this enhancement could be strong when $Fr\ll1$.

Finally, consider the limit $Fr\to 0$ where the relative velocities in the vertical direction are deterministic and given by
\begin{equation}
\begin{split}
{w}_3^p(t) =  -\Delta St Fr^{-1},
\end{split}
\label{w3Fr0}
\end{equation}
which is simply the difference in the Stokes settling velocities of the two particles (i.e. the differential settling velocity), in dimensionless form. However, in the horizontal directions, if $R_\lambda\gg1$ the motion remains turbulent even in the limit $Fr\to 0$ since
\begin{equation}
\begin{split}
{w}_1^p(t) = &\frac{1}{St_2} \int_0^t {\rm e}^{-(t-s)/St_2}\Delta {u}_1^p(s) \,ds + \frac{\Delta St}{St_2} \int_0^t {\rm e}^{-(t-s)/St_2} {a}_1^p(s)\,ds.
\end{split}
\label{w1Fr0}
\end{equation}
The effect of gravity is implicit in (\ref{w1Fr0}), and in the limit $Fr\to 0$ the large settling velocities of the particles causes them to experience $\Delta {u}_1^p$ as a white noise signal \citep{bec14b}. As such, the limit $Fr\to 0$ is subtle in the bidisperse case: In the vertical direction, ${w}_3^p(t) =  -\Delta St Fr^{-1}$ when $Fr\to 0$, which applies to quiescent and turbulent flows. However, for the horizontal direction, in a quiescent flow $w^p_1(t)=0\,\forall Fr$, but in a turbulent flow, $w^p_1(t)\neq 0$ in the limit $Fr\to 0$. As a consequence of this, $\lim_{Fr\to0}\bm{w}^p(t)\not\to-\Delta St Fr^{-1}\bm{e}_{\mathfrak{g}} $ in a turbulent flow.

Furthermore, provided $R_\lambda <\infty$, then in the limit $Fr\to0$ \eqref{aperp} becomes
\begin{align}
\frac{\langle a^p_1(t)a^p_1(t)\rangle}{a_\eta^2}=\Bigg(\frac{u'}{St_1 u_\eta}\Bigg)^2\sim \frac{R_\lambda}{St_1^{2}} ,
\end{align}
which is independent of $Fr$. This predicts that even in the limit $Fr\to0$, turbulence can lead to horizontal accelerations for the inertial particles that are not only finite, but $\gg a_\eta$.

%
\subsection{Spatial Clustering}
To consider the spatial clustering of settling, bidisperse particles, we begin by defining the Probability Density Function (PDF) for $\bm{r}^p(t),\bm{w}^p(t)$ in the phase-space $\bm{r},\bm{w}$ 
\begin{align}
\mathcal{P}(\bm{r},\bm{w},t)\equiv\langle\delta(\bm{r}^p(t)-\bm{r})\delta(\bm{w}^p(t)-\bm{w})\rangle,
\end{align}
for which the evolution equation is 
\begin{align}
\partial_t\mathcal{P}=-\bm{\nabla_r\cdot}\mathcal{P}\bm{w}-\bm{\nabla_w\cdot}\mathcal{P}\langle\dot{\bm{w}}^p(t)\rangle_{\bm{r},\bm{w}},\label{PDFrw}
\end{align}
where $\bm{\nabla_r}\equiv\partial/\partial\bm{r}$ and $\bm{\nabla_w}\equiv\partial/\partial\bm{w}$.

The spatial clustering of the particles is described by the marginal distribution
\begin{align}
\varphi(\bm{r},t)\equiv\int_{\mathbb{R}^3}\mathcal{P}(\bm{r},\bm{w},t)\,d\bm{w}.
\end{align}
By multiplying the stationary form of \eqref{PDFrw} by $\bm{w}$ and then integrating over all $\bm{w}$, we can obtain an equation for $\partial_t\varphi\langle{\bm{w}}^p(t)\rangle_{\bm{r}}$, where $\varphi\langle{\bm{w}}^p(t)\rangle_{\bm{r}}\equiv\int_{\mathbb{R}^3}\bm{w}\mathcal{P}\,d\bm{w}$ and $\langle\cdot\rangle_{\bm{r}}$ denotes an ensemble average conditioned on $\bm{r}^p(t)=\bm{r}$. Then, inserting the equation of motion for $\dot{\bm{w}}^p(t)$, and considering the horizontal component $\partial_t\varphi\langle w_1^p(t)\rangle_{\bm{r}}$, we obtain an equation governing the steady-state form of $\varphi$
\begin{align}
\begin{split}
{0}=-St_2\langle w^p_1(t)\bm{w}^p(t)\rangle_{\bm{r}}\bm{\cdot\nabla_r}\varphi-St_2\varphi\bm{\nabla_r\cdot}\langle w^p_1(t)\bm{w}^p(t)\rangle_{\bm{r}}+\varphi\langle\Delta{u}_1^p(t)\rangle_{\bm{r}}+\Delta St \varphi\langle{a}_1^p(t)\rangle_{\bm{r}},
\end{split}
\label{ADFeq}
\end{align}
where we have used $\langle {w}_1^p(t)\rangle_{\bm{r}}=0$ due to statistical stationarity and isotropy of the system in the horizontal plane. Note that although ${a}_1^p(t)$ is not a function of $\bm{r}^p(t)$, $\langle{a}_1^p(t)\rangle_{\bm{r}}\neq\langle{a}_1^p(t)\rangle$ because $\bm{r}^p(t)$ is a functional of ${a}_1^p(t)$. 

In previous works, we have have studied \eqref{ADFeq} in detail for the monodisperse case $\Delta St=0$ \citep{bragg14b,bragg14d,bragg14e}. In that case we have demonstrated that clustering (i.e. non-uniform $\varphi$) arises through $-St_2\varphi\bm{\nabla_r\cdot}\langle w^p_1(t)\bm{w}^p(t)\rangle_{\bm{r}}<{0}$, corresponding to a net drift of the particle-pair towards each other. When $St_2\ll1$, this inward drift arises from the preferential sampling of strain over rotation along their trajectory \citep{chun05}, associated with the centrifuge effect \citep{maxey87}. However, when $St_2\geq O(1)$, a fundamentally different clustering mechanism begins to dominate: Particle-pairs that are approaching carry a memory of more energetic turbulence along their path-history than particles that are separating, and this asymmetry gives rise to a net inward drift leading to clustering.

The clustering of bidisperse particles for $Fr=\infty$ has been considered previously in theoretical studies, and quantitative predictions have been developed \citep{chun05,zaichik06a}. Here we will simply attempt to qualitatively consider how gravity affects the clustering of bidisperse particles, leaving a quantitative theoretical analysis for future work.

Under standard closure assumptions, the terms $\langle\Delta{u}_1^p(t)\rangle_{\bm{r}}$ and $\langle{a}_1^p(t)\rangle_{\bm{r}}$ can be modeled as diffusive fluxes of the form \citep{chun05,zaichik06a} 
\begin{align}
\varphi\langle\Delta{u}_1^p(t)\rangle_{\bm{r}}&\approx -\bm{\mathcal{D}}^{\Delta u}\bm{\cdot\nabla_r}\varphi,\label{Duclos}\\
\Delta St\varphi\langle{a}_1^p(t)\rangle_{\bm{r}}&\approx -(\Delta St)^2\bm{\mathcal{D}}^{a}\bm{\cdot\nabla_r}\varphi,\label{aclos}
\end{align}
where $\bm{\mathcal{D}}^{\Delta u}$ and $\bm{\mathcal{D}}^{a}$ are diffusion tensors that depend upon the properties of $\Delta{u}_1^p(t)$ and ${a}_1^p(t)$, respectively. Their particular form is not important for this qualitative discussion; the important point is that $\bm{\mathcal{D}}^{\Delta u}$ decreases with decreasing $\|\bm{r}\|$, whereas $\bm{\mathcal{D}}^{a}$ does not depend upon $\bm{r}$. Using \eqref{Duclos} and \eqref{aclos} in \eqref{ADFeq} we obtain
\begin{align}
\bm{0}=-\Big(St_2\langle w^p_1(t)\bm{w}^p(t)\rangle_{\bm{r}}+\bm{\mathcal{D}}^{\Delta u}+(\Delta St)^2\bm{\mathcal{D}}^{a}\Big)\bm{\cdot\nabla_r}\varphi-St_2\varphi\bm{\nabla_r\cdot}\langle w^p_1(t)\bm{w}^p(t)\rangle_{\bm{r}}.\label{ADFeq2}
\end{align}
We first consider bidisperse clustering for $Fr=\infty$. In this case, the main effect of bidispersity on the clustering is through the additional diffusion contribution $(\Delta St)^2\bm{\mathcal{D}}^{a}$. In the limit of small-separations, $\|\bm{\mathcal{D}}^{a}\|\gg\|\bm{\mathcal{D}}^{\Delta u}\|$, and the enhanced diffusion due to the acceleration contribution leads to a reduction in the clustering as $|\Delta St|$ is increased, and to a plateau of  $\varphi$ at small separations. This reduced clustering due to the particle acceleration was captured in the theoretical analysis by \cite{chun05} for the regime $St_2\ll1$ and $|\Delta St|\ll 1$. This reduced clustering due to the particle accelerations carries over to the $|\Delta St|\geq O(1)$ regime. However, in addition to this, in the regime $|\Delta St|\geq O(1)$ the clustering will also be affected implicity through the effect of $a^p_1(t)$ on $\langle w^p_1(t)\bm{w}^p(t)\rangle_{\bm{r}}$ and $\bm{\nabla_r\cdot}\langle w^p_1(t)\bm{w}^p(t)\rangle_{\bm{r}}$. As discussed earlier, bidispersity leads both to an enhancement of $\bm{w}^p(t)$, and also causes $\bm{w}^p(t)$ to depend weakly upon separation when $|\Delta St|\geq O(1)$, i.e. where the acceleration effect dominates the behavior of $\bm{w}^p(t)$. These effects lead to an enhancement of $\langle w^p_1(t)\bm{w}^p(t)\rangle_{\bm{r}}$ and a reduction of $\bm{\nabla_r\cdot}\langle w^p_1(t)\bm{w}^p(t)\rangle_{\bm{r}}$, and thus to further suppression of clustering when $|\Delta St|\geq O(1)$. For these reasons, we expect the clustering of monodisperse particles to be stronger than that for bidisperse particles, something that has already been observed in previous DNS studies \citep[e.g.][]{zww01,pan11}.

We now turn to consider the effect of gravity on the clustering. In the monodisperse case, the effect of gravity on the clustering was found to be non-trivial, with gravity leading to either reduced or enhanced clustering depending upon $St,Fr$ \citep{bec14b,gustavsson14,ireland16b}. The theoretical explanation given for this in \cite{ireland16b} can be summarized conceptually as follows: Gravity leads to a reduction in the correlation time of the flow seen by the inertial particles. This reduces the effectiveness of the centrifuge mechanism, and also reduces the path-history effect since it reduces the temporal span over which the particles are affected by their past interaction with the turbulence. Both of these effects lead to reduced clustering. Now, although the path-history effect generates strong clustering over a range of $St$, when $St$ is large enough the path-history effect begins to destroy the clustering since the particle relative motion becomes increasingly ballistic. However, since gravity reduces the path-history effect, one has to go to larger $St$ before the path-history effect begins to destroy the clustering. As a result, at large enough $St$, gravity has the effect of enhancing clustering relative to the $Fr=\infty$ case.

In the bidisperse case, the clustering enhancement due to gravity may occur when $St_2\geq O(1)$ if $|\Delta St|$ is sufficiently small and $Fr$ is not too small. However, in general, we expect that gravity will reduce the clustering, and not enhance it. This is because, as explained earlier, as $Fr$ is decreased, $\bm{a}^p(t)$ increases, and $\bm{w}^p(t)$ both increases and becomes less dependent upon the particle separation. Thus, in terms of \eqref{ADFeq2}, decreasing $Fr$ both enhances the diffusion contributions $\langle w^p_1(t)\bm{w}^p(t)\rangle_{\bm{r}}$ and $\bm{\mathcal{D}}^{a}$, and reduces the drift $\bm{\nabla_r\cdot}\langle w^p_1(t)\bm{w}^p(t)\rangle_{\bm{r}}$, leading to a suppresion of clustering.


\section{Computational Details}\label{CompD}
DNS of isotropic turbulence are performed on a three-dimensional periodic cube using a pseudospectral method. 
The cubic domain of length $\mathscr{L}$ is uniformly discretized using $N^3$ grid points. The fluid velocity field $\bm{u}(\bm{x},t)$ 
is obtained by solving the incompressible Navier-Stokes equation
\begin{eqnarray}
\partial_t\bm{u} + {\bm{\omega}} \times \bm{u} 
+ \bm{\nabla}\left ( \frac{p}{\rho_f} + \frac{\|\bm{u}\|^{2}}{2}  \right )  = \nu \bm{\nabla}^2 \bm{u} + \bm{f}, \label{eq:ns}
\end{eqnarray}
where $\bm{\omega} \equiv \bm{\nabla} \times \mathbf{u}$ is the vorticity, $\rho_f$ is the fluid
density, $p$ is the pressure (determined using $\bm{\nabla} \bcdot \bm{u} = 0$), $\nu$ is the kinematic viscosity and  $\bm{f}$ is the external forcing applied to generate statistically stationary turbulence.
A deterministic forcing scheme is used for $\bm{f}$ \citep {witkowska97}, where the energy dissipated during one time step is resupplied to the 
wavenumbers with magnitude $\kappa=\sqrt{2}$. Time integration is performed through a second-order, explicit Runge-Kutta scheme with aliasing errors removed by means of a combination of 
spherical truncation and phase-shifting. 
\begin{table}
\centering
\renewcommand{\arraystretch}{1.1}
\setlength{\tabcolsep}{32pt}
  \begin{tabular}{cccc}
    $\mathrm{Parameter}$ & $\mathrm{DNS}$~I  & $\mathrm{DNS}$~II & $\mathrm{DNS}$~III \\
       $N$ & 128 & 128 & 1024 \\
       $R_\lambda$ &  93 &  94 & 90 \\
       $Fr$ & $\infty$ & 0.3 & 0.052 \\
       $\mathscr{L}$ & 2$\upi$ & 2$\upi$ & 16$\upi$\\       
       $\nu$ & 0.005 &  0.005 & 0.005 \\
       $\epsilon$ & 0.324 &  0.332 & 0.257 \\
       $l$ & 1.48 & 1.49 & 1.47 \\
       $l/\eta$ & 59.6 & 60.4 & 55.6 \\
       $u'$ & 0.984 & 0.996 & 0.912 \\
       $u'/u_\eta$ & 7.92 & 8.12 & 4.82 \\
       $T_L$ & 1.51 &  1.50 & 1.61 \\
       $T_L/\tau_\eta$ & 12.14 & 12.24 & 11.52 \\
       $\kappa_{{\rm max}}\eta$ & 1.5 &  1.48 & 1.61\\
       $N_p$ & 262,144 & 262,144 & 16,777,216\\
  \end{tabular}
  \caption{Flow parameters in DNS of isotropic turbulence (arbitrary units).
          $N$ is the number of grid points in each direction, 
          $R_\lambda \equiv u'\lambda/\nu$ is the Taylor micro-scale
          Reynolds number,
          $Fr \equiv \epsilon^{3/4}/(\nu^{1/4}\mathfrak{g})$ is the Froude number, $\mathscr{L} $ is the domain size, $\nu$ is the fluid kinematic viscosity, $\epsilon \equiv 2\nu \int_0^{\kappa_{\rm max}}\kappa^2 E(\kappa) {\rm d}\kappa $ is the mean
          turbulent kinetic energy dissipation rate,
          $l \equiv 3\upi/(2k)\int_0^{\kappa_{\rm max}}E(\kappa)/\kappa {\rm d}\kappa $  is the integral length scale, $\eta \equiv \nu^{3/4}/\epsilon^{1/4}$ is the Kolmogorov length scale, 
           $u' \equiv \sqrt{(2k/3)}$ is the fluid r.m.s. fluctuating 
          velocity, $k$ is the turbulent kinetic energy, 
          $u_\eta$ is the Kolmogorov velocity scale, 
          $T_L \equiv l/u^\prime$ is the large-eddy turnover
          time, $\tau_\eta \equiv \sqrt{(\nu/\epsilon)}$ is the Kolmogorov time scale, 
          $\kappa_{\rm max}$ is the maximum
          resolved wavenumber, and $N_p$ is  
          the number of particles per Stokes number.}
  {\label{tab:parameters}}
\end{table}
Particles are tracked in the flow field using \eqref{eq:part2} for their equation of motion. Fifteen particle classes are simulated with Stokes number $St$ ranging from 0 to 3. In \eqref{eq:part2}, $\bm{u}(\bm{x}^p(t),t)$ 
is the fluid velocity at the particle position, and this must be evaluated by interpolating the fluid velocity at the surrounding grid points
to $\bm{x}^p(t)$. In this study we use an $8^{th}$-order, B-spline interpolation method which provides a good balance between high-accuracy and efficiency \citep[see][]{ireland13}. Further details on all aspects of the computational methods can be found in \cite{ireland13}. 

Since we want to explore the role of gravity on the small-scale motion of the bidisperse particles, we must consider the choice of $Fr$ for the DNS. Observations have shown that $\epsilon$ can vary by orders of magnitude in clouds \citep{prupp97}, with corresponding variations in $Fr$. Therefore, in addition to the zero gravity case $Fr=\infty$, we follow \cite{ireland16a} and consider $Fr= 0.3, 0.052$, which may be considered to be representative of strongly turbulent cumulonimbus clouds and weakly turbulent stratiform clouds, respectively \citep{pinsky07}. 

One of the difficulties in simulating very small values of $Fr$ is that the use of periodic boundary conditions in the DNS can artificially influence the motion of inertial particles if the DNS box length $\mathscr{L}$ is too small. In particular, the use of periodic boundary conditions is problematic if the time it takes the settling particles to traverse the distance $\mathscr{L}$ is $\leq O(T_L)$. This issue was explored in detail in \cite{ireland16a} and it was found that box sizes much larger than the standard $\mathscr{L}=2\pi$ are needed when $Fr$ is very small. For example, in \cite{ireland16a} it was found that for $R_\lambda\approx 90$, $\mathscr{L}=16\pi$ was necessary, which, with the resolution constraints for accurately resolving the small-scales requires $N=1024$. Such requirements place significant limitations on the value of $R_\lambda$ that can be simulated. In this study we will consider $R_\lambda\approx 90$, where $\mathscr{L}=16\pi$ is used for $Fr=0.052$, whereas the standard size $\mathscr{L}=2\pi$ is used for the larger values $Fr=\infty, 0.3$. Although the higher-order statistics of the particle motion are strongly affected by $R_\lambda$, our recent studies in \citet{ireland16a,ireland16b} showed that at the small-scales, the lower-order statistics (e.g. relative velocity variances, collision rates etc), were insensitive to $R_\lambda$ over the range $90\lesssim R_\lambda\lesssim 600$ when $St\leq O(1)$. This justifies the use of the low $R_\lambda$ in this study, but points to the need for future studies to explore the role of $R_\lambda$ on the higher-order statistics of settling, bidisperse particles at the small-scales of turbulence.

Details of the DNS are summarized in Table~\ref{tab:parameters}.

\section{Results and discussion}\label{RD}
\subsection{Relative Velocities}
We begin by considering results for the relative velocities of the particles. The parallel (longitudinal) and perpendicular (transverse) projections of $\bm{w}^p(t)$ are defined as $w^p_\parallel(t)\equiv \bm{e}^p_\parallel(t)\bm{\cdot}\bm{w}^p(t)$ and $w^p_\perp(t)\equiv \bm{e}^p_\perp(t)\bm{\cdot}\bm{w}^p(t) $, where $\bm{e}^p_\parallel(t),\bm{e}^p_\perp(t)$ are unit vectors parallel and perpendicular to $\bm{r}^p(t)$, respectively. 

\begin{figure}\vspace{0.1in}
\psfrag{a}[cc][2]{$r/\eta$}
  \psfrag{b}[cc][2]{$S_{2\parallel}/u_\eta^2$}
    \psfrag{TT}[cc][2]{Increasing $St_2$}
    \centering
    \begin{subfigure}[b]{0.5\textwidth}
        \centering
        \hspace{-0.0in}
        \includegraphics[width=0.9\textwidth]{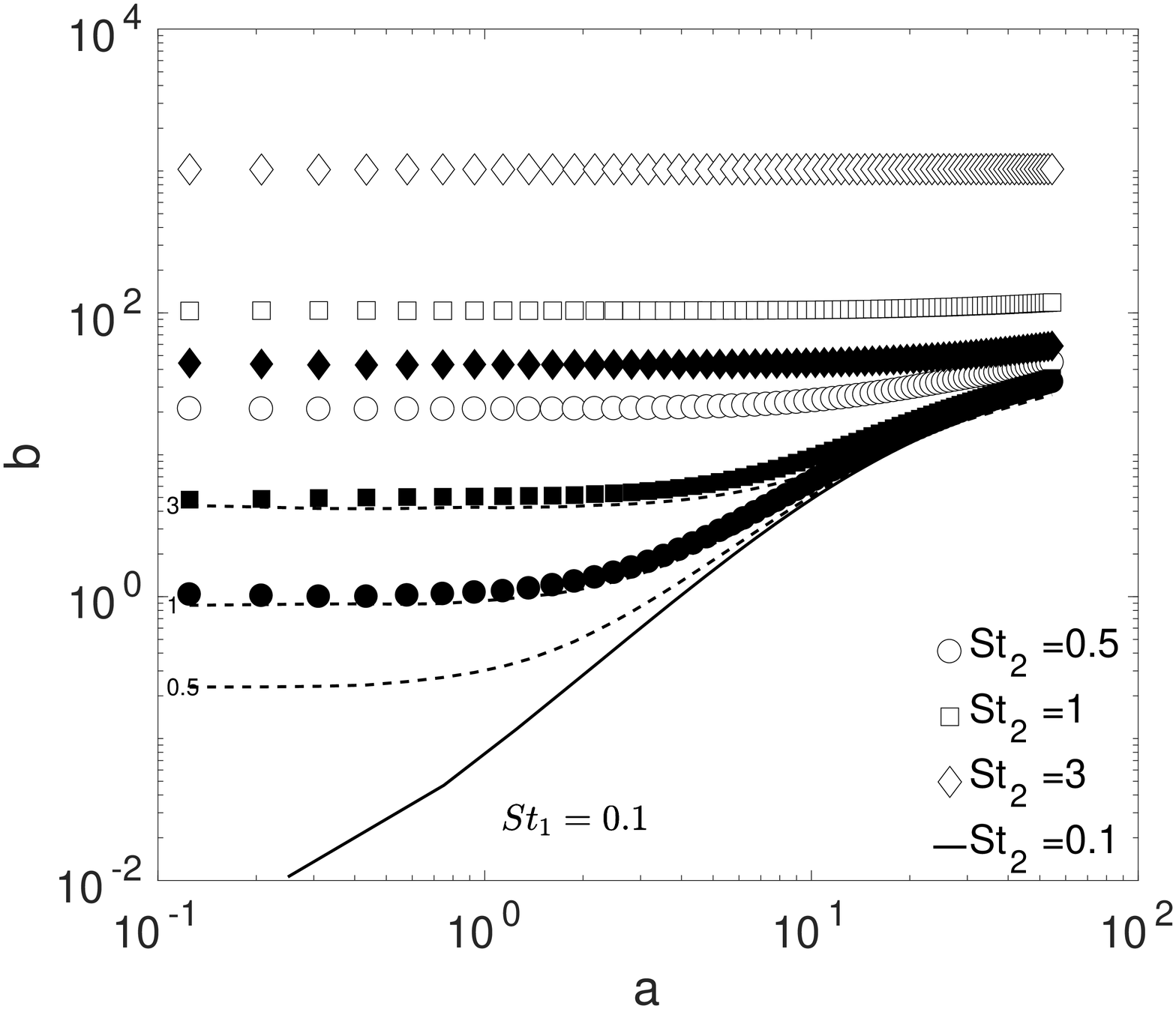}
        \caption{}
    \end{subfigure}%
    ~
    \begin{subfigure}[b]{0.5\textwidth}
        \centering
	    \includegraphics[width=0.9\textwidth]{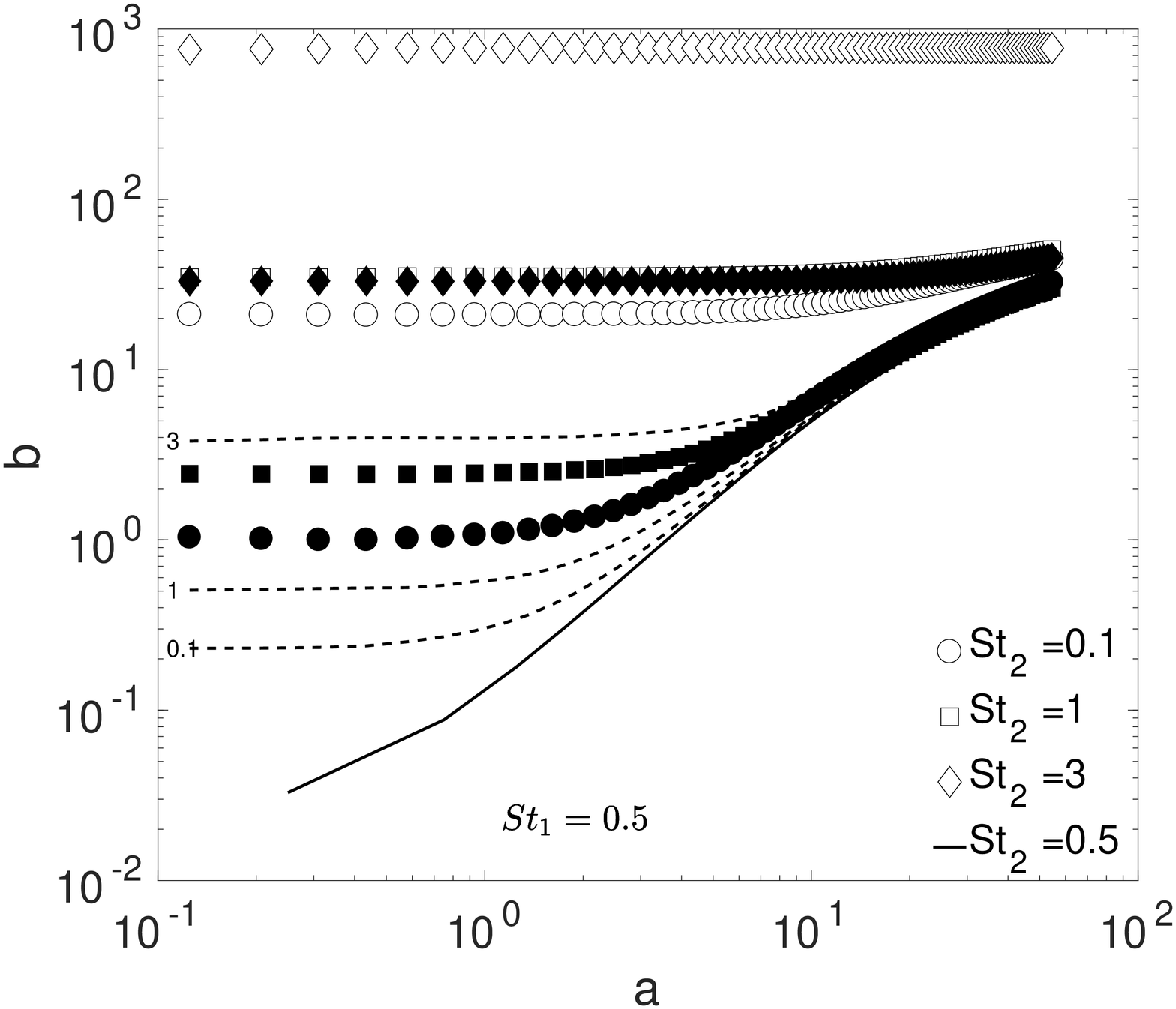}        
	    \caption{}
    \end{subfigure}

\vspace{0.1in}
    \begin{subfigure}[b]{0.5\textwidth}
        \centering
		\includegraphics[width=0.9\textwidth]{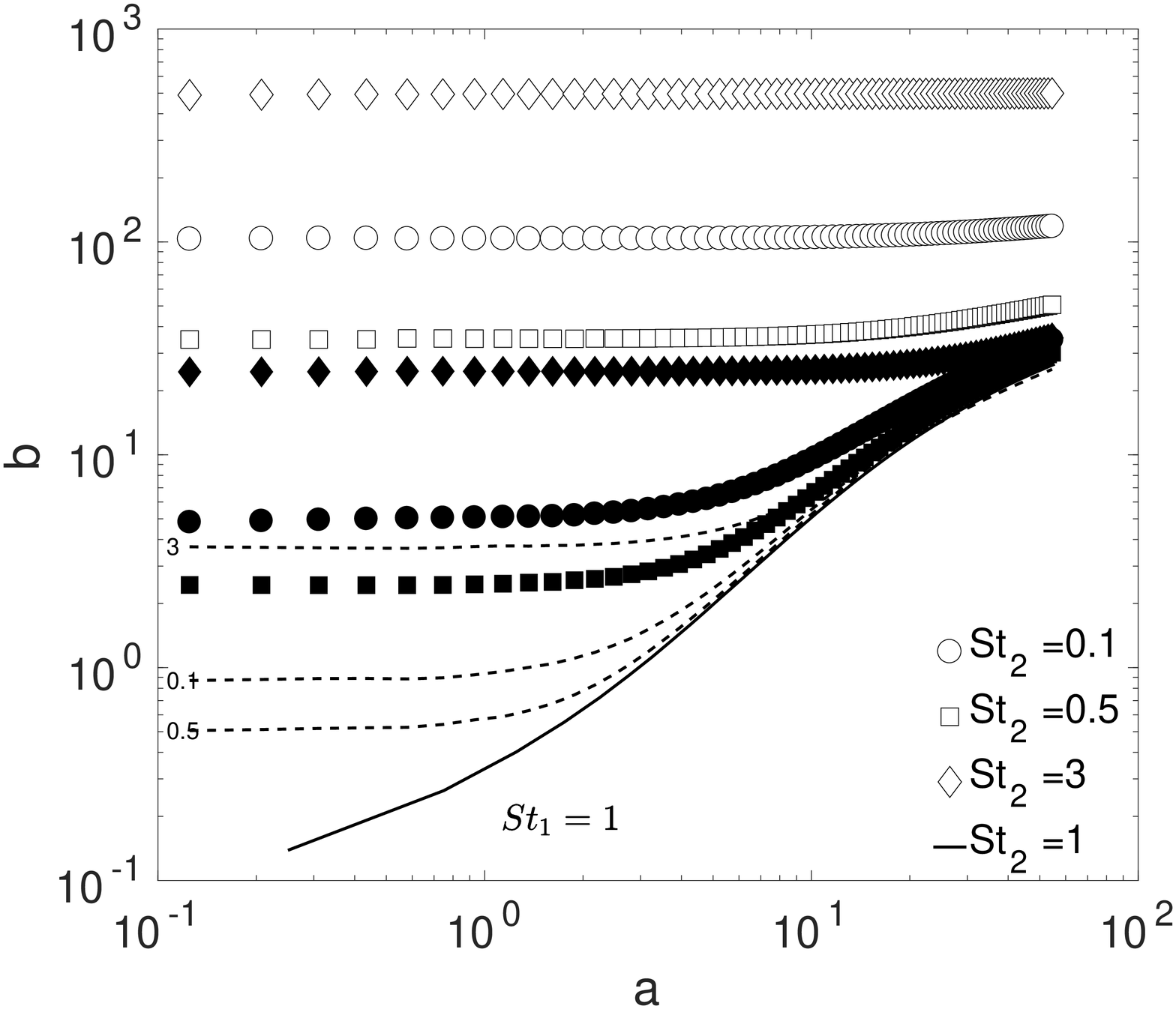}        
		\caption{}
    \end{subfigure}%
    ~
    \begin{subfigure}[b]{0.5\textwidth}
        \centering
         \includegraphics[width=0.9\textwidth]{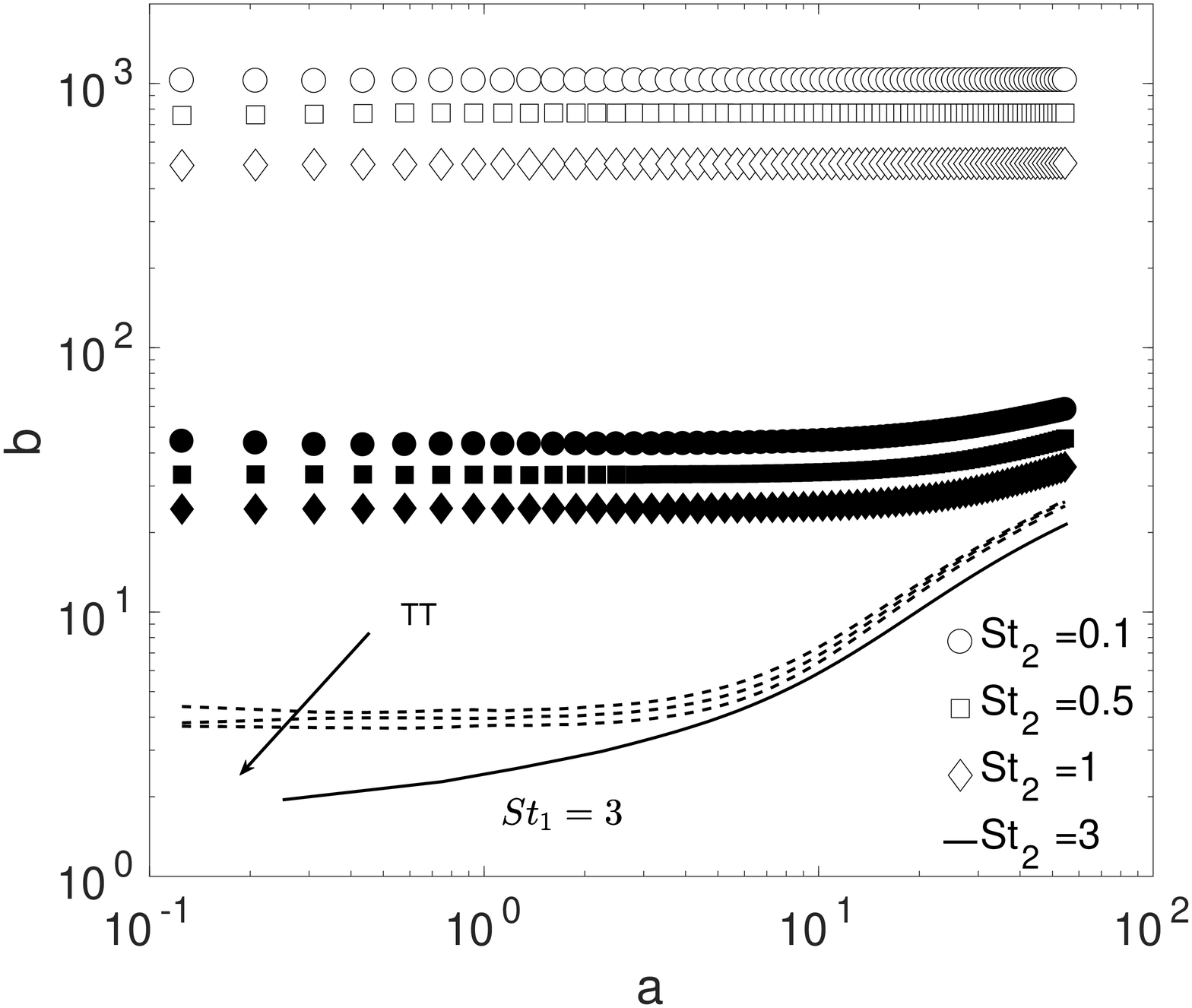}    
          \caption{}
    \end{subfigure}

      \caption{DNS results for $S_{2\parallel}/u_\eta^2$ as a function of $r/\eta$ for different $St_1, St_2, Fr$ combinations. Dashed lines correspond to $Fr=\infty$, filled symbols to $Fr=0.3$, open symbols to $Fr=0.052$, and solid lines correspond to the monodisperse case with $Fr=\infty$. The numbers next to the dashed lines indicate the $St_2$ value that they correspond to.}
  \label{fig:par_var_vs_r}
\end{figure}

Figure~\ref{fig:par_var_vs_r} shows the DNS results for the second order, longitudinal velocity structure function $S_{2\parallel}(r)\equiv \langle[w^p_\parallel(t)]^2 \rangle_r$, where $\langle \cdot \rangle_r$ denotes an ensemble average conditioned on $\|\bm{r}^p(t)\| = r$. We also include the monodisperse results with $Fr=\infty$ for comparison. It can be seen that the bidisperse values, with and without gravity, are bounded from below by the monodisperse values. This is in agreement with previous studies that considered bidisperse particle motion without gravitational effects \citep{zww01,zaichik06a,zaichik09b,pan10,pan14}. We also observe in figure~\ref{fig:par_var_vs_r} that $S_{2\parallel}(r)$ becomes independent of $r$ at sufficiently small separations. Both of these effects arise because of the acceleration contribution to the bidisperse particle motion, as explained in \S\ref{theory}. 

\begin{figure}\vspace{0.1in}
\psfrag{a}[cc][2]{$r/\eta$}
  \psfrag{b}[cc][2]{$S_{2\perp}/u_\eta^2$}
      \psfrag{TT}[cc][2]{Increasing $St_2$}
    \centering
    \begin{subfigure}[b]{0.5\textwidth}
         \psfrag{L}[cc][2]{$St_1 = 0.4$}
        \centering
        \hspace{-0.0in}
        \includegraphics[width=0.9\textwidth]{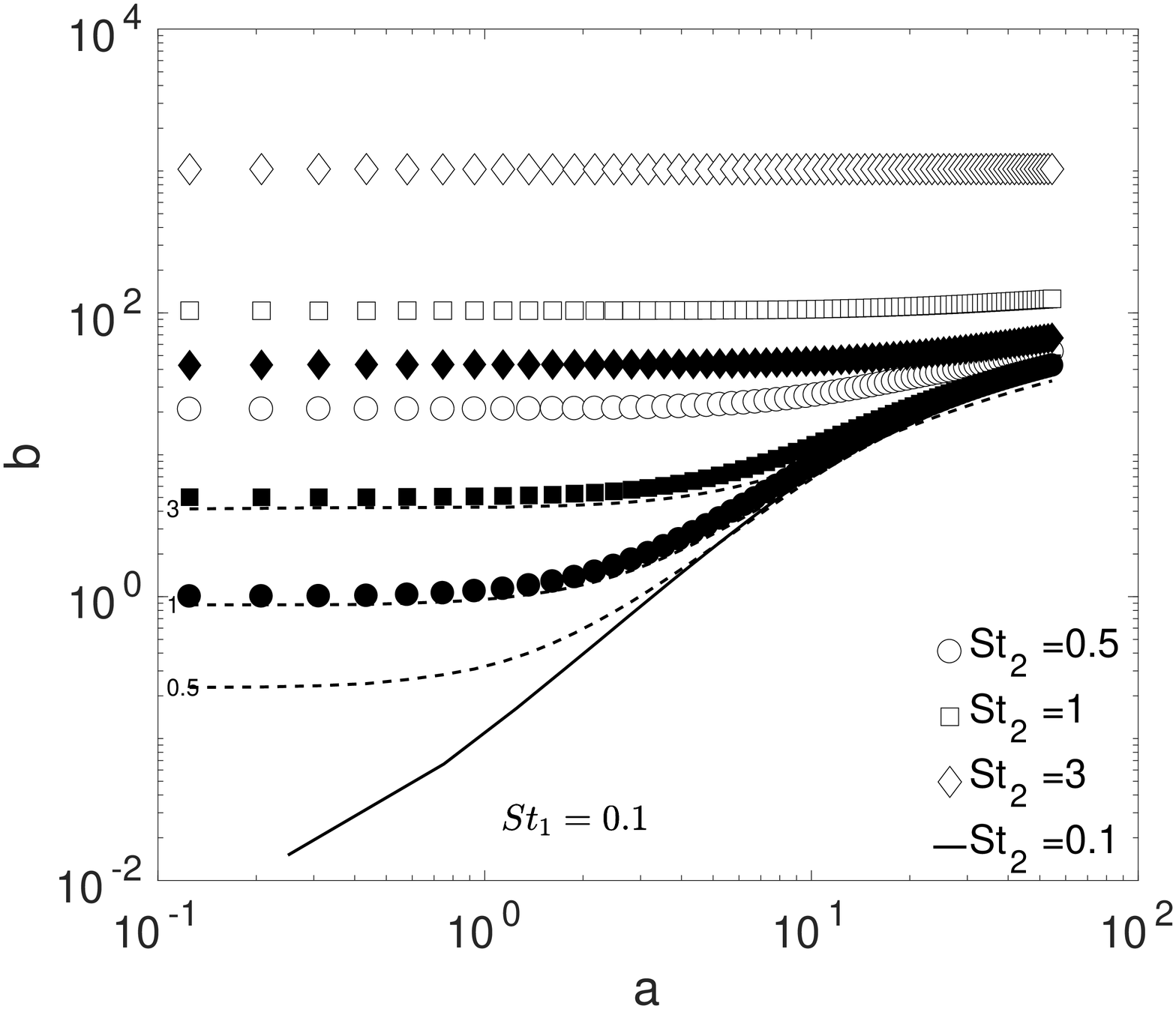}
        \caption{}
    \end{subfigure}%
    ~
    \begin{subfigure}[b]{0.5\textwidth}
     \psfrag{L}[cc][2]{$St_1 = 0.5$}
        \centering
	    \includegraphics[width=0.9\textwidth]{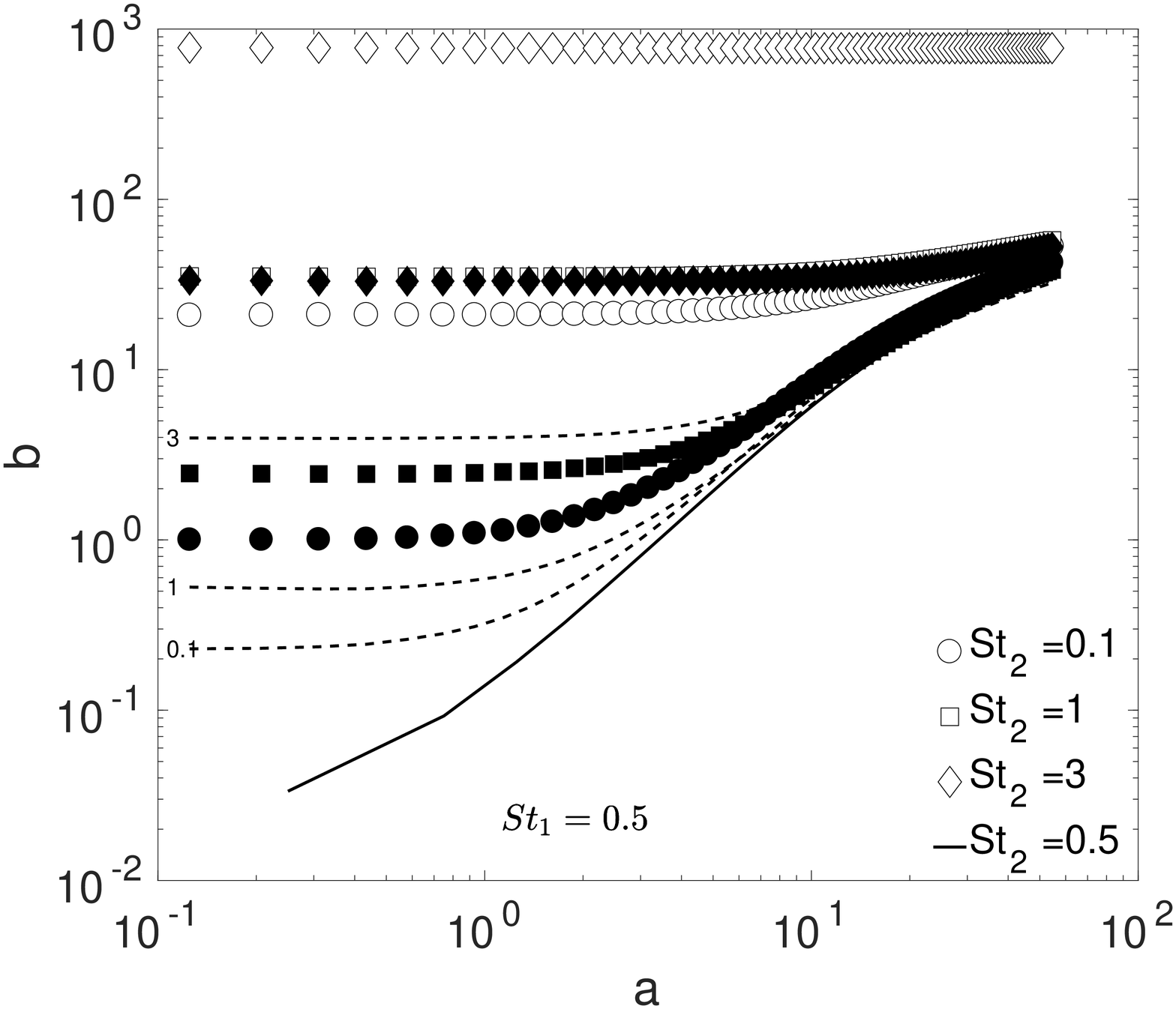}        
	    \caption{}
    \end{subfigure}

\vspace{0.1in}
    \begin{subfigure}[b]{0.5\textwidth}
     \psfrag{L}[cc][2]{$St_1 = 1$}
        \centering
		\includegraphics[width=0.9\textwidth]{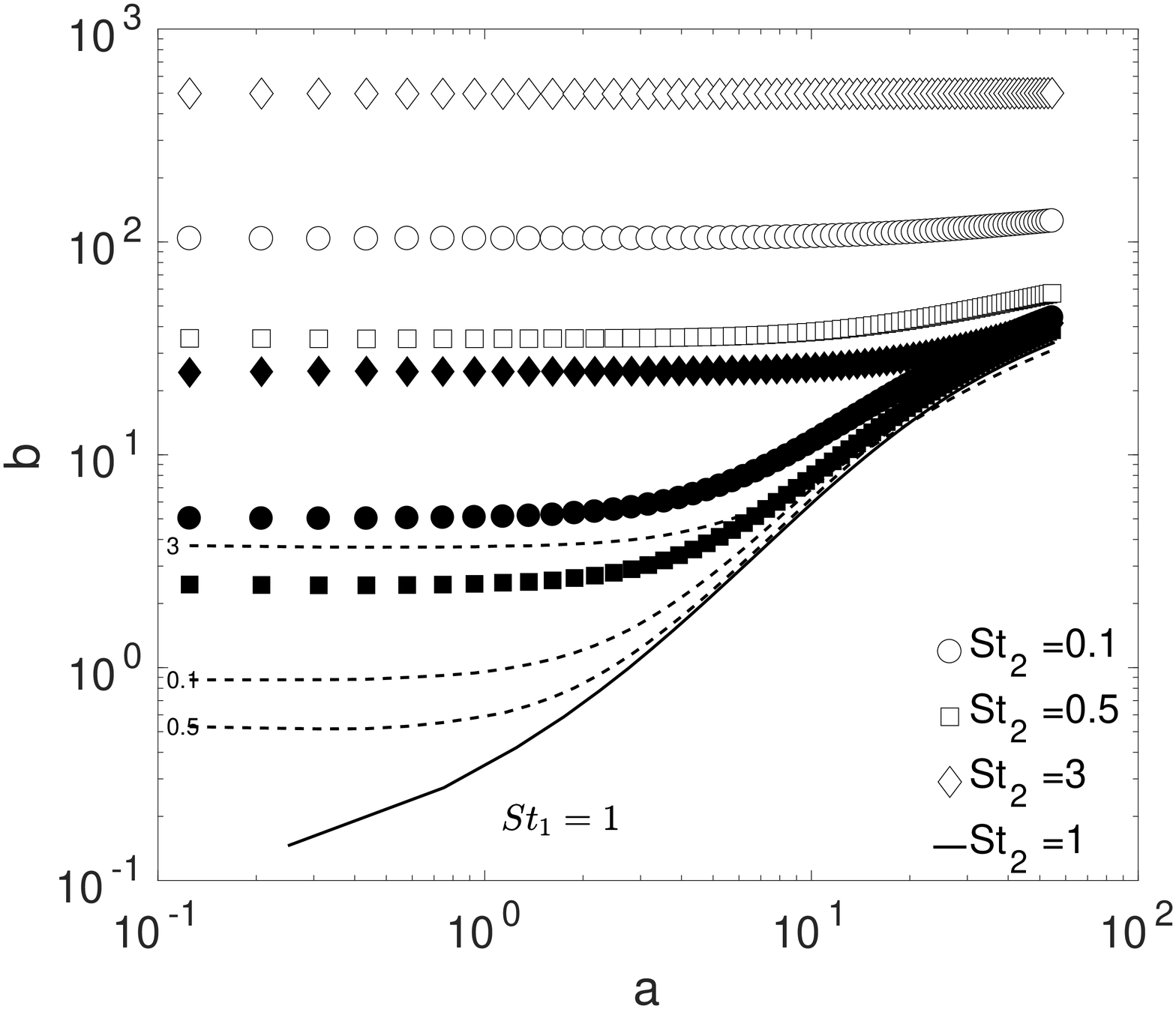}        
		\caption{}
    \end{subfigure}%
    ~
    \begin{subfigure}[b]{0.5\textwidth}
     \psfrag{L}[cc][2]{$St_1 = 3$}
        \centering
         \includegraphics[width=0.9\textwidth]{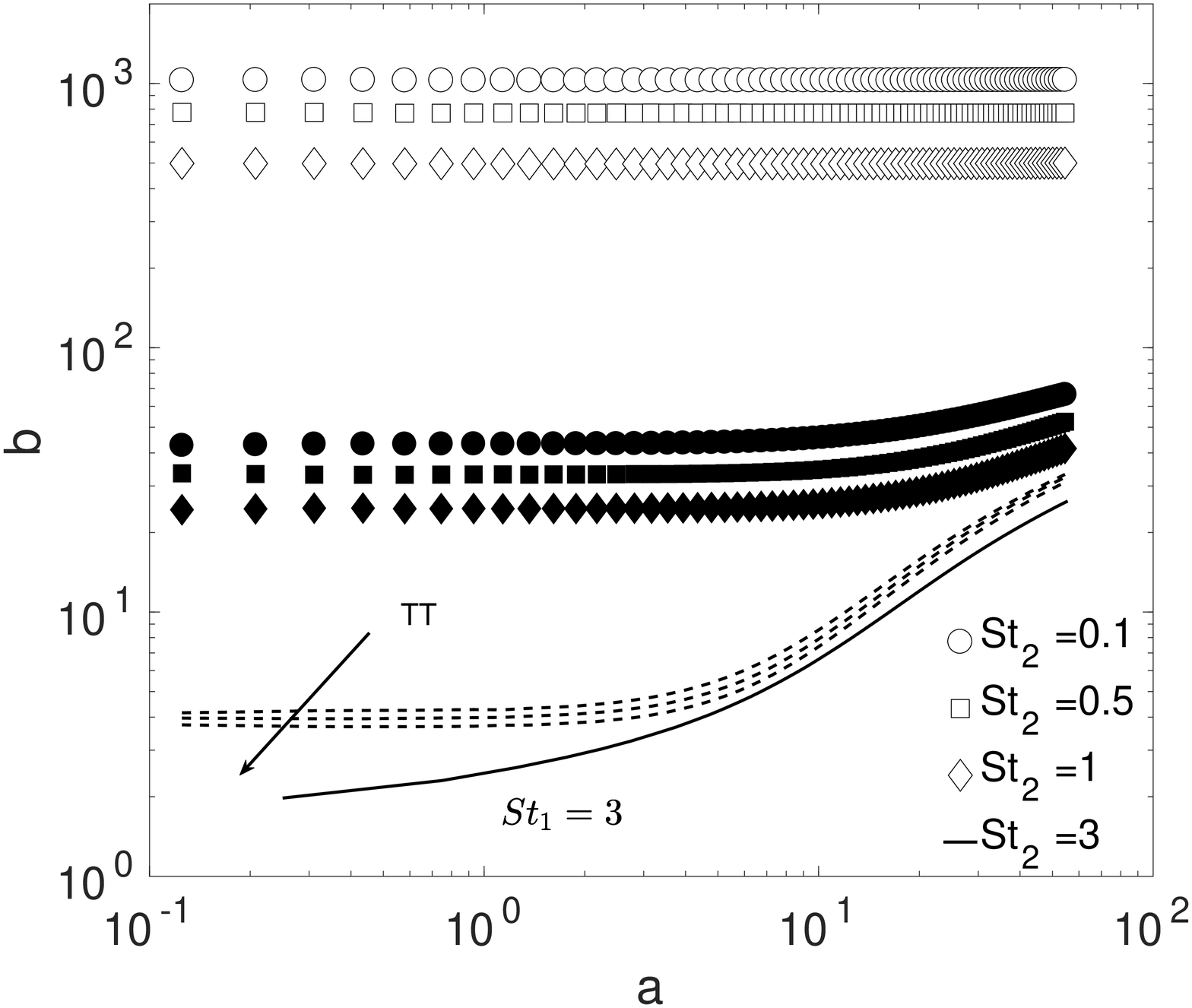}    
          \caption{}
    \end{subfigure}

      \caption{DNS results for $S_{2\perp}/u_\eta^2$ as a function of $r/\eta$ for different $St_1, St_2, Fr$ combinations. Dashed lines correspond to $Fr=\infty$, filled symbols to $Fr=0.3$, open symbols to $Fr=0.052$, and solid lines correspond to the monodisperse case with $Fr=\infty$. The numbers next to the dashed lines indicate the $St_2$ value that they correspond to.}
  \label{fig:per_var_vs_r}
\end{figure}

The enhancement of the relative velocities of bidisperse particles at small separations grows as $Fr$ is reduced and/or $|\Delta St|$ is increased. For sufficiently large $r$, the effect of the bidispersity weakens, corresponding to the dominance of the first term on the rhs of \eqref{eq:sol_w} at these separations. The results for $S_{2\perp}(r)\equiv \langle[w^p_\perp(t)]^2 \rangle_r,$ are shown in figure~\ref{fig:per_var_vs_r}, which are quantitatively similar to those for $S_{2\parallel}(r)$. 

\FloatBarrier

\begin{figure}\vspace{0.1in}
\psfrag{a}[cc][2]{$|w_\parallel |/u_\eta$}
  \psfrag{b}[cc][2]{$\mathcal{P}(|w_\parallel |\vert r)$}
    \centering
    \begin{subfigure}[b]{0.5\textwidth}
         \psfrag{L}[cc][2]{$St_1 = 0.4$}
        \centering
        \hspace{-0.0in}        
        \includegraphics[width=\textwidth]{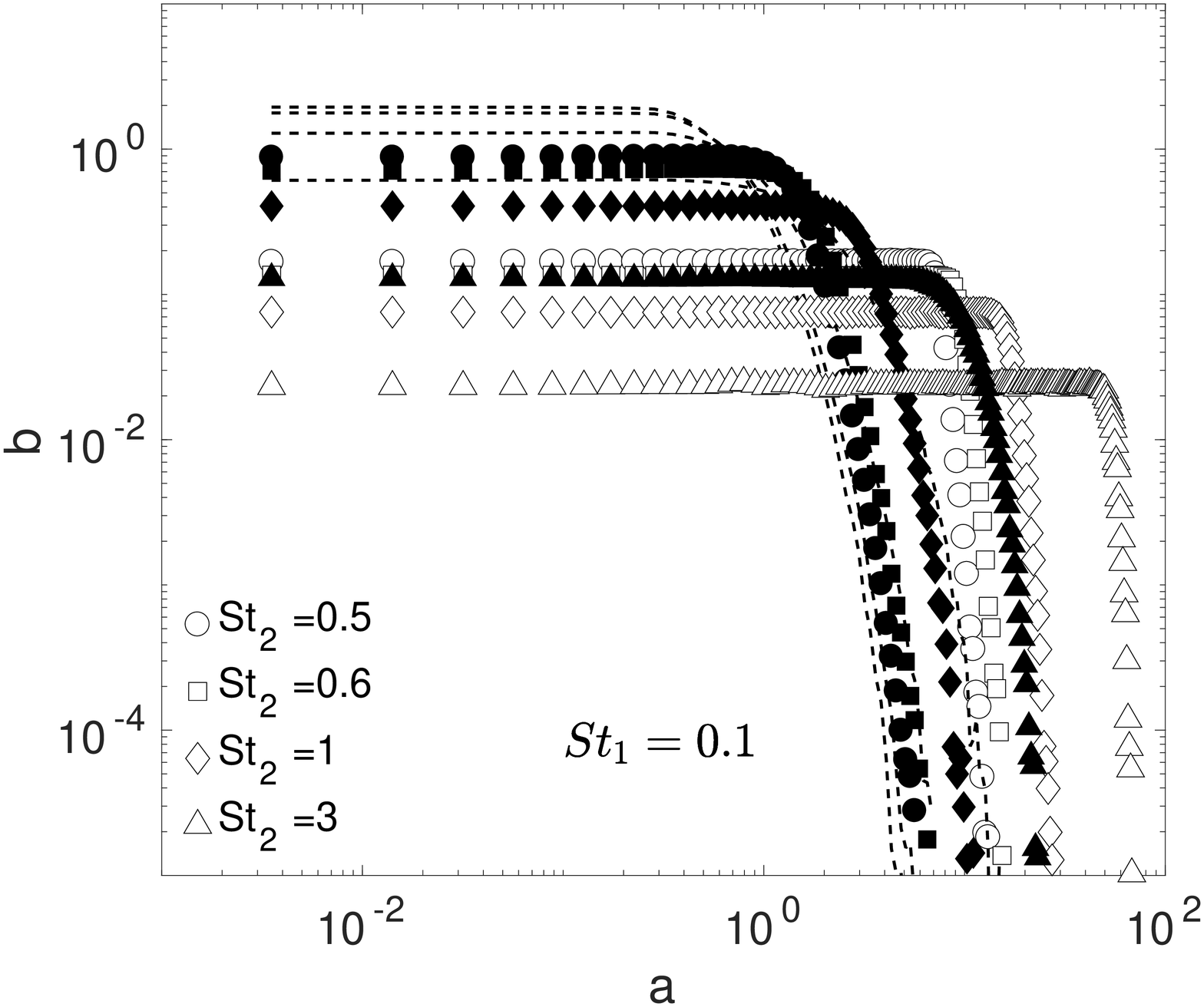}
        \caption{}
    \end{subfigure}%
    ~
    \begin{subfigure}[b]{0.5\textwidth}
     \psfrag{L}[cc][2]{$St_1 = 0.5$}
        \centering
	    \includegraphics[width=\textwidth]{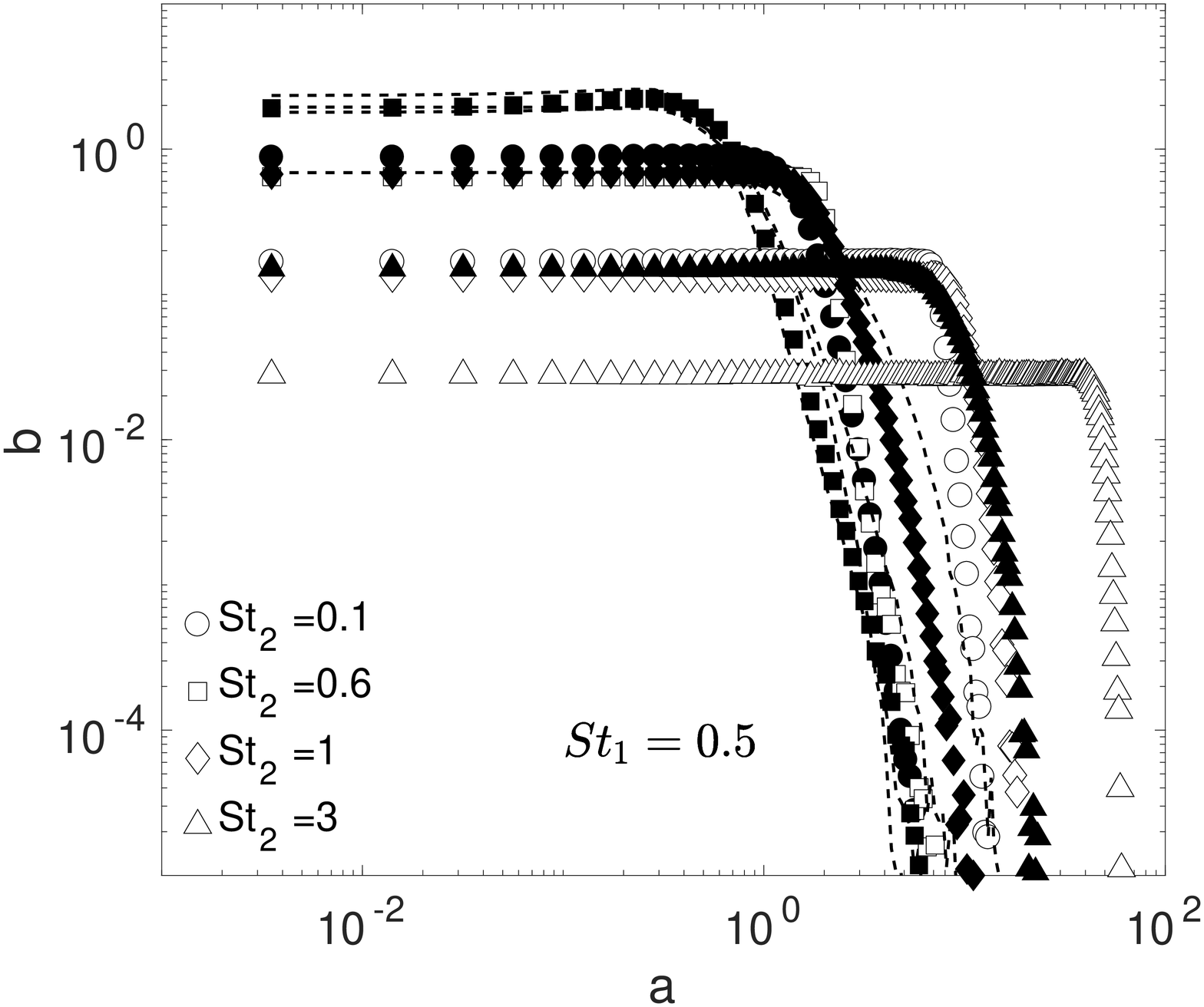}      
	    \caption{}
    \end{subfigure}

\vspace{0.13in}
    \begin{subfigure}[b]{0.5\textwidth}
     \psfrag{L}[cc][2]{$St_1 = 1$}
        \centering
		\includegraphics[width=\textwidth]{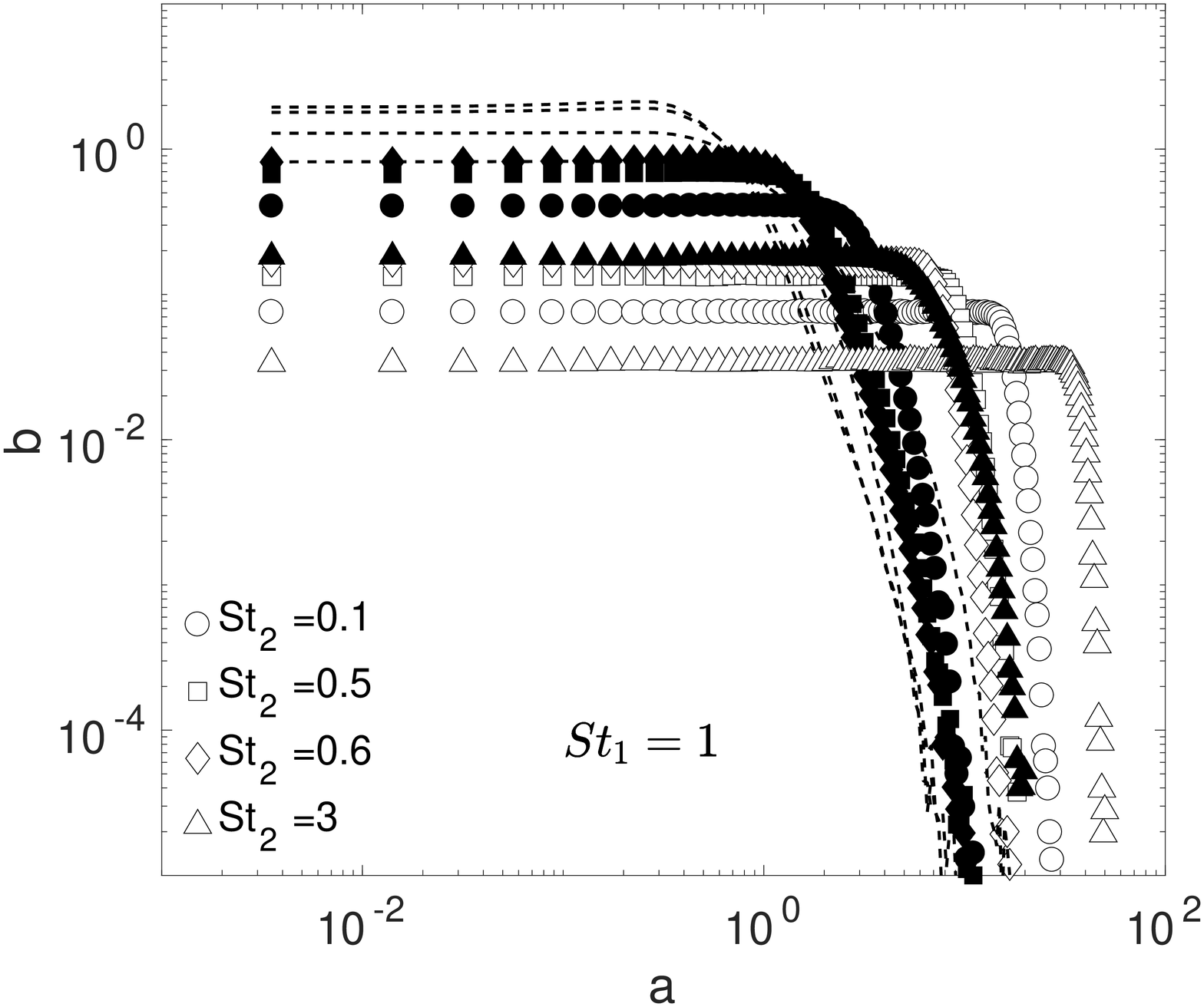}        
		\caption{}
    \end{subfigure}%
    ~
    \begin{subfigure}[b]{0.5\textwidth}
     \psfrag{L}[cc][2]{$St_1 = 3$}
        \centering
         \includegraphics[width=\textwidth]{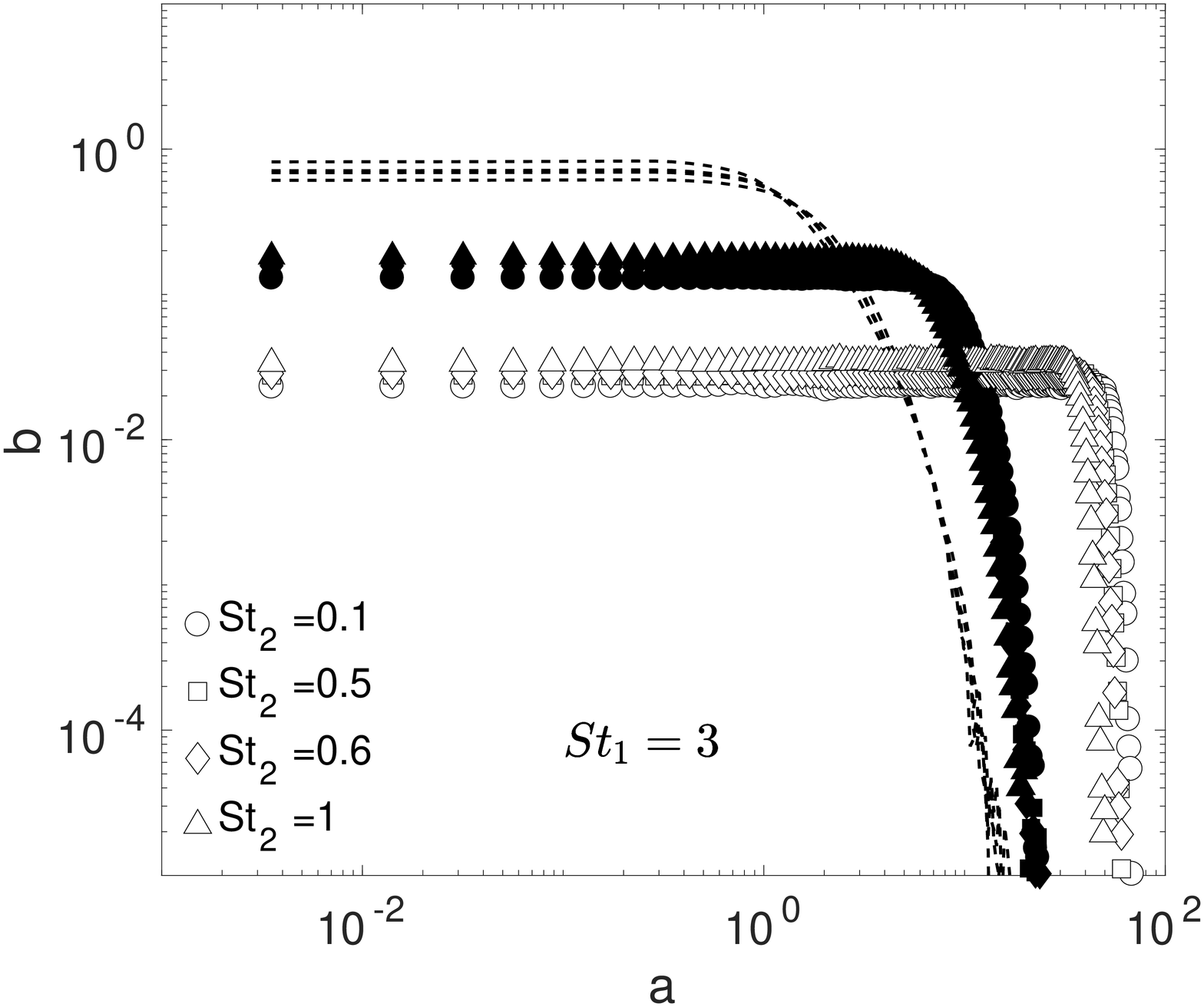}    
          \caption{}
    \end{subfigure}

      \caption{DNS results for $\mathcal{P}(| w_\parallel | |r)$ as a function of $|w_\parallel |/u_\eta$ for different $St_1, St_2, Fr$ combinations, with $r = \eta$. Dashed lines correspond to $Fr=\infty$, filled symbols to $Fr=0.3$, and open symbols to $Fr=0.052$.}
  \label{fig:par_pdf}
\end{figure}

In figure~\ref{fig:par_pdf} we show results for $\mathcal{P}(\vert w_\parallel\vert\vert r)$, the PDF of the absolute value of the longitudinal relative velocities at separation $r$. As observed by \cite{pan14}, for $Fr=\infty$, increasing $|\Delta St|$ leads to a broadening of the PDF, corresponding to the enhanced relative velocities through the acceleration effect. Decreasing $Fr$ leads to further broadening of the PDF. Most strikingly, when $\Delta St$ is sufficiently large, the PDF is almost constant up to $| w_\parallel |\gg u_\eta$ and then suddenly reduces. In \S\ref{theory} we showed that reducing $Fr$ can enhance $\bm{w}^p(t)$ for bidisperse particles in two ways; explicitly through the differential settling contribution $u_\eta\Delta St Fr^{-1}\bm{e}_{\mathfrak{g}}$, and implicitly through the way that gravity enhances the primary particle acceleration $\bm{a}^p$. Since the orientation of $\bm{r}^p(t)$ is time dependent, the effect of gravity on the statistics of the parallel and perpendicular projections of $\bm{w}^p(t)$ is not entirely clear, and will involve a mixture of these explicit and implicit effects.

\FloatBarrier

In order to untangle the explicit and implicit effects of gravity on the statistics of $\bm{w}^p(t)$ and gain further insight, we consider the second order structure functions based on the Cartesian components of $\bm{w}^p(t)$. That is, we consider ${\langle [w^p_3(t)]^2 \rangle_r}$ and ${\langle [w^p_1(t)]^2 \rangle_r}$, corresponding to the velocities in the vertical and horizontal directions, respectively (recall also that ${\langle [w^p_1(t)]^2 \rangle_r}={\langle [w^p_2(t)]^2 \rangle_r}$ due to the axisymmetry of the statistics). 

\begin{figure}\vspace{0.1in}
\psfrag{a}[cc][2]{$r/\eta$}
  \psfrag{b}[cc][2]{$\langle [w^p_{1,3}(t)]^2\rangle _r/u_\eta^2$}
    \centering
    \begin{subfigure}[b]{0.5\textwidth}
         \psfrag{L}[cc][2]{$St_1 = 0.4$}
        \centering
        \hspace{-0.0in}
        \includegraphics[width=\textwidth]{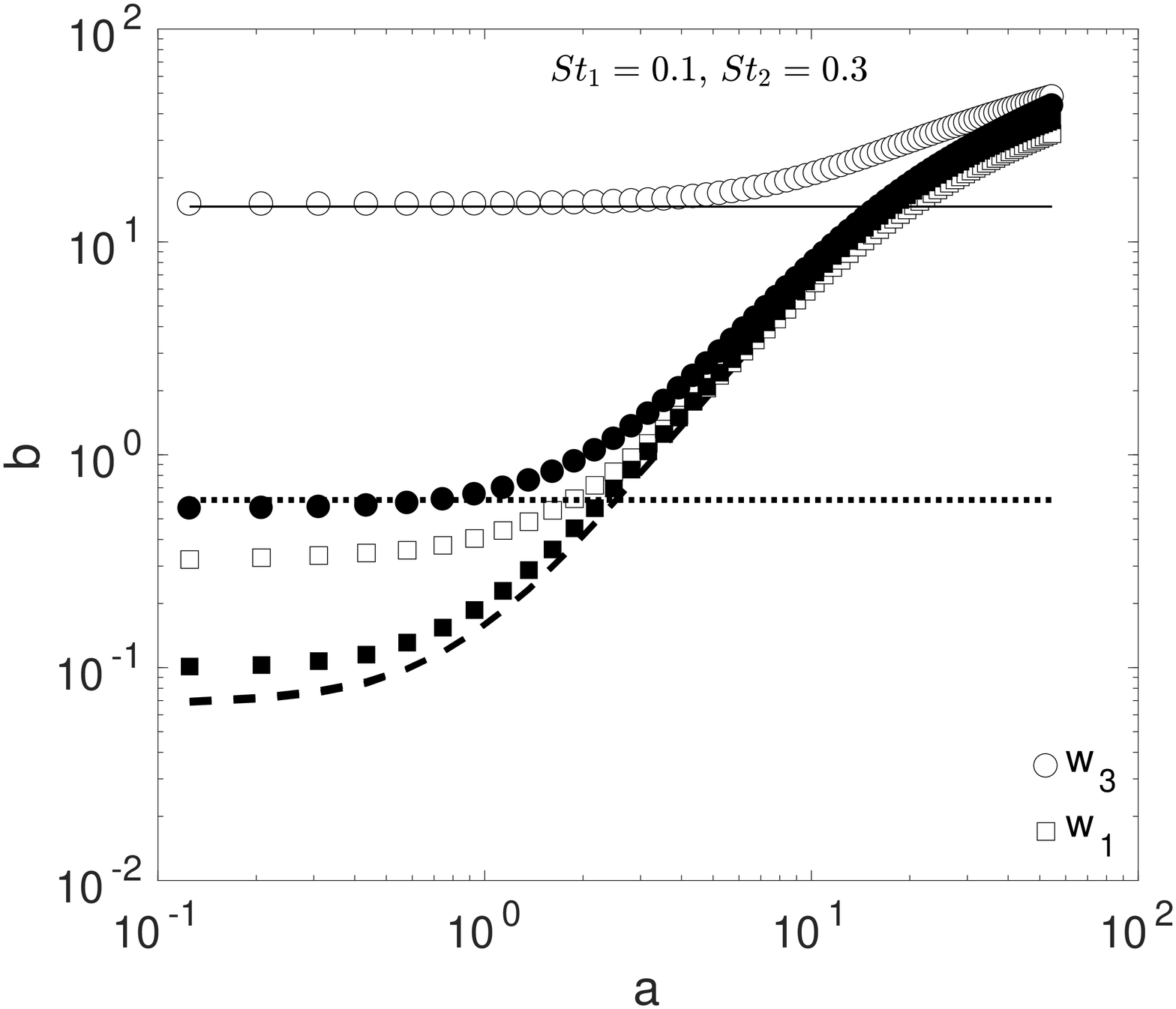}
        \caption{}
    \end{subfigure}%
    ~
    \begin{subfigure}[b]{0.5\textwidth}
     \psfrag{L}[cc][2]{$St_1 = 0.5$}
        \centering
	    \includegraphics[width=\textwidth]{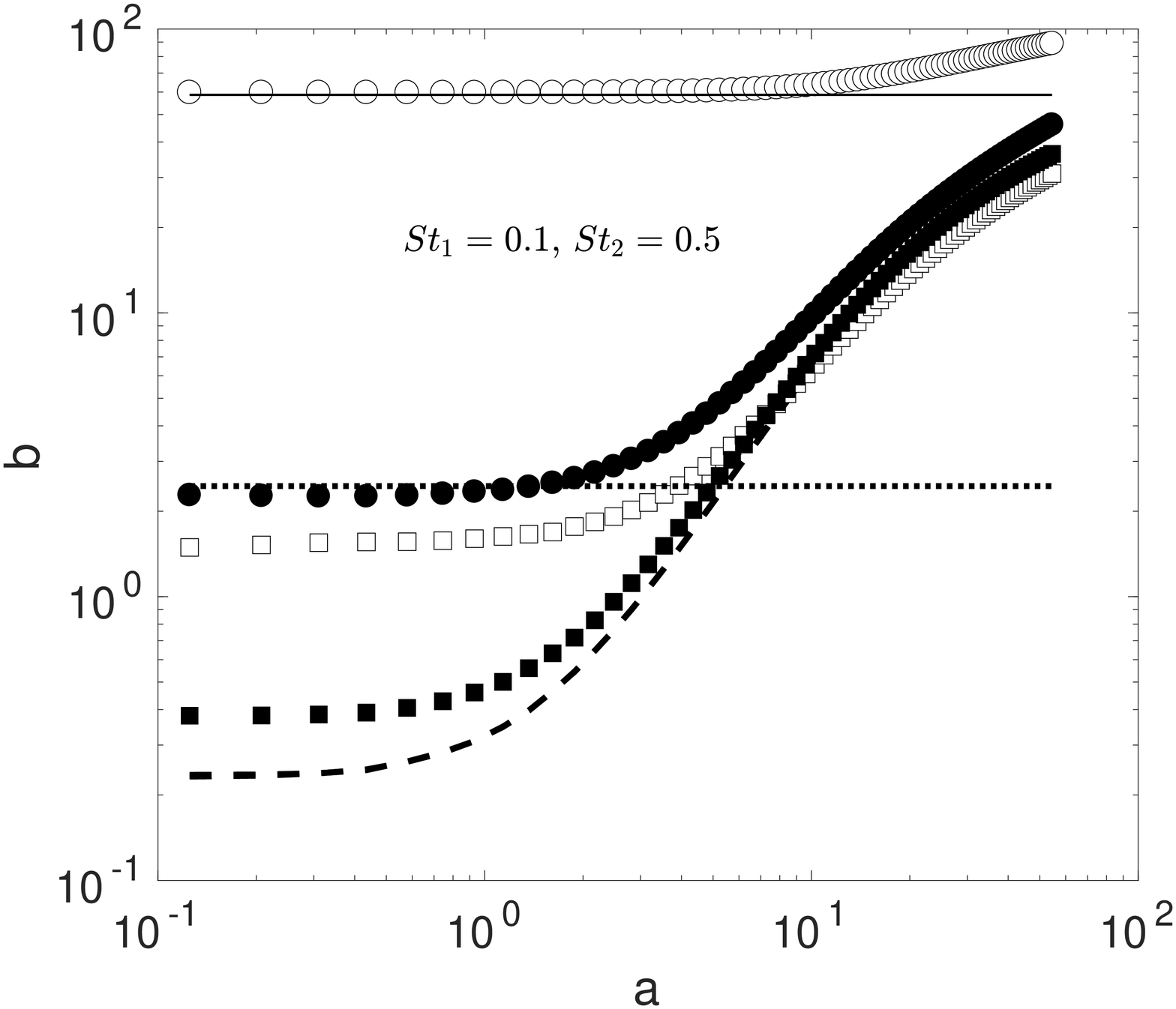}        
	    \caption{}
    \end{subfigure}

\vspace{0.1in}
    \begin{subfigure}[b]{0.5\textwidth}
     \psfrag{L}[cc][2]{$St_1 = 1$}
        \centering
		\includegraphics[width=\textwidth]{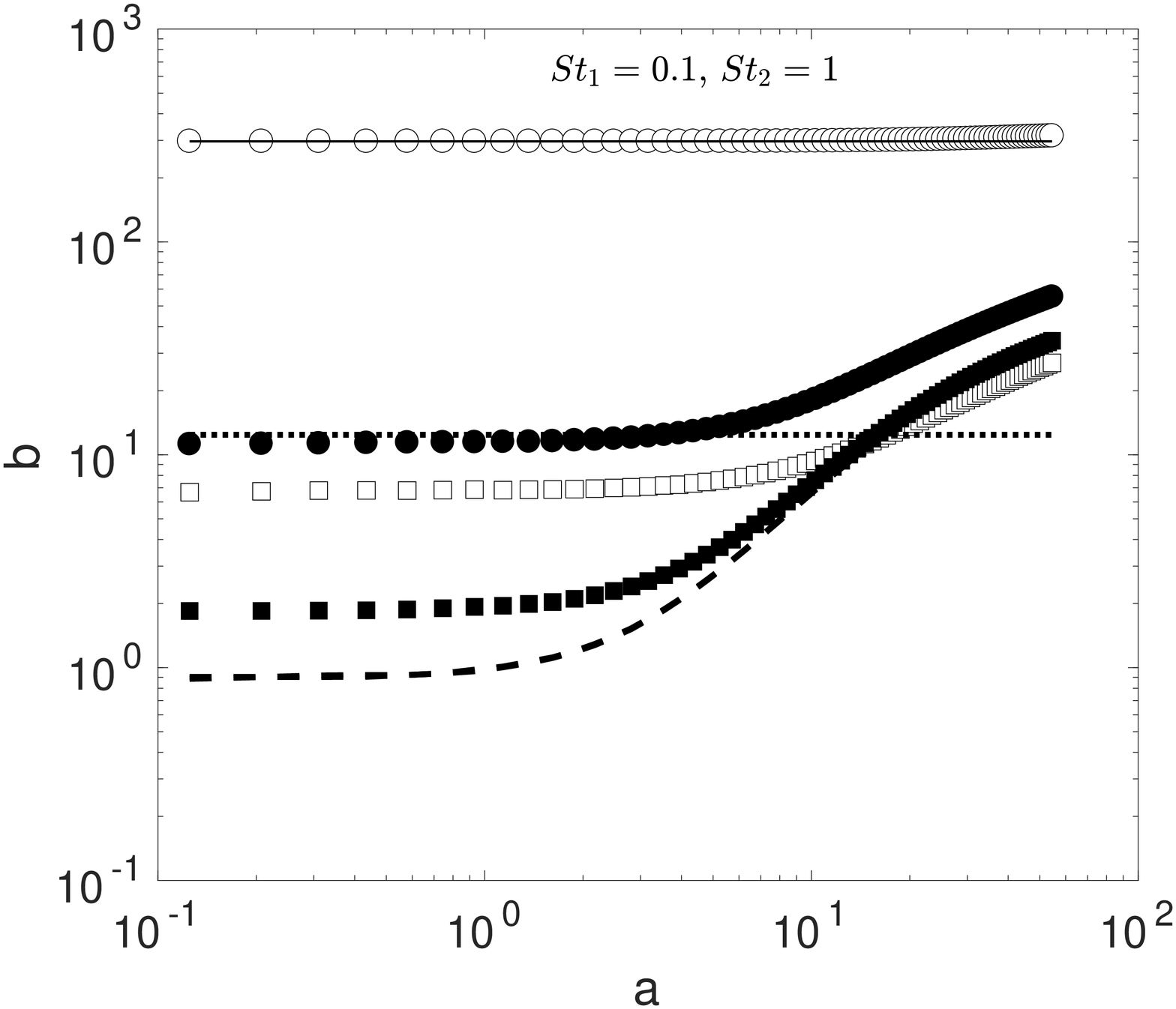}        
		\caption{}
    \end{subfigure}%
    ~
    \begin{subfigure}[b]{0.5\textwidth}
     \psfrag{L}[cc][2]{$St_1 = 3$}
        \centering
         \includegraphics[width=\textwidth]{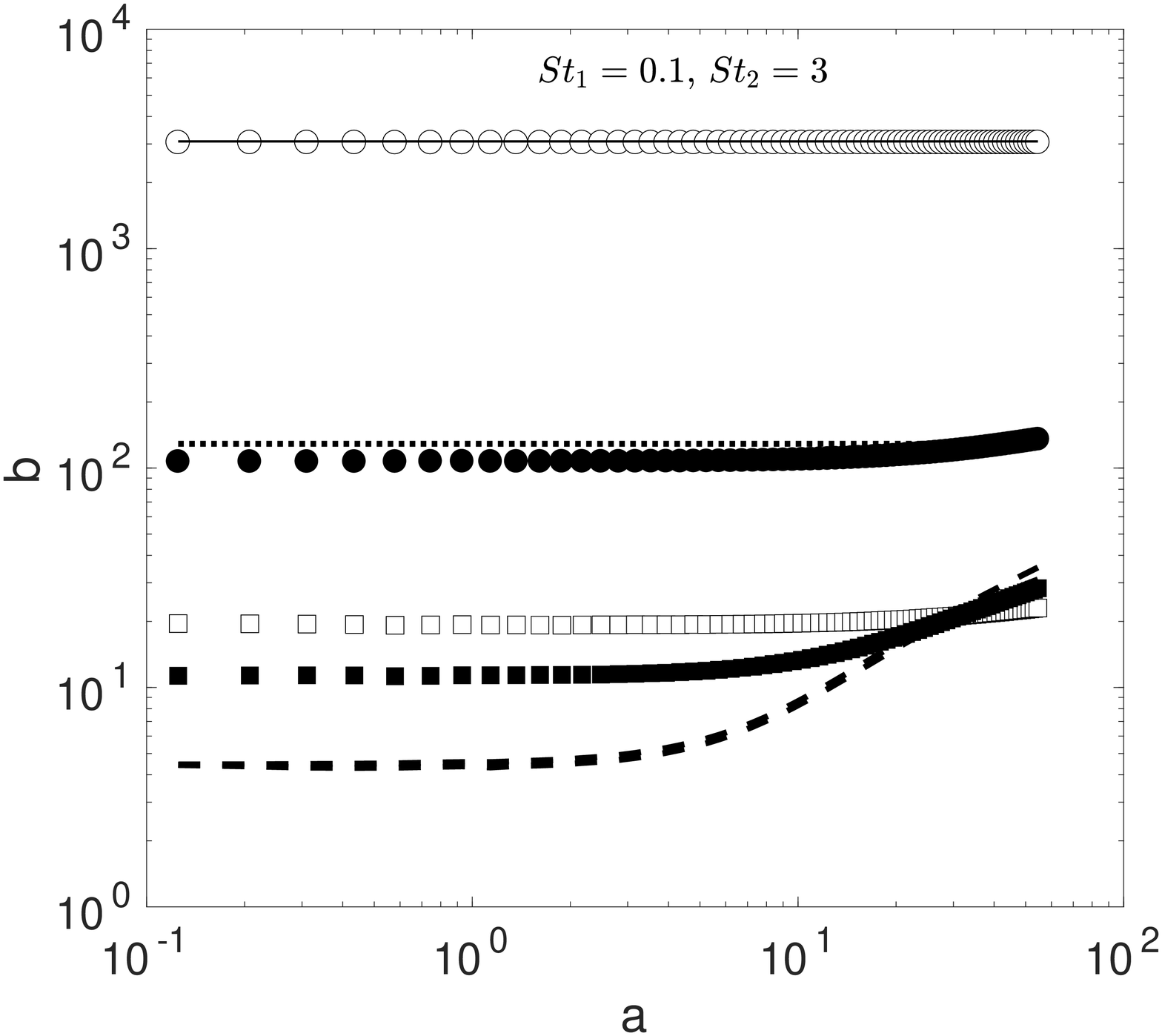}    
          \caption{}
    \end{subfigure}

      \caption{DNS results for $\langle [w^p_{1,3}(t)]^2\rangle _r/u_\eta^2$ as a function of $r/\eta$ for $St_1=0.1$ and different $St_2, Fr$ combinations. Circles correspond to the vertical velocities and squares correspond to the horizontal velocities. Dashed lines correspond to $Fr=\infty$, filled symbols to $Fr=0.3$, open symbols to $Fr=0.052$. The dotted and solid horizontal lines correspond to \eqref{w3w3} for $Fr=0.3$ and $Fr=0.052$, respectively.}
  \label{fig:cart_var_vs_r}
\end{figure}

\begin{figure}\vspace{0.1in}
\psfrag{a}[cc][2]{$r/\eta$}
  \psfrag{b}[cc][2]{$\langle [w^p_{1,3}(t)]^2\rangle _r/u_\eta^2$}
    \centering
    \begin{subfigure}[b]{0.5\textwidth}
         \psfrag{L}[cc][2]{$St_1 = 0.4$}
        \centering
        \hspace{-0.0in}
        \includegraphics[width=\textwidth]{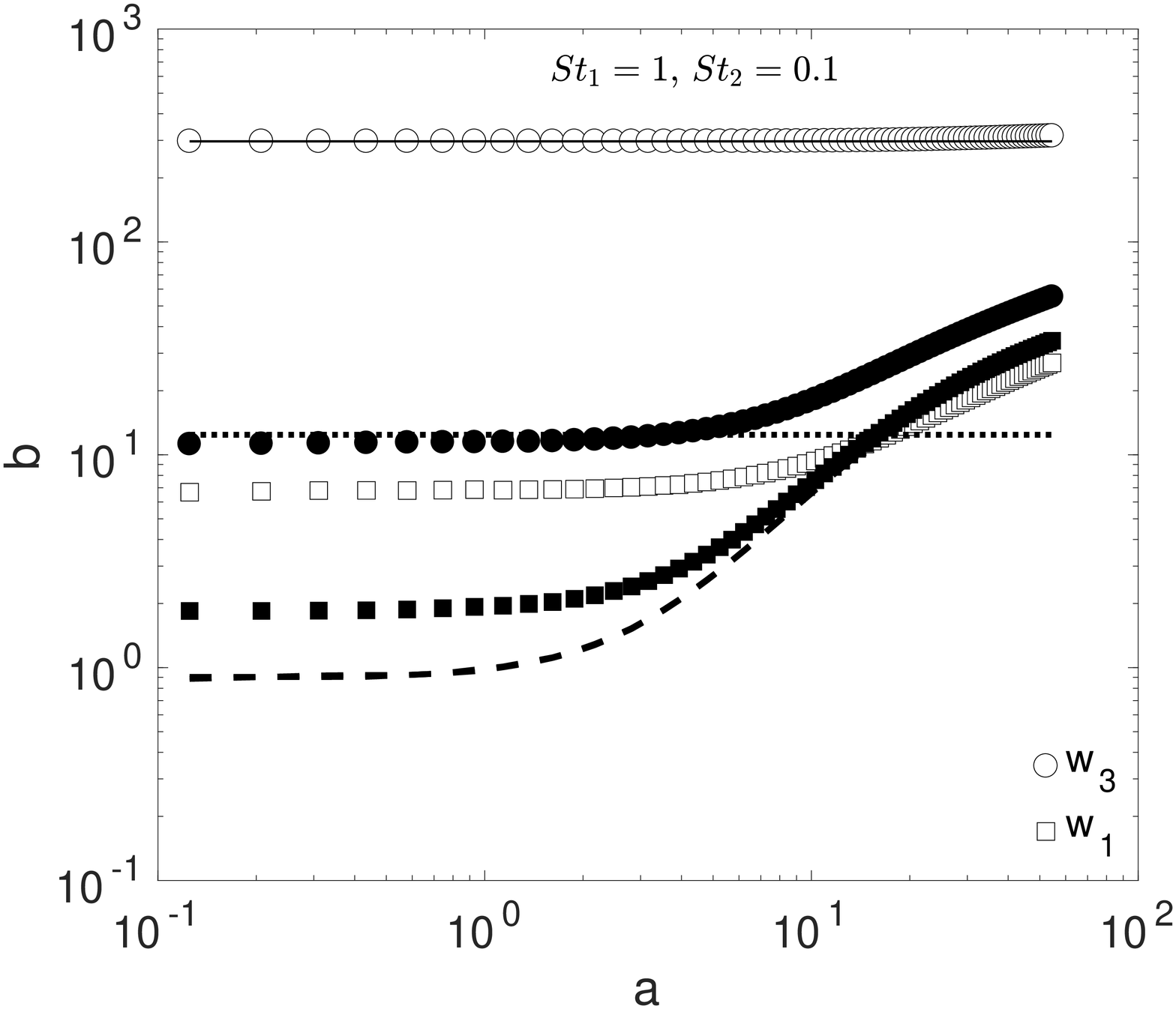}
        \caption{}
    \end{subfigure}%
    ~
    \begin{subfigure}[b]{0.5\textwidth}
     \psfrag{L}[cc][2]{$St_1 = 0.5$}
        \centering
	    \includegraphics[width=\textwidth]{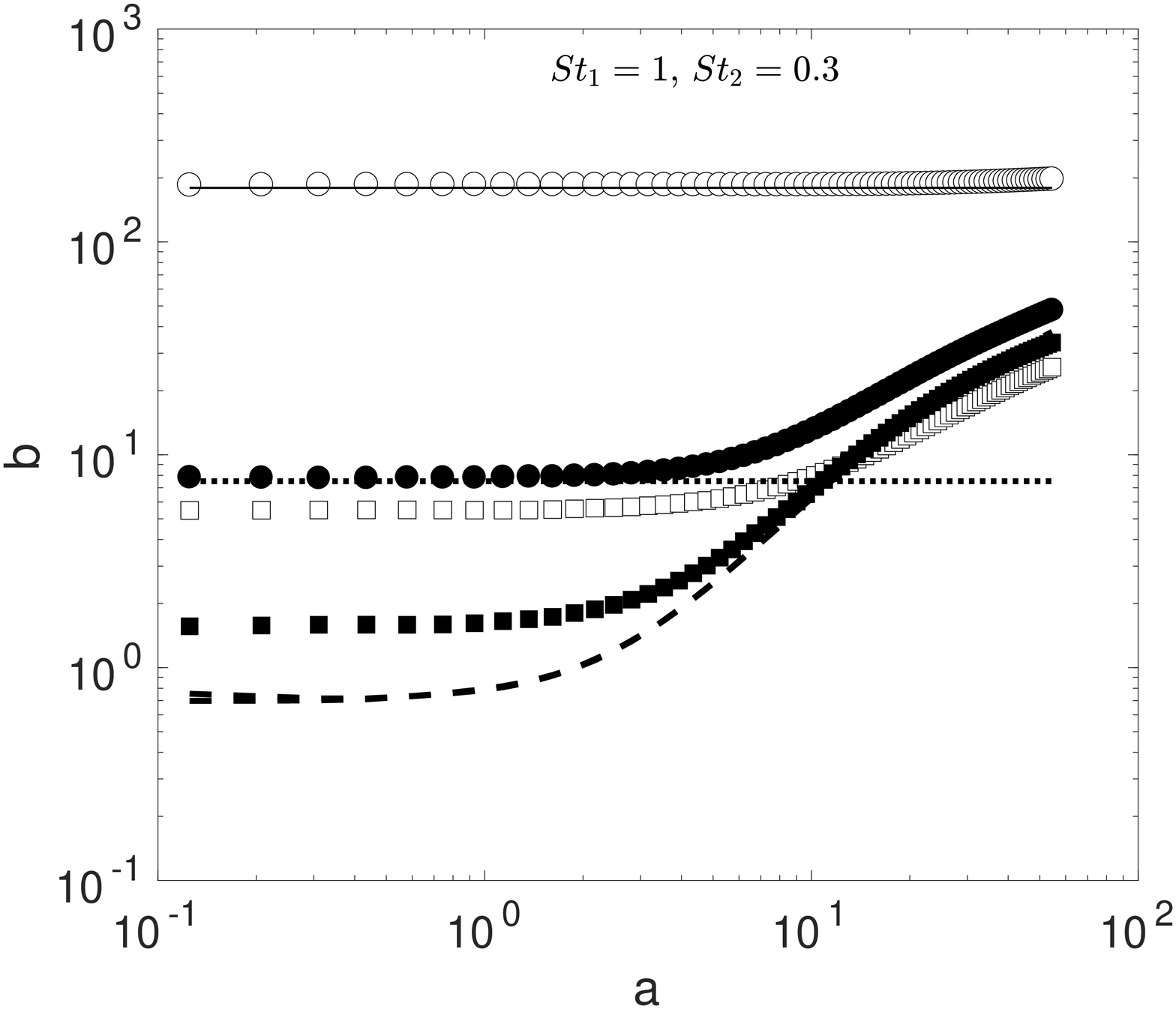}        
	    \caption{}
    \end{subfigure}

\vspace{0.1in}
    \begin{subfigure}[b]{0.5\textwidth}
     \psfrag{L}[cc][2]{$St_1 = 1$}
        \centering
		\includegraphics[width=\textwidth]{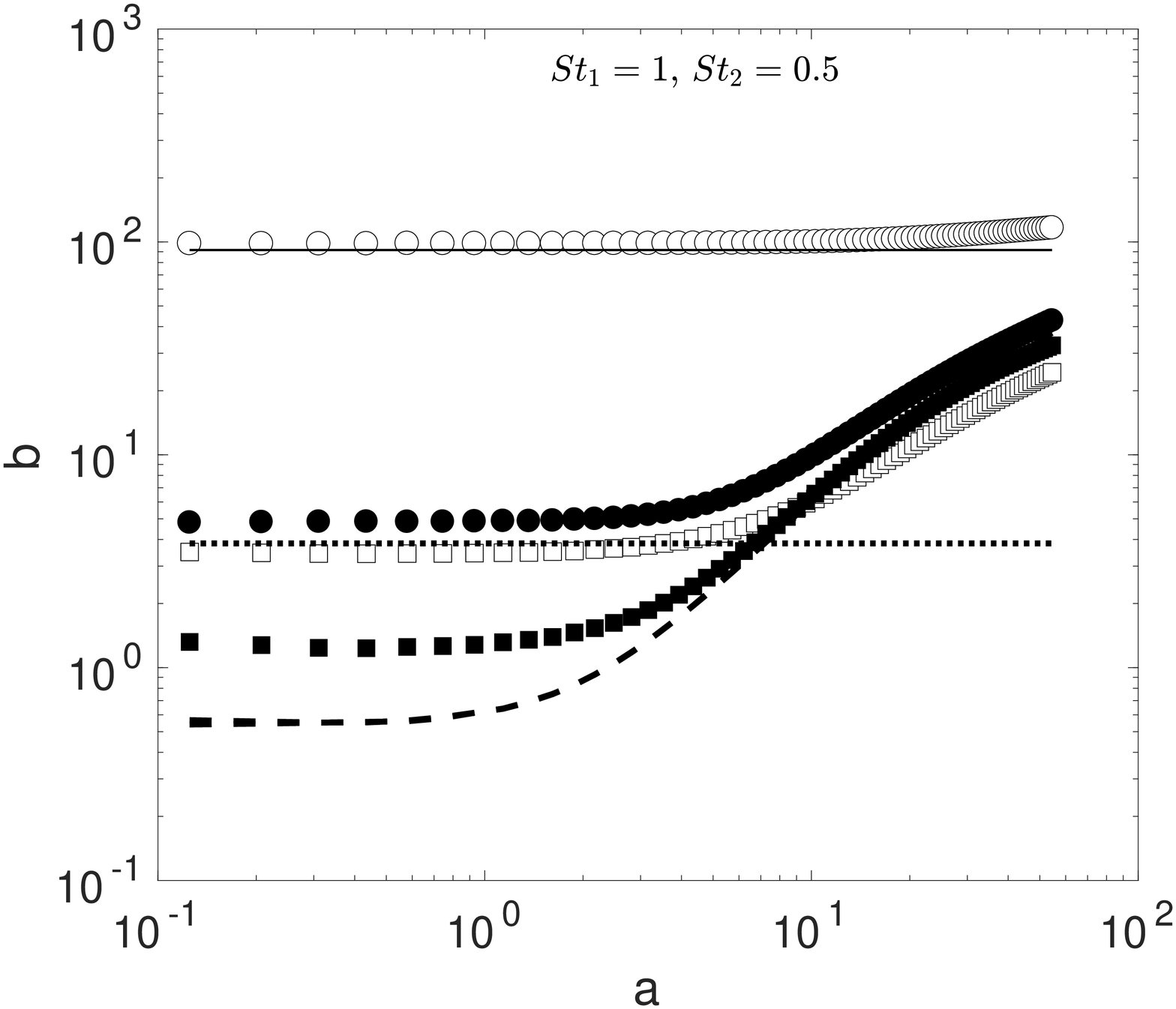}        
		\caption{}
    \end{subfigure}%
    ~
    \begin{subfigure}[b]{0.5\textwidth}
     \psfrag{L}[cc][2]{$St_1 = 3$}
        \centering
         \includegraphics[width=\textwidth]{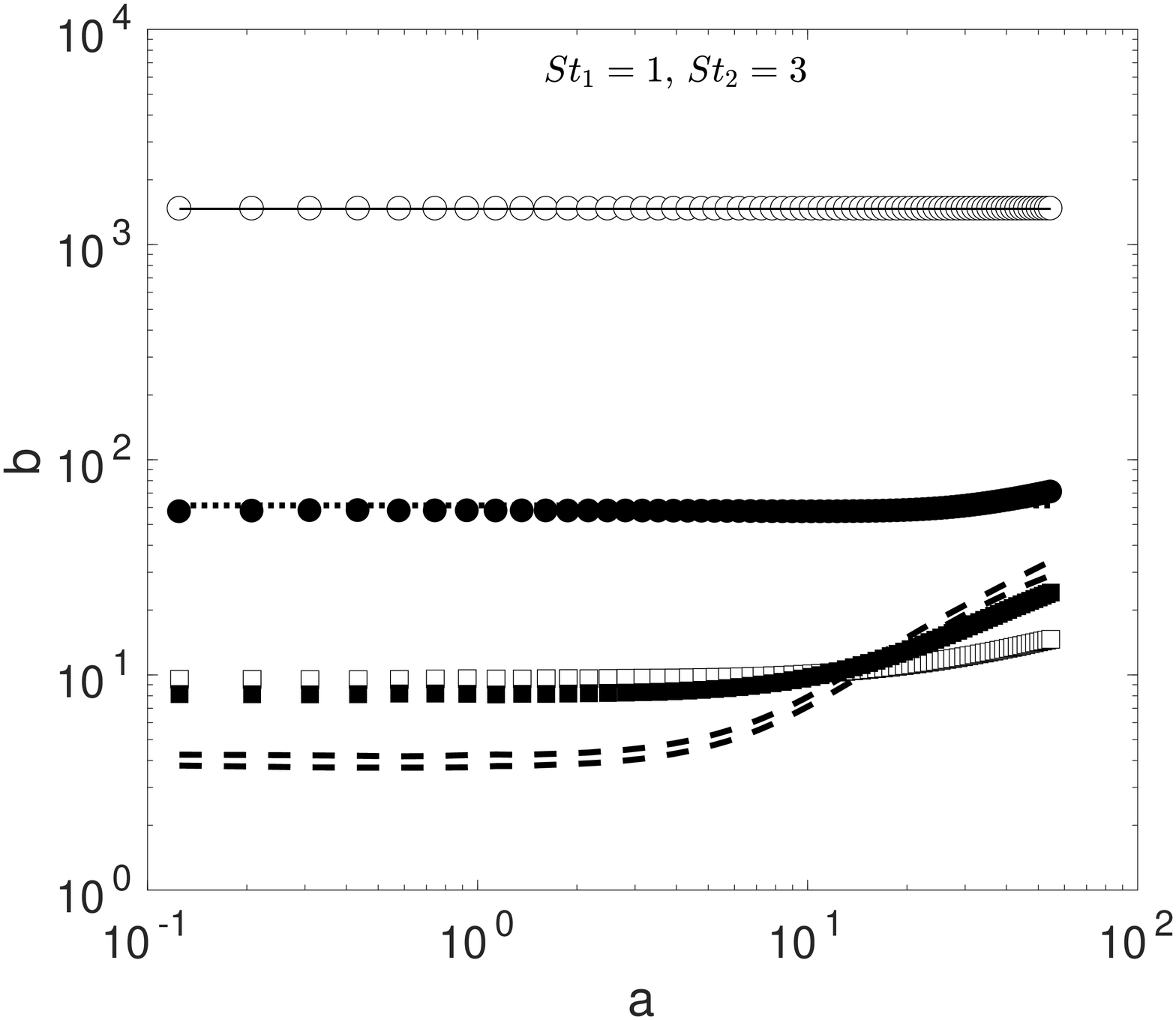}    
          \caption{}
    \end{subfigure}

      \caption{DNS results for $\langle [w^p_{1,3}(t)]^2\rangle _r/u_\eta^2$ as a function of $r/\eta$ for $St_1=1$ and different $St_2, Fr$ combinations. Circles correspond to the vertical velocities and squares correspond to the horizontal velocities. Dashed lines correspond to $Fr=\infty$, filled symbols to $Fr=0.3$, open symbols to $Fr=0.052$. The dotted and solid horizontal lines correspond to \eqref{w3w3} for $Fr=0.3$ and $Fr=0.052$, respectively.}
  \label{fig:cart_var_vs_r_st1_1}
\end{figure}

The results are shown in figures~\ref{fig:cart_var_vs_r} and~\ref{fig:cart_var_vs_r_st1_1}, where we also show the $Fr\to0$ prediction 
\begin{align}
\langle [w^p_3(t)]^2 \rangle_r=\Big(u_\eta\Delta St Fr^{-1}\Big)^2,\label{w3w3}
\end{align}
which follows from \eqref{w3Fr0}. When $Fr=0.3$, $\langle [w^p_3(t)]^2 \rangle_r$ is well approximated by \eqref{w3w3}, with some small discrepancies for certain $St_1, St_2$ combinations. However, for $Fr=0.052$, \eqref{w3w3} is in almost perfect agreement with the data for each of the $St_1, St_2$ combinations shown. When gravity is active, $\langle [w^p_3(t)]^2 \rangle_r>\langle [w^p_1(t)]^2 \rangle_r$, and the difference grows with increasing $|\Delta St|$ and decreasing $Fr$. The analysis in \S\ref{theory} shows why; as $|\Delta St|$ increases and $Fr$ decreases, the contribution to ${\langle [w^p_3(t)]^2 \rangle_r}$ from the differential settling velocity grows, yet this contribution is absent from ${\langle [w^p_1(t)]^2 \rangle_r}$. The results in figures~\ref{fig:cart_var_vs_r} and~\ref{fig:cart_var_vs_r_st1_1} also show that the implicit effect of gravity can lead to strong enhancements of ${\langle [w^p_1(t)]^2 \rangle_r}$, compared with the $Fr=\infty$ case, and in some cases the increase corresponds to an order of magnitude. This strong implicit effect of gravity on the horizontal relative velocities could have important implications for understanding and modeling the small-scale mixing and dispersion of bidisperse particles that are settling under gravity.

\FloatBarrier


\FloatBarrier



\begin{figure}\vspace{0.1in}
\psfrag{a}[cc][2]{$|w_{1,3}|/u_\eta$}
  \psfrag{b}[cc][2]{$\mathcal{P}(|w_{1,3}|\vert r)$}
    \centering
    \begin{subfigure}[b]{0.5\textwidth}
         \psfrag{L}[cc][2]{$St_1 = 0.4$}
        \centering
        \hspace{-0.0in}
        \includegraphics[width=\textwidth]{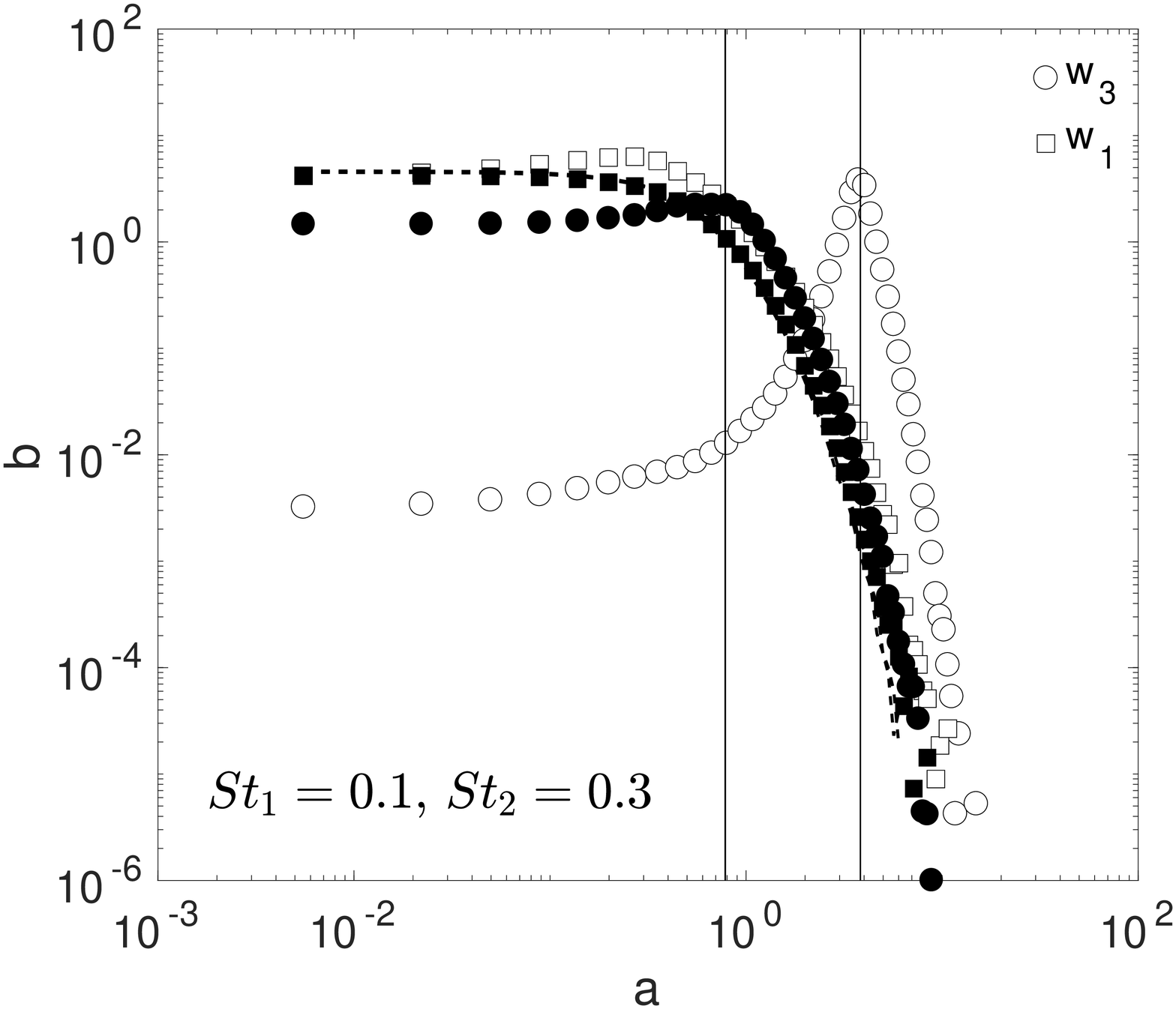}
        \caption{}
    \end{subfigure}%
    ~
    \begin{subfigure}[b]{0.5\textwidth}
     \psfrag{L}[cc][2]{$St_1 = 0.5$}
        \centering
	    \includegraphics[width=\textwidth]{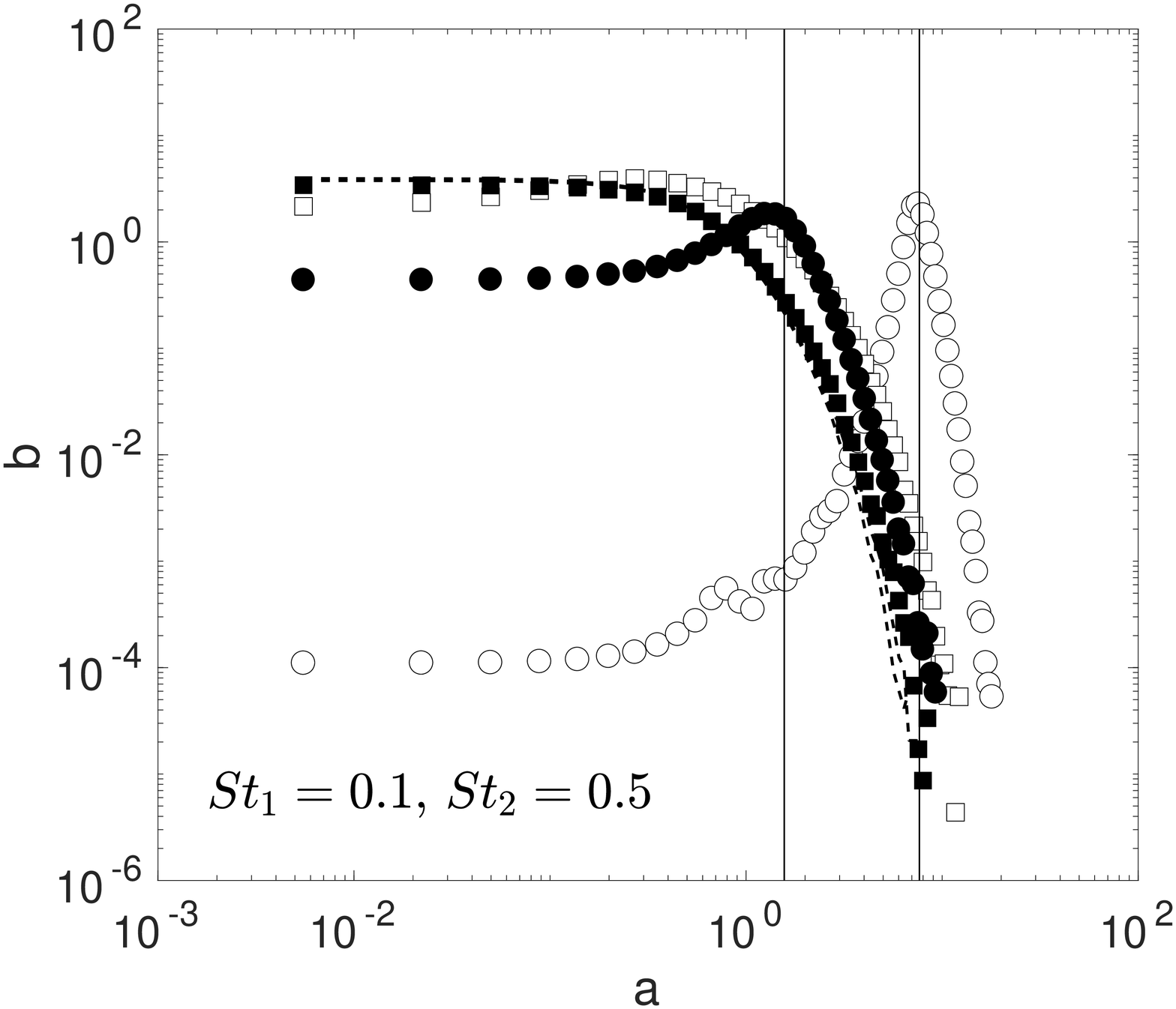}        
	    \caption{}
    \end{subfigure}

\vspace{0.1in}
    \begin{subfigure}[b]{0.5\textwidth}
     \psfrag{L}[cc][2]{$St_1 = 1$}
        \centering
		\includegraphics[width=\textwidth]{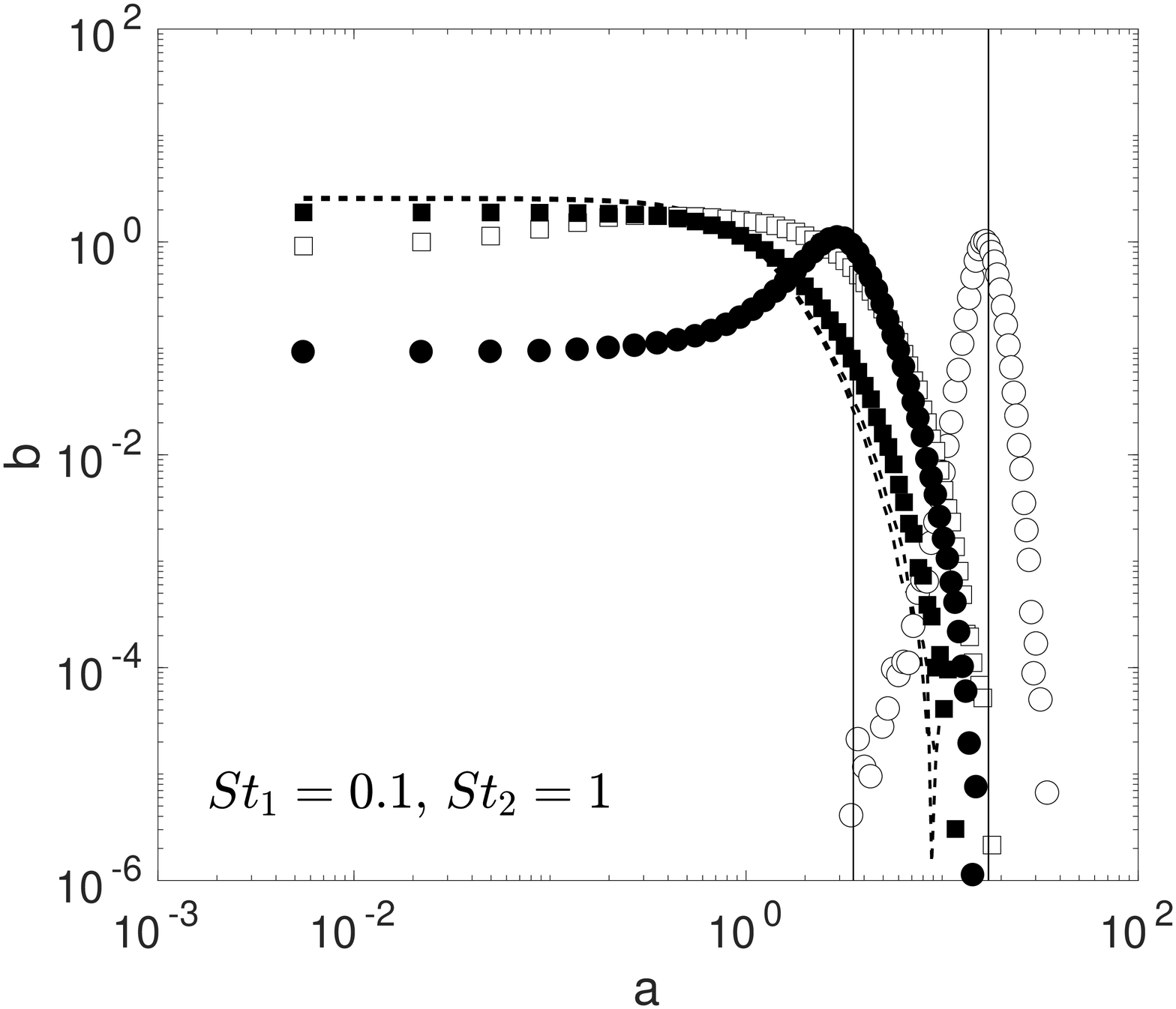}        
		\caption{}
    \end{subfigure}%
    ~
    \begin{subfigure}[b]{0.5\textwidth}
     \psfrag{L}[cc][2]{$St_1 = 3$}
        \centering
         \includegraphics[width=\textwidth]{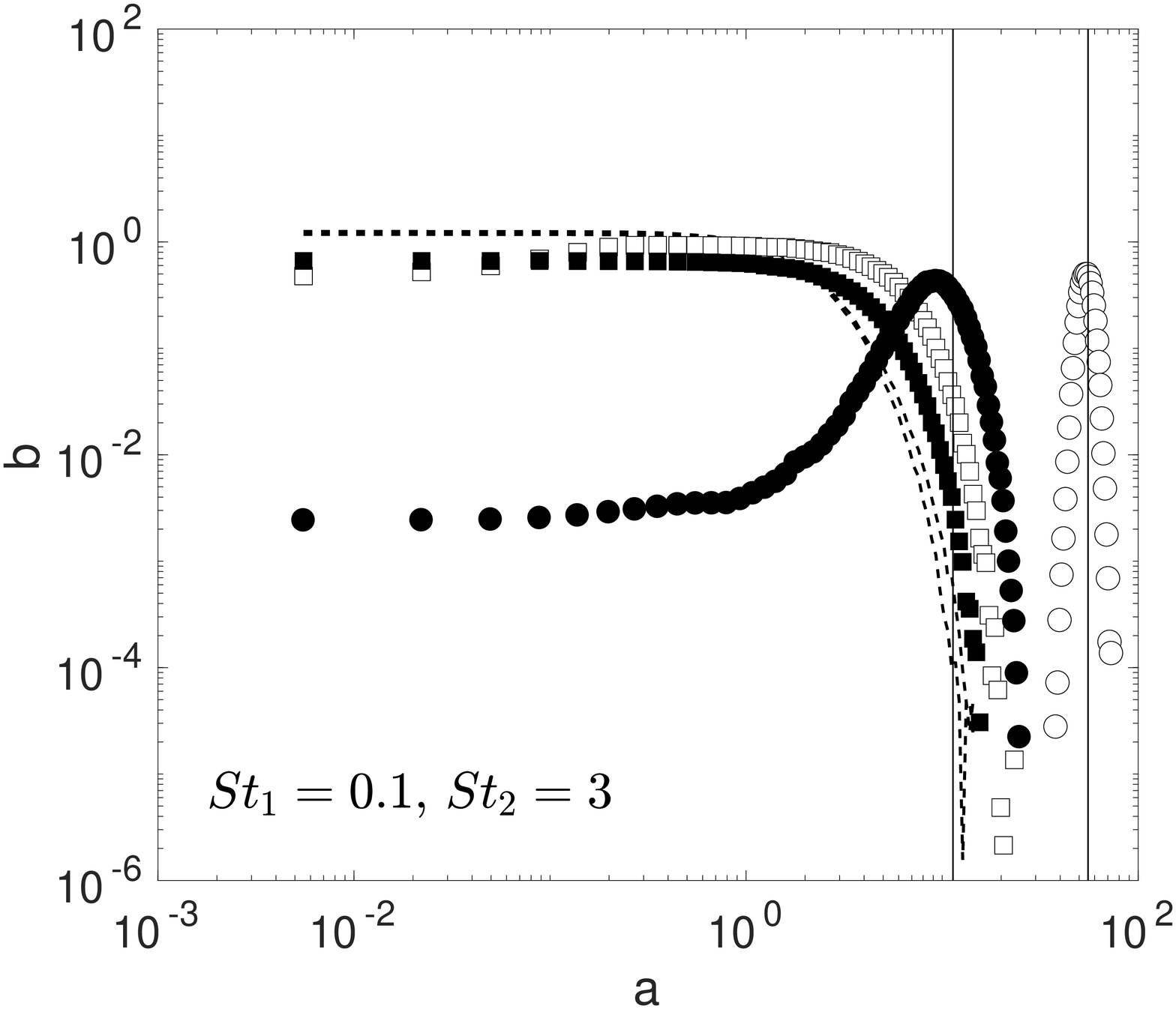}    
          \caption{}
    \end{subfigure}

      \caption{DNS results for $\mathcal{P}(|w_{1}|\vert r)$ and $\mathcal{P}(|w_{3}|\vert r)$ as a function of $|w_{1,3}|/u_\eta$ for $St_1=0.1$, different $St_2, Fr$ combinations, and with $r = \eta$. Circles correspond to the vertical velocities and squares correspond to the horizontal velocities. Dashed lines correspond to $Fr=\infty$, filled symbols to $Fr=0.3$, open symbols to $Fr=0.052$. The vertical lines correspond to $u_\eta |\Delta St| Fr^{-1}$ for $Fr=0.3$ and $Fr=0.052$.}
  \label{fig:cart_pdf}
\end{figure}


\begin{figure}\vspace{0.1in}
\psfrag{a}[cc][2]{$|w_{1,3}|/u_\eta$}
  \psfrag{b}[cc][2]{$\mathcal{P}(|w_{1,3}|\vert r)$}
    \centering
    \begin{subfigure}[b]{0.5\textwidth}
         \psfrag{L}[cc][2]{$St_1 = 0.4$}
        \centering
        \hspace{-0.0in}
        \includegraphics[width=\textwidth]{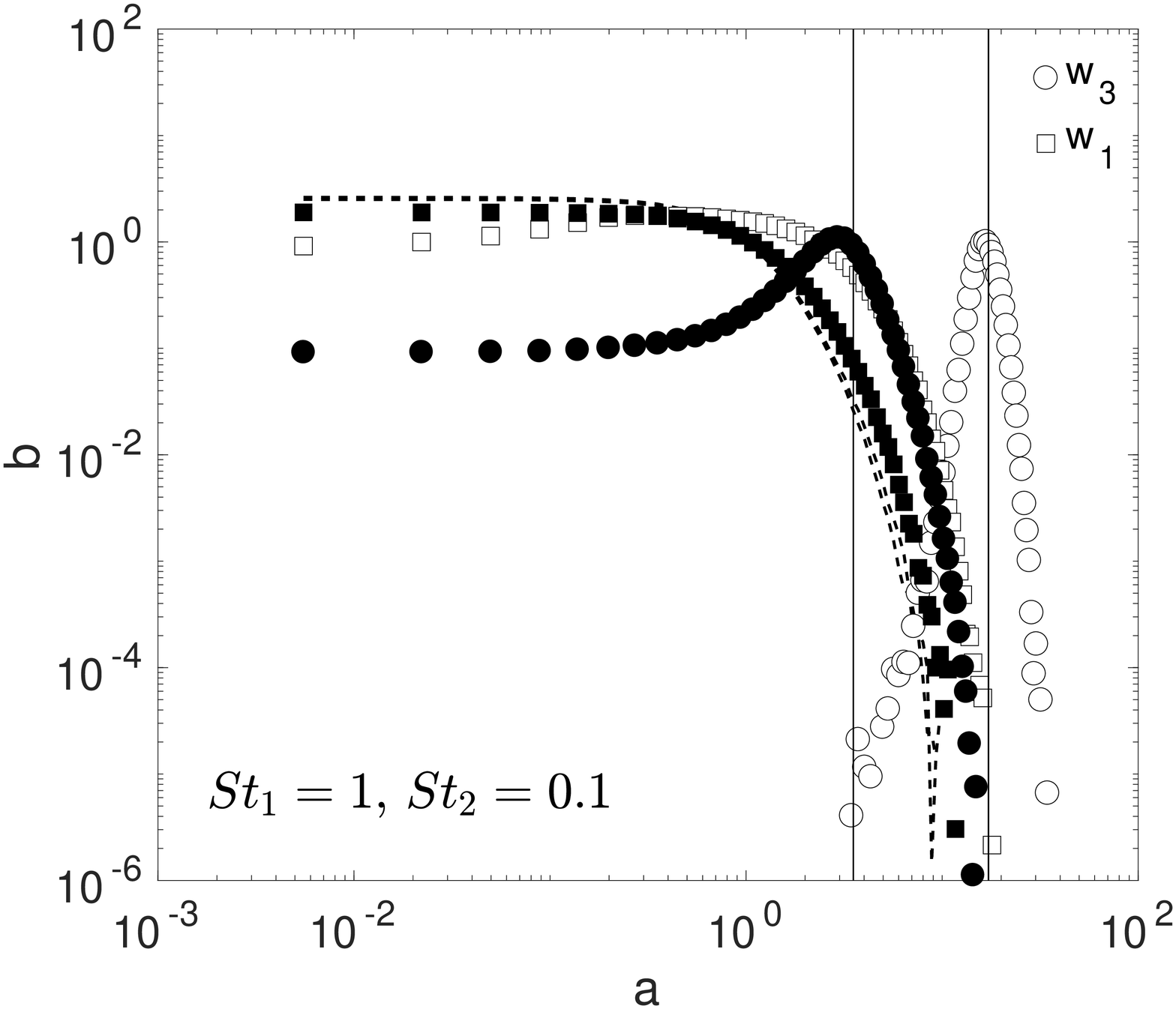}
        \caption{}
    \end{subfigure}%
    ~
    \begin{subfigure}[b]{0.5\textwidth}
     \psfrag{L}[cc][2]{$St_1 = 0.5$}
        \centering
	    \includegraphics[width=\textwidth]{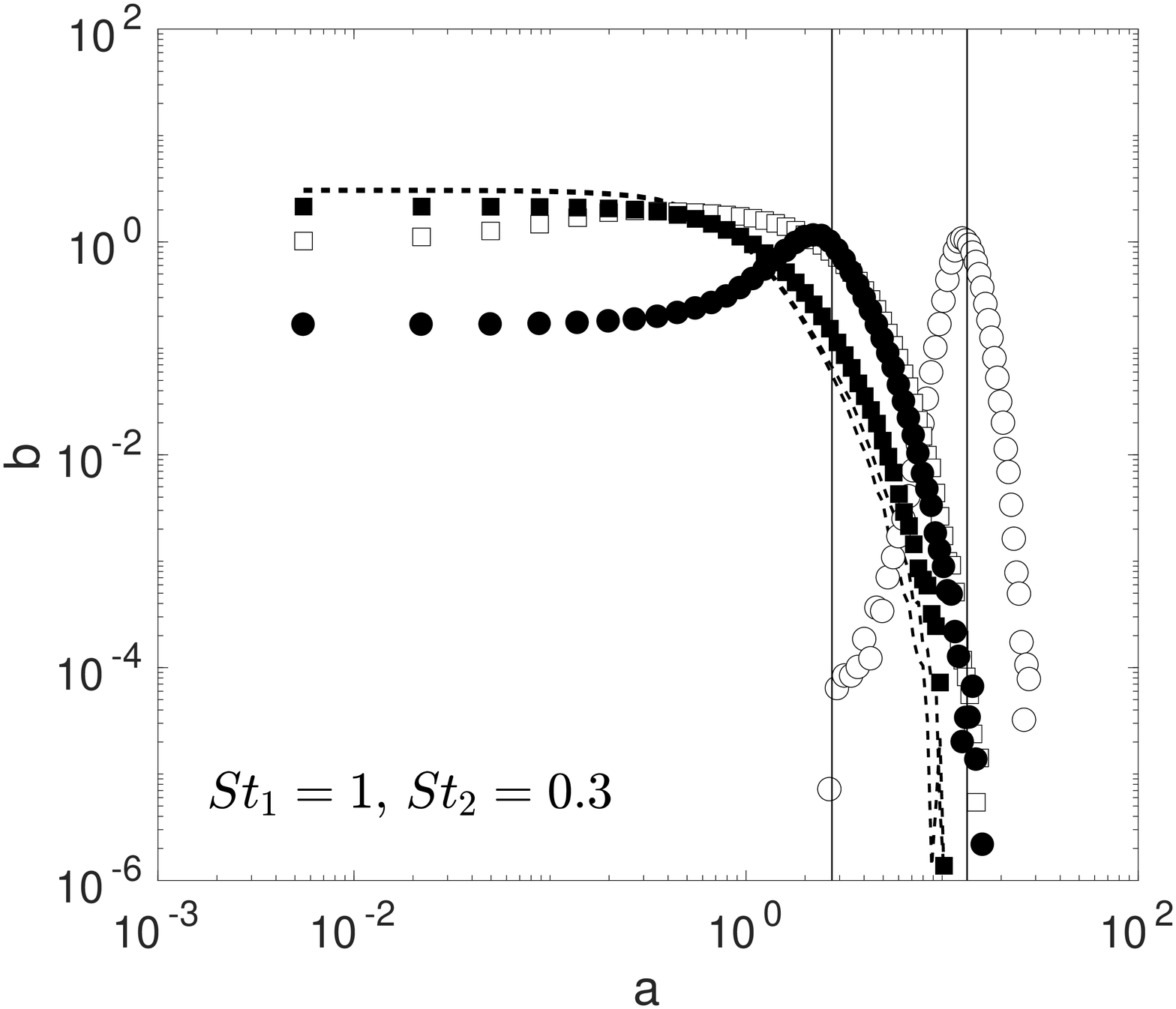}        
	    \caption{}
    \end{subfigure}

\vspace{0.1in}
    \begin{subfigure}[b]{0.5\textwidth}
     \psfrag{L}[cc][2]{$St_1 = 1$}
        \centering
		\includegraphics[width=\textwidth]{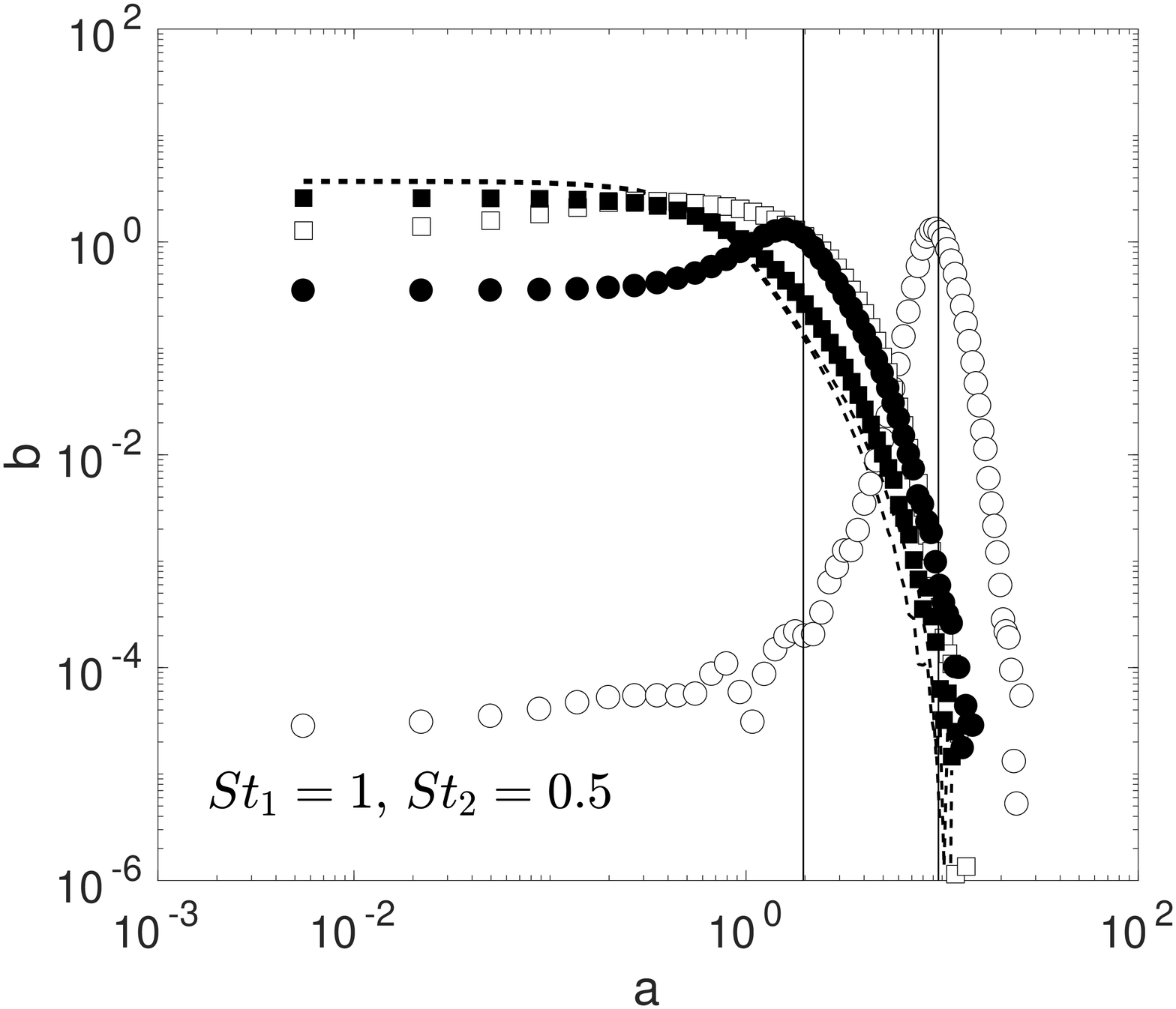}        
		\caption{}
    \end{subfigure}%
    ~
    \begin{subfigure}[b]{0.5\textwidth}
     \psfrag{L}[cc][2]{$St_1 = 3$}
        \centering
         \includegraphics[width=\textwidth]{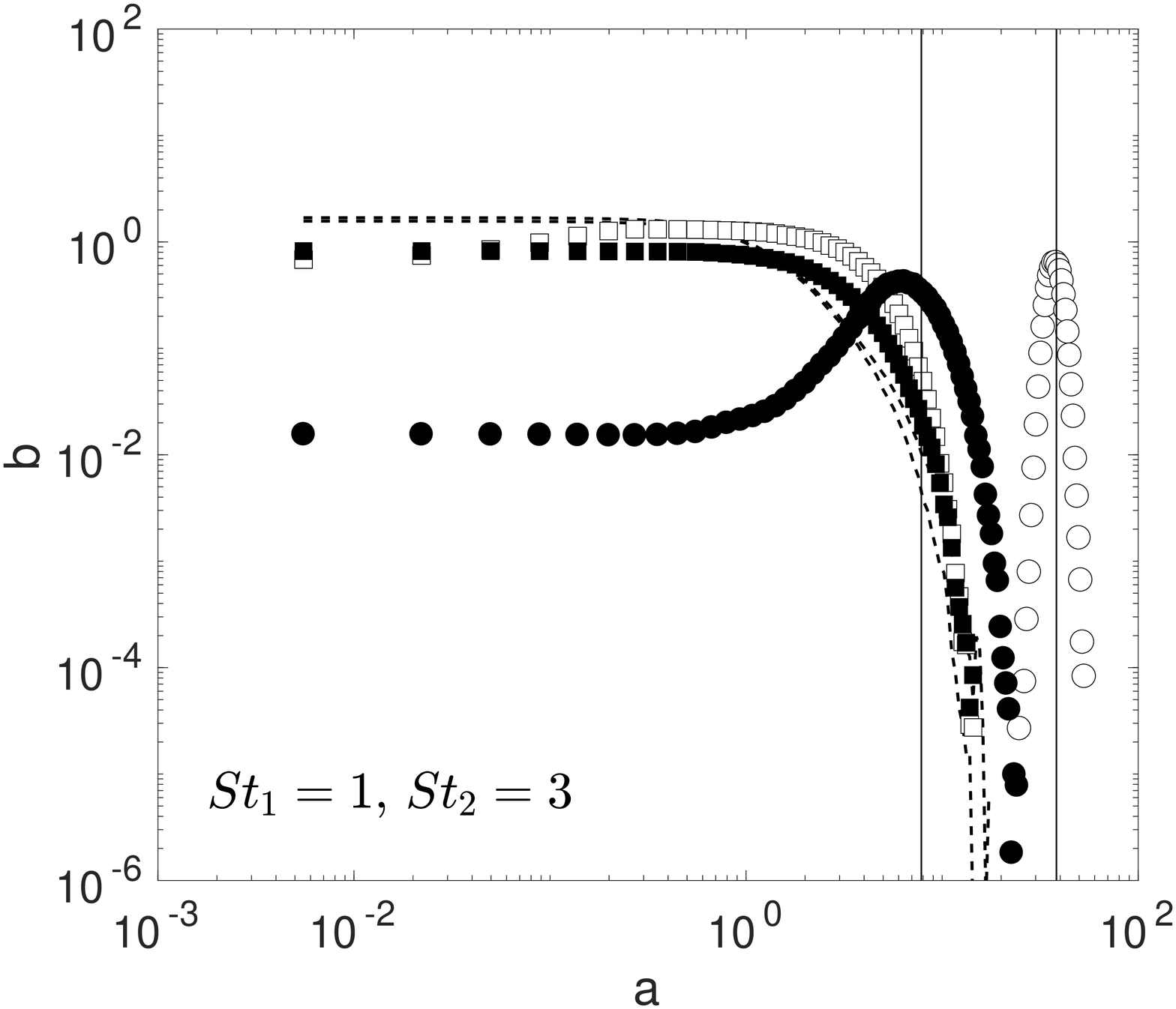}    
          \caption{}
    \end{subfigure}

      \caption{DNS results for $\mathcal{P}(|w_{1}|\vert r)$ and $\mathcal{P}(|w_{3}|\vert r)$ as a function of $|w_{1,3}|/u_\eta$ for $St_1=1$, different $St_2, Fr$ combinations, and with $r\in[0,2]\eta$. Circles correspond to the vertical velocities and squares correspond to the horizontal velocities. Dashed lines correspond to $Fr=\infty$, filled symbols to $Fr=0.3$, open symbols to $Fr=0.052$. The vertical lines correspond to $u_\eta |\Delta St| Fr^{-1}$ for $Fr=0.3$ and $Fr=0.052$.}
  \label{fig:cart_pdf_st1_1}
\end{figure}

In figures~\ref{fig:cart_pdf} and \ref{fig:cart_pdf_st1_1} we show results for $\mathcal{P}(|w_1|\vert r)$ and $\mathcal{P}(|w_3|\vert r)$, the PDFs of the absolute value of the horizontal and vertical relative velocities at separation $r$, respectively. As expected, the results show that gravity affects $\mathcal{P}(|w_1|\vert r)$ and $\mathcal{P}(|w_3|\vert r)$ in very different ways, since gravity only plays an explicit role in the vertical direction. Consistent with the results for the structure functions and the arguments of \S\ref{theory}, the PDFs for the horizontal velocities shift towards larger values, corresponding to the enhancement arising from the implicit effect of gravity. For the vertical velocities, the mode of the PDF is close to $u_\eta|\Delta St |Fr^{-1}$, which is indicated by the vertical lines in figures~\ref{fig:cart_pdf} and \ref{fig:cart_pdf_st1_1}. However, only for $|\Delta St|>2$ and $Fr=0.052$ does the PDF approach the asymptotic behavior 
\begin{align}
\lim_{Fr\to0}\mathcal{P}(|w_3|\vert r)\to \delta\Big(u_\eta|\Delta St| Fr^{-1}-|w_3|\Big).\label{PDFga}
\end{align}
When $Fr=0.052$, the explicit effect of gravity was shown to dominate $\langle [w^p_3(t)]^2 \rangle_r$, even when $|\Delta St|<1$. However, the results in figures~\ref{fig:cart_pdf} and \ref{fig:cart_pdf_st1_1} show that for the same parameter range, $\mathcal{P}(|w_3|\vert r)$ is not dominated by the gravity asymptote, implying that $\langle [w^p_3(t)]^n\rangle_r=(u_\eta\Delta St Fr^{-1})^n$ can fail for $n\neq 2$. We would also expect the deviation from (\ref{PDFga}) to increase as $R_\lambda$ is increased due to the enhanced intermittency in the turbulence, since the probability of the terms involving $\Delta{u}_3^p$ and ${a}_3^p$ becoming comparable to $u_\eta\Delta St Fr^{-1}$ in the equation for $w^p_3(t)$ would increase. This implies that in applications such as clouds where $R_\lambda=O(10^4)$ \citep{grabowski13}, the asymptotic behavior (\ref{PDFga}) may never be realized due to the extremely small values of $Fr$ that would be required. This could have important implications for understanding and modeling the vertical mixing of settling, bidisperse particles in a variety of turbulent flows.

\FloatBarrier

\subsection{Spatial Clustering} \label{sec:clustering}
We now turn to consider the spatial clustering of the particles, which may be quantified using the Radial Distribution Function (RDF), defined as \citep{mcquarrie}
\begin{align}
g(r)\equiv \frac{N_i/\delta V_i}{N_1N_2/V}.
\end{align}
If we define the number of satellite particles (with Stokes number $St_2$) found at a distance $r$ from a given primary particle (with Stokes number $St_1$) to be $N'_i$, then $N_i$ is the sum of $N'_i$ over all primary particles in the system. $\delta V_i$ is the volume of an elemental spherical shell with surface at
a distance $r$ from the primary particle, $V$ is the total volume of the system, and $N_1$ and $N_2$ are the total numbers of primary and satellite particles in the domain, respectively. Note that the RDF is the same (to within a constant) as $\|\bm{r}\|^{-2}$ times the spherical average of the PDF $\varphi(\bm{r})$ introduced in \S\ref{theory}.

\begin{figure}\vspace{0.1in}
\psfrag{a}[cc][2]{$r/\eta$}
  \psfrag{b}[cc][2]{$g(r)$}
    \centering
    \begin{subfigure}[b]{0.5\textwidth}
         \psfrag{L}[cc][2]{$St_1 = 0.4$}
        \centering
        \hspace{-0.0in}
        \includegraphics[width=\textwidth]{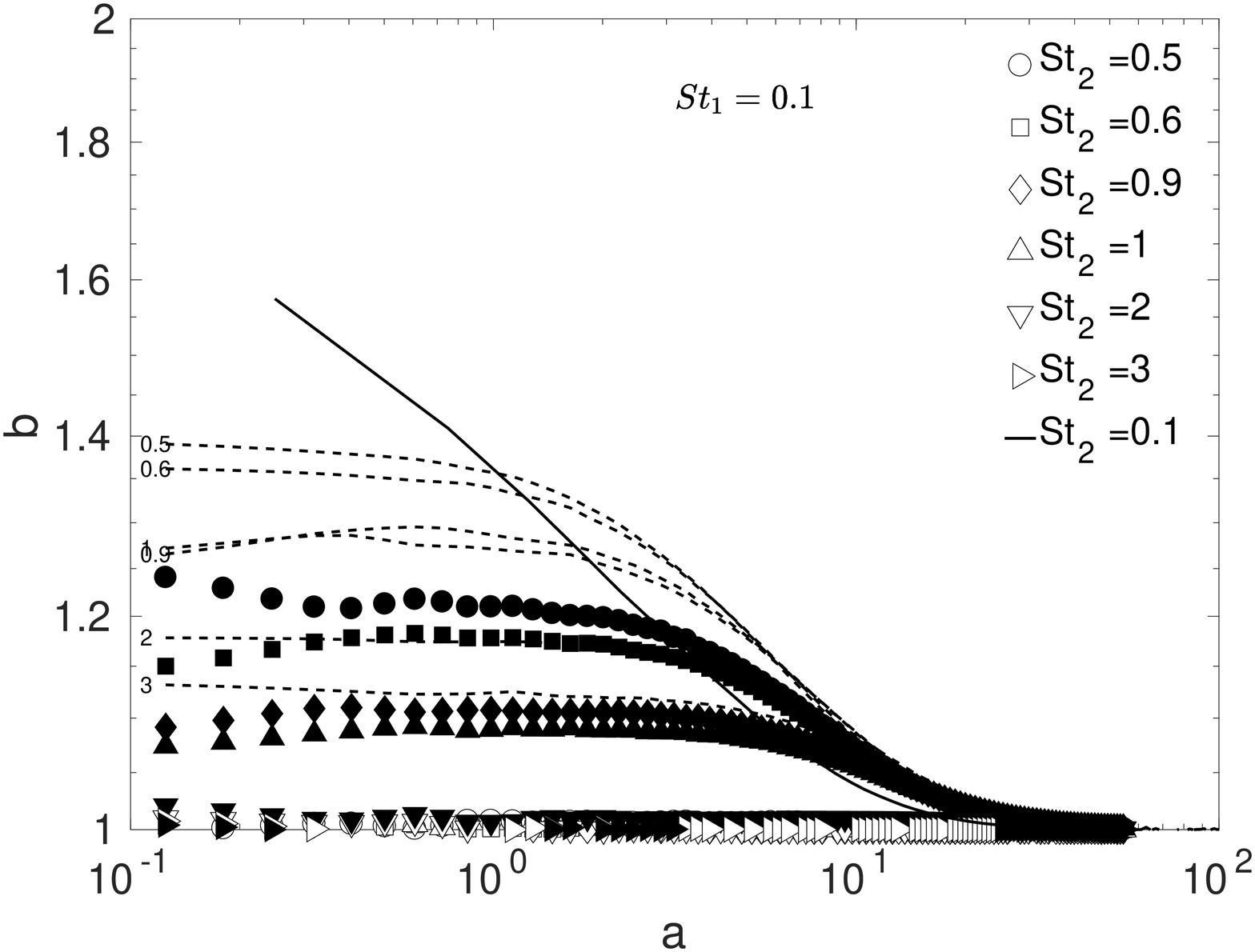}
        \caption{}
    \end{subfigure}%
    ~
    \begin{subfigure}[b]{0.5\textwidth}
     \psfrag{L}[cc][2]{$St_1 = 0.5$}
        \centering
        \hspace{-0.0in}
	    \includegraphics[width=\textwidth]{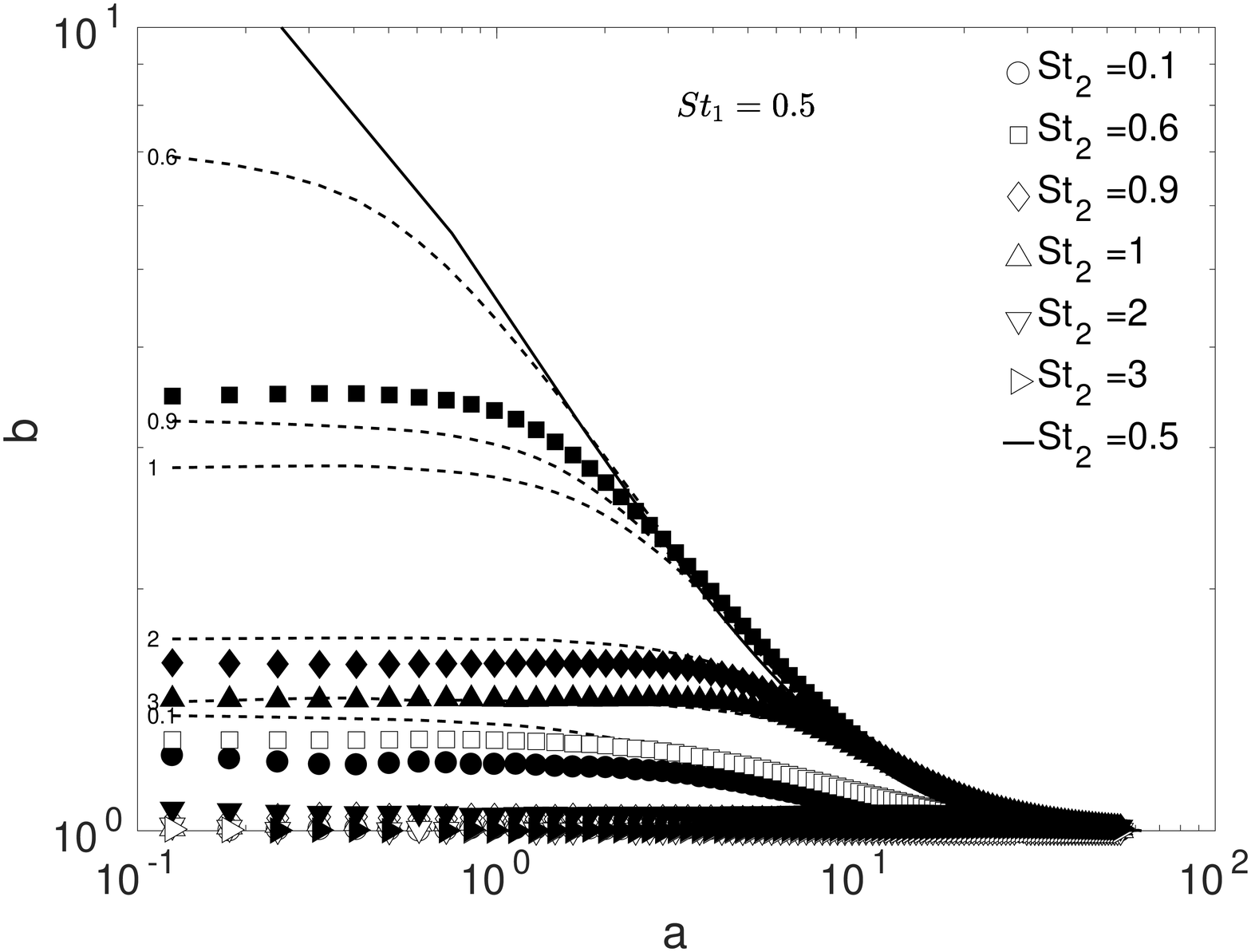}        
	    \caption{}
    \end{subfigure}

\vspace{0.1in}
    \begin{subfigure}[b]{0.5\textwidth}
     \psfrag{L}[cc][2]{$St_1 = 1$}
        \centering
		\includegraphics[width=\textwidth]{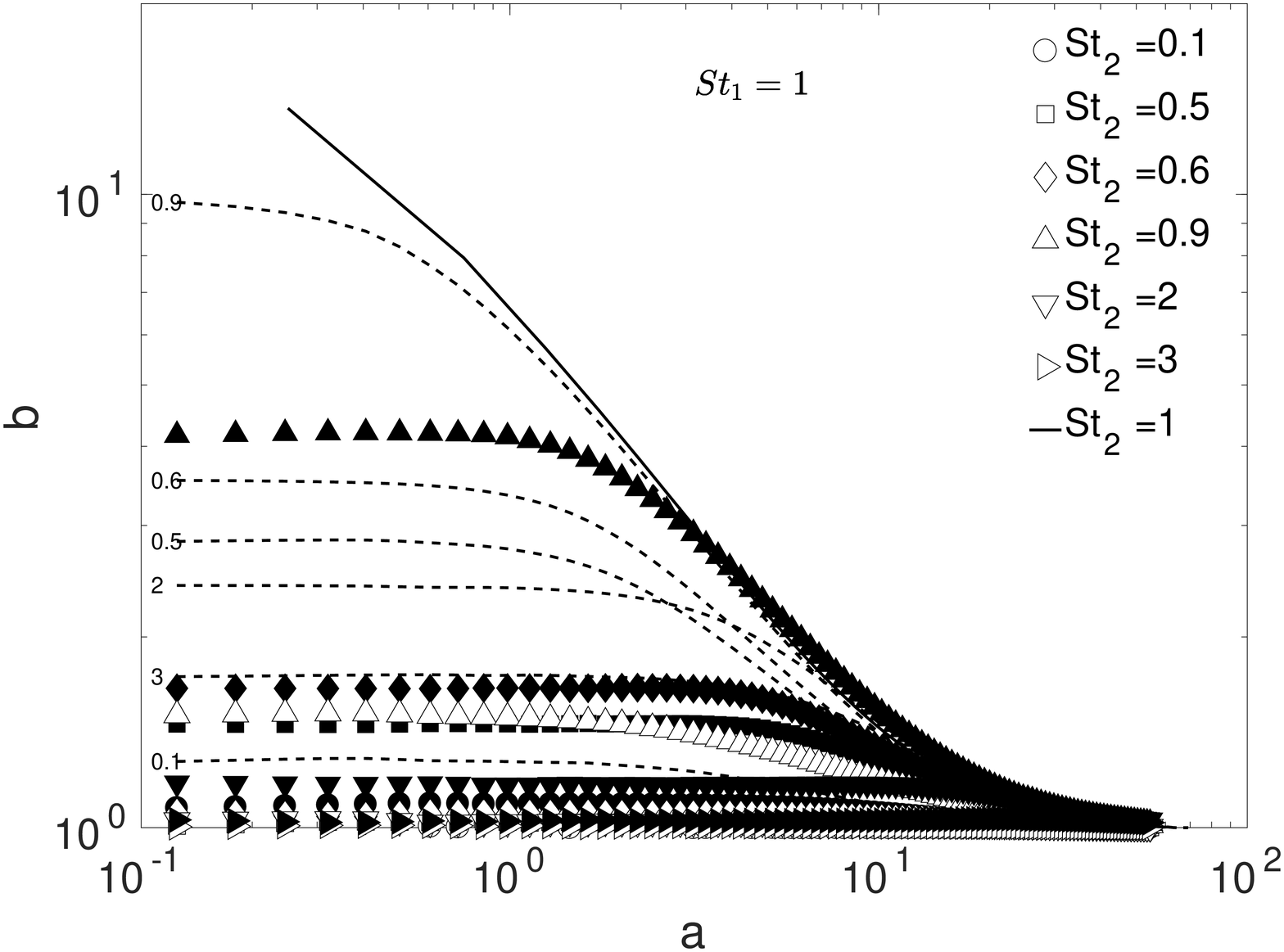}        
		\caption{}
    \end{subfigure}%
    ~
    \begin{subfigure}[b]{0.5\textwidth}
     \psfrag{L}[cc][2]{$St_1 = 3$}
        \centering
         \includegraphics[width=\textwidth]{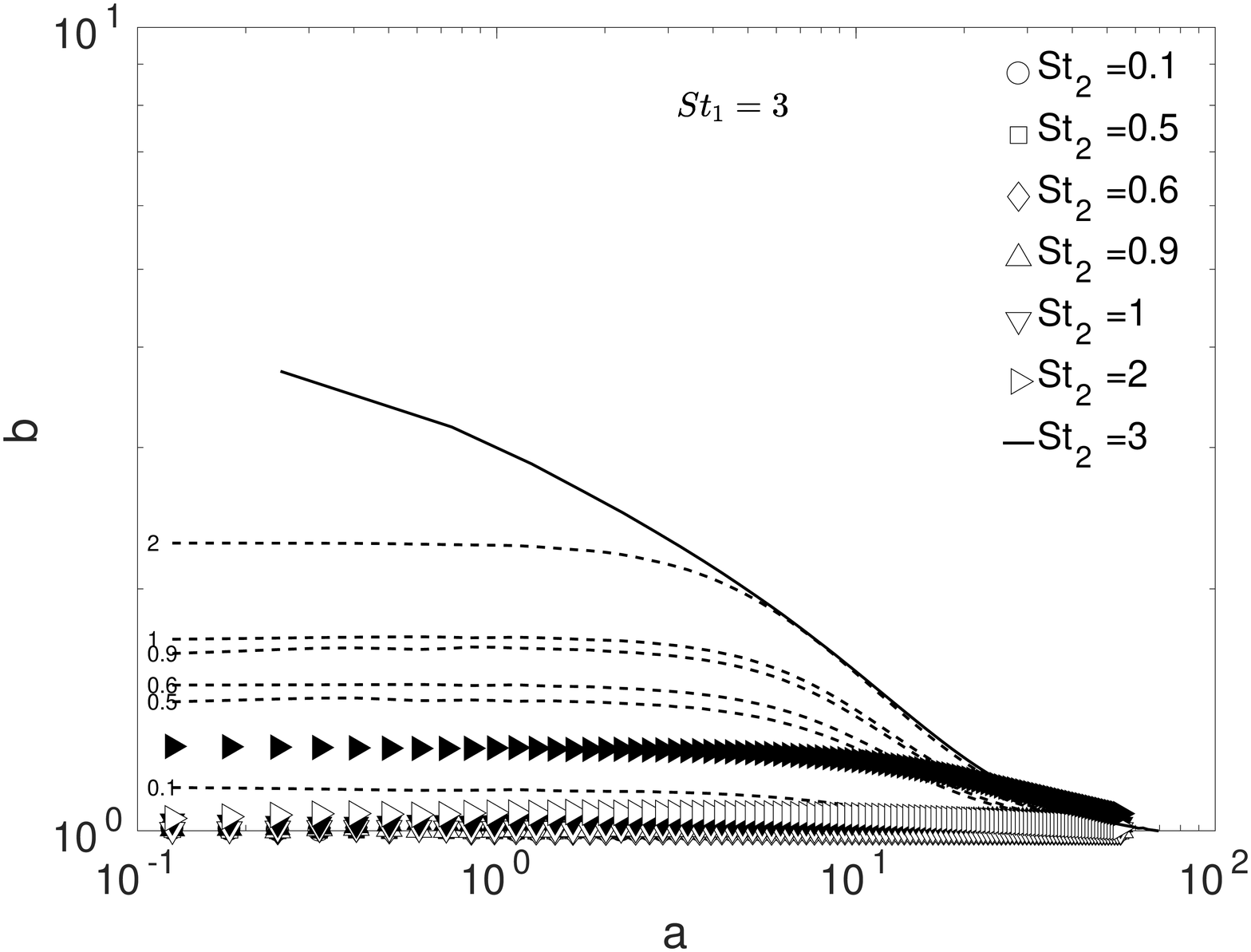}    
          \caption{}
    \end{subfigure}

      \caption{DNS results for $g(r)$ as a function of $r/\eta$ for different $St_1,St_2, Fr$ combinations.  Dashed lines correspond to $Fr=\infty$, filled symbols to $Fr=0.3$, open symbols to $Fr=0.052$, and solid lines correspond to the monodisperse case with $Fr=\infty$. The numbers next to the dashed lines indicate the $St_2$ value that they correspond to.}
  \label{fig:rdf_vs_r}
\end{figure}
The results in figure~\ref{fig:rdf_vs_r} show that in general, increasing $|\Delta St |$ reduces the RDF, and that the clustering is significantly reduced for $r/\eta \ll1$ even when $|\Delta St|\ll1$. These results for $Fr=\infty$ are consistent with previous investigations \citep{zww01,pan11,chun05}. This reduction is consistent with the explanations in \S\ref{theory}, namely, that the acceleration contribution to the relative motion of bidisperse particles leads to an enhanced particle-pair diffusion and reduced inward drift, leading to reduced clustering, and to a plateau of the RDF at small separations. 


We also note that although the RDF at $r/\eta\leq 1$ is always largest for $\Delta St=0$, i.e. the monodisperse case, the same is not always true at larger separations (e.g. see figure~\ref{fig:rdf_vs_r} (a)). Over the $St_1, St_2$ range we are here considering, $|\Delta St |>0$ monotonically increases the particle-pair diffusion tensor, and therefore, this behavior must be caused by bidispersity leading to an enhancement of the inward drift term $St_2\varphi\bm{\nabla_r\cdot}\langle w^p_1(t)\bm{w}^p(t)\rangle_{\bm{r}}$ in \eqref{ADFeq2} in certain regimes of $St_1, St_2$.

In agreement with the arguments in \S\ref{theory}, the results in figure~\ref{fig:rdf_vs_r} also show that decreasing $Fr$ further suppresses the clustering of bidisperse particles. Indeed, even for $|\Delta St |= 0.1$, gravity is very effective at suppressing the clustering. 


\begin{figure}\vspace{0.1in}
\psfrag{a}[cc][2]{$St_2$}
  \psfrag{b}[cc][2]{$g(r)$}
    \centering
    \begin{subfigure}[b]{0.5\textwidth}
         \psfrag{L}[cc][2]{$St_1 = 0.4$}
        \centering
        \hspace{-0.0in}
        \includegraphics[width=0.9\textwidth]{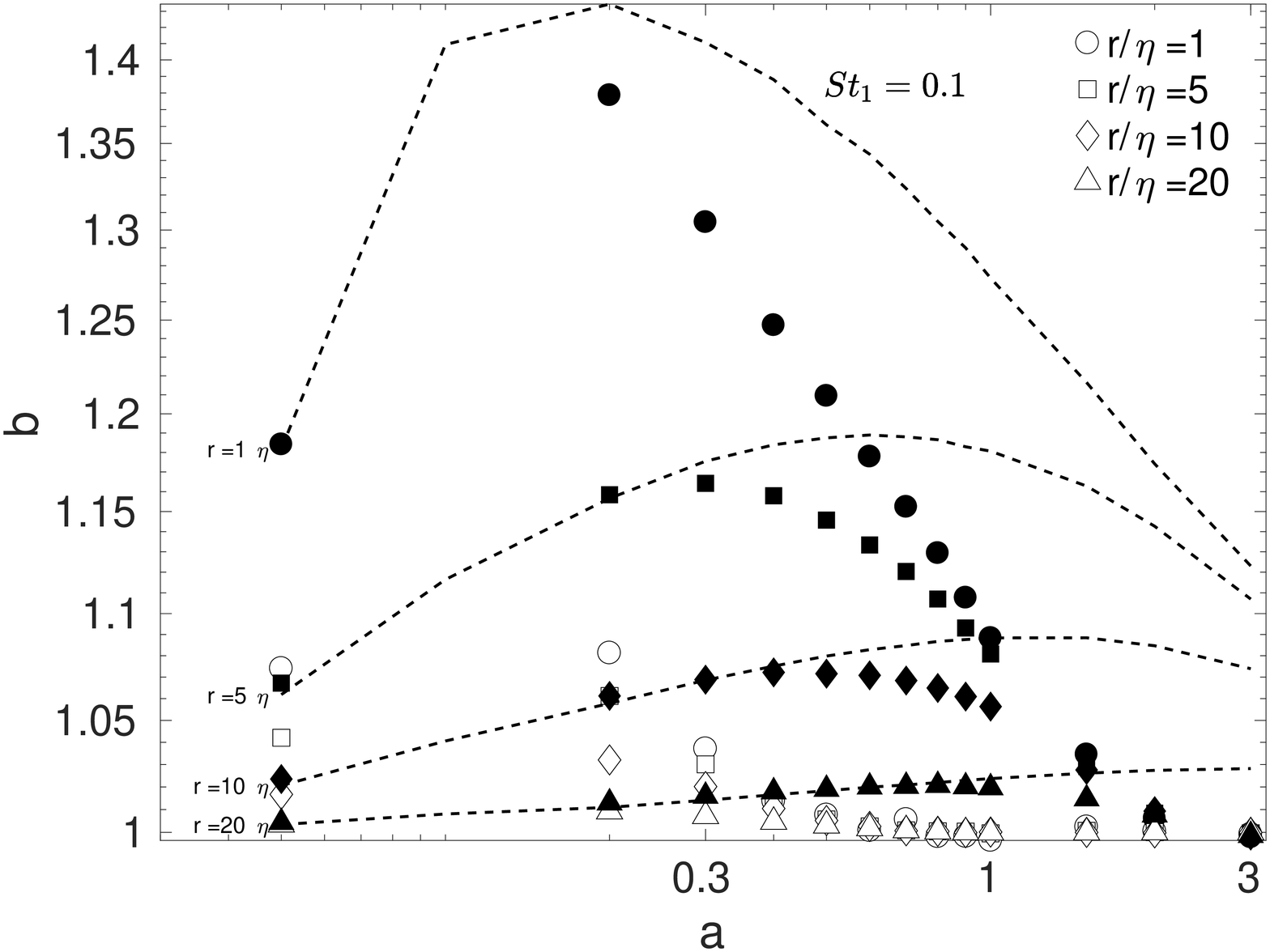}
        \caption{}
    \end{subfigure}%
    ~
    \begin{subfigure}[b]{0.5\textwidth}
     \psfrag{L}[cc][2]{$St_1 = 0.5$}
        \centering
	    \includegraphics[width=0.9\textwidth]{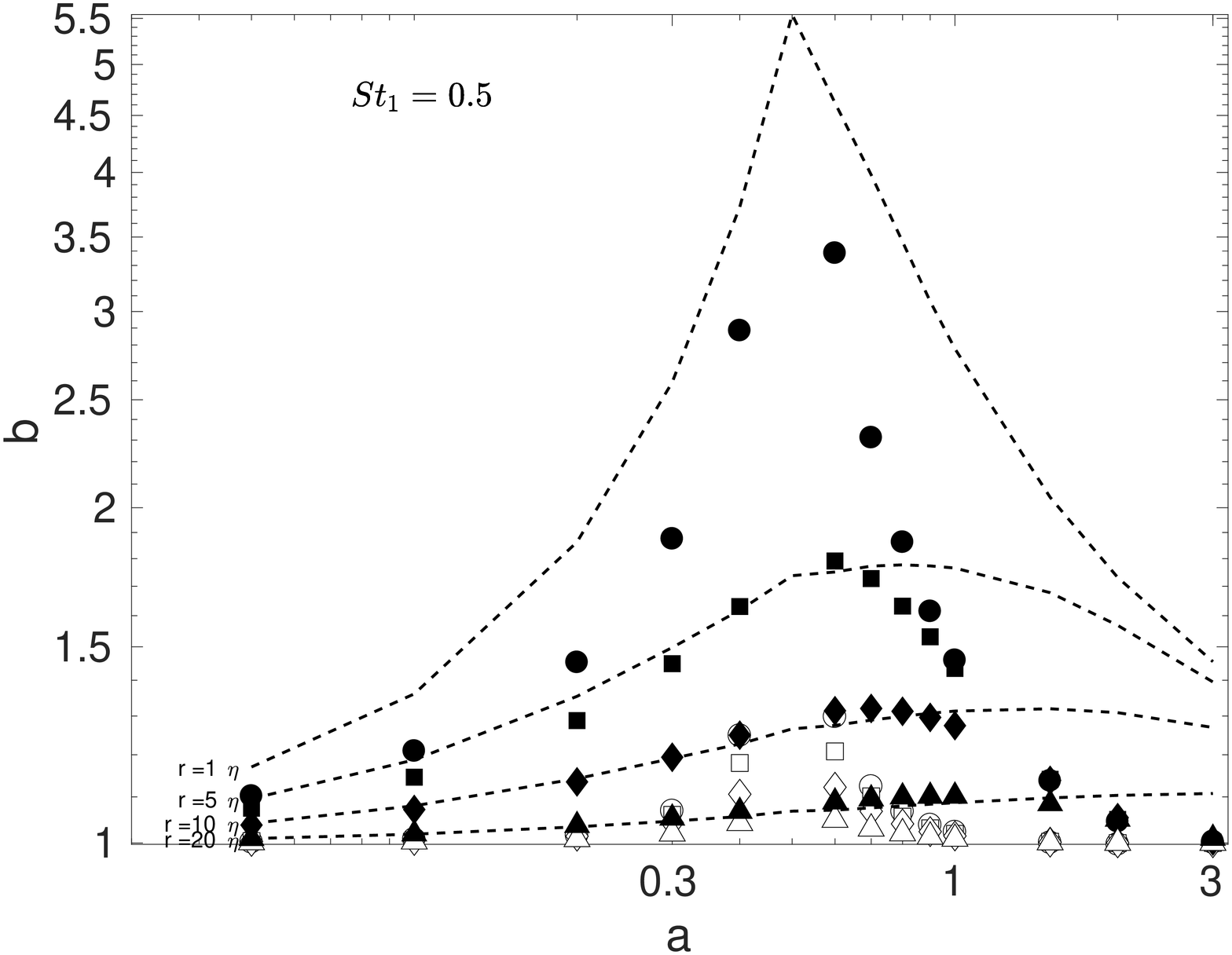}        
	    \caption{}
    \end{subfigure}

\vspace{0.1in}
    \begin{subfigure}[b]{0.5\textwidth}
     \psfrag{L}[cc][2]{$St_1 = 1$}
        \centering
		\includegraphics[width=0.9\textwidth]{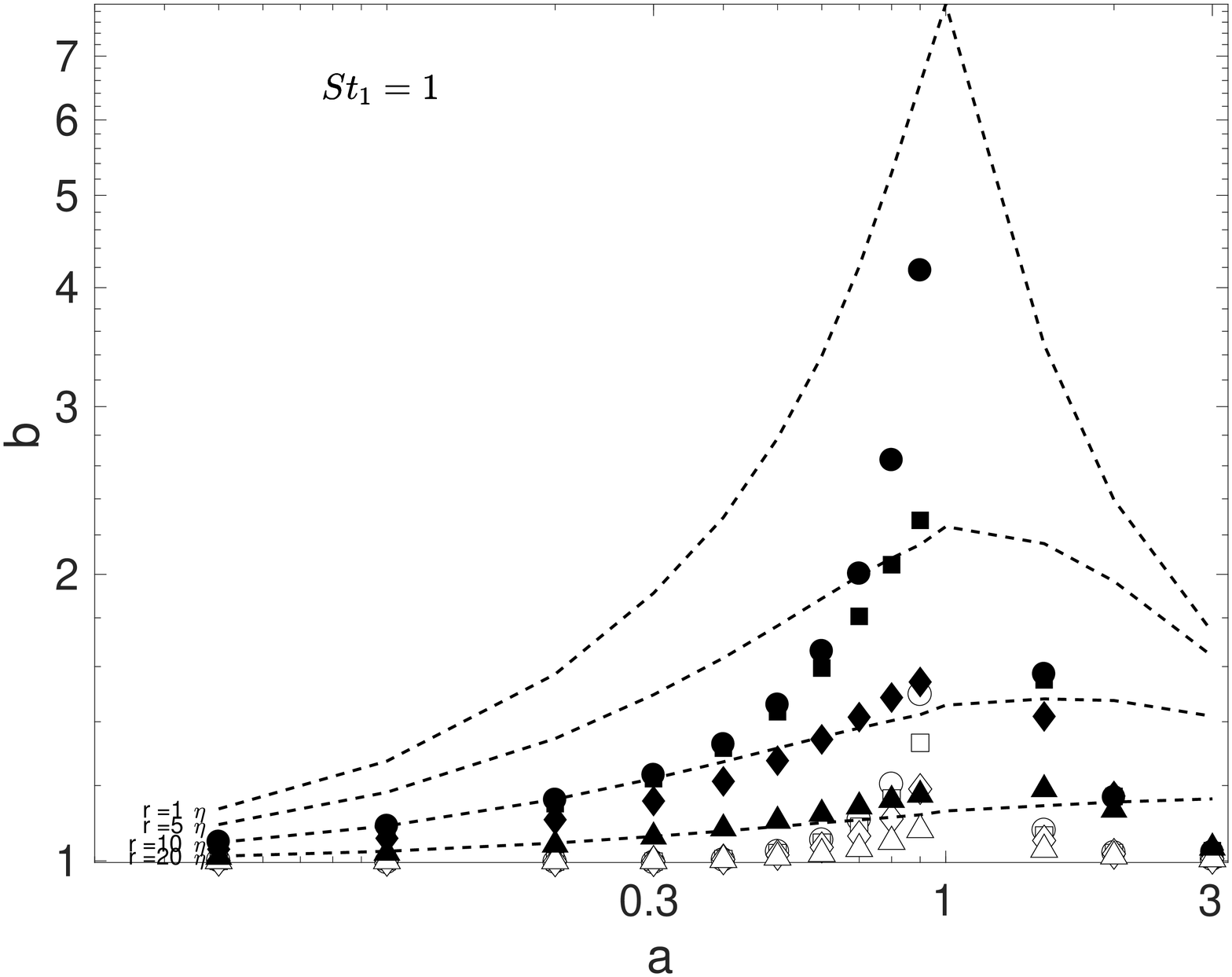}        
		\caption{}
    \end{subfigure}%
    ~
    \begin{subfigure}[b]{0.5\textwidth}
     \psfrag{L}[cc][2]{$St_1 = 3$}
        \centering
         \includegraphics[width=0.9\textwidth]{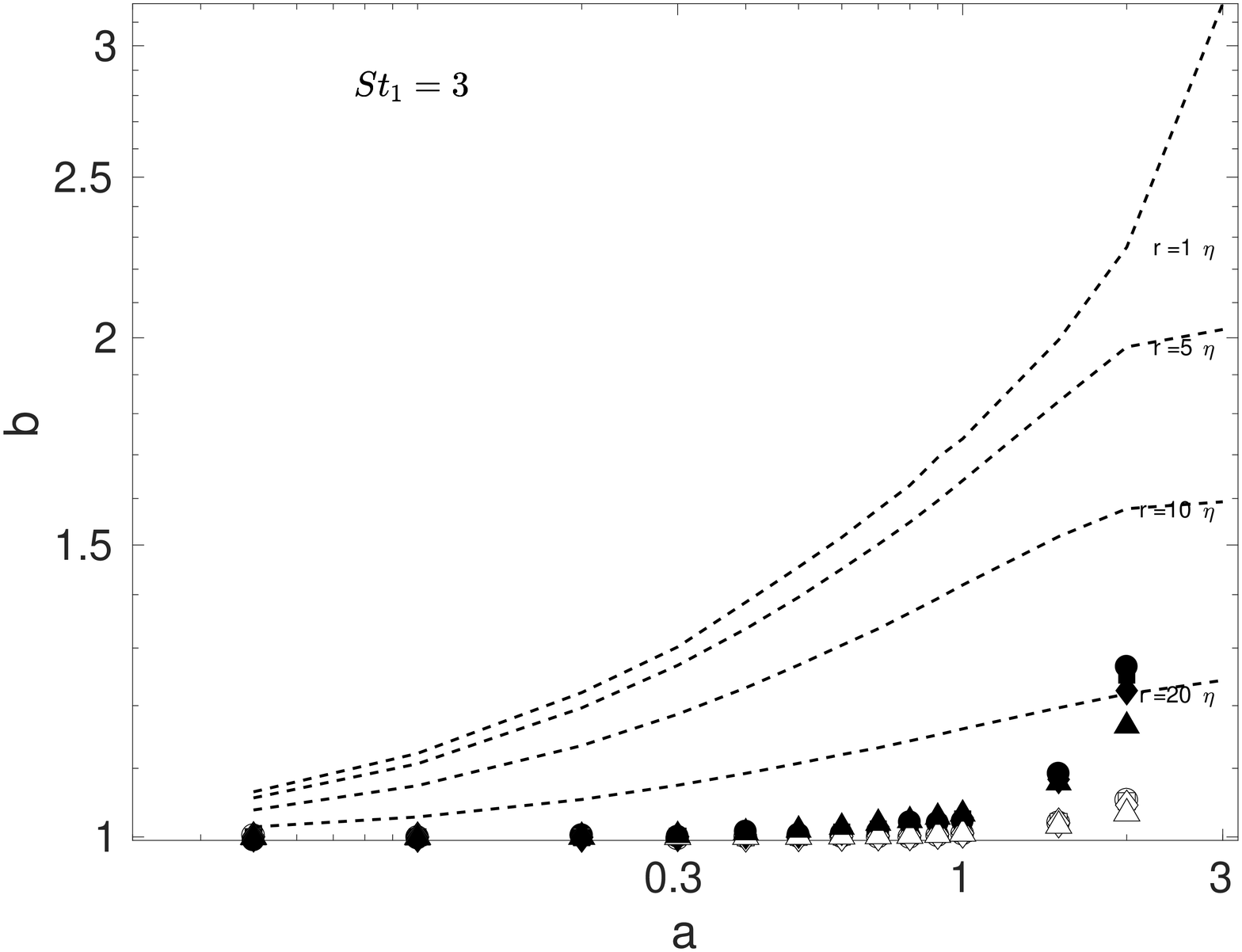}    
          \caption{}
    \end{subfigure}

      \caption{DNS results for $g(r)$ as a function of $St_2$, at specific values of $r/\eta$ and for different $St_1, Fr$ combinations.  Dashed lines correspond to $Fr=\infty$, filled symbols to $Fr=0.3$, open symbols to $Fr=0.052$.}
  \label{fig:rdf_vs_st}
\end{figure}
To see more clearly the effect of bidispersity and gravity on the RDF, in figure~\ref{fig:rdf_vs_st} we plot the RDF for a given $St_1$ as a function of $St_2$ at four different separations. As discussed earlier, the results show that the bidisperse RDFs generally peak when the difference between $St_1$ and $St_2$ is smallest. Figure~\ref{fig:rdf_vs_st} also shows more clearly how gravity systematically suppresses the clustering over the range of $St, \Delta St, Fr$ considered. This is in contrast to the monodisperse case for which it has been observed that gravity can actually enhance clustering in certain regimes of $St, Fr$ \citep{bec14b,gustavsson14,ireland16b}. We expect that this enhancement could also be observed for bidisperse particles when $St_2> 1$ if $| \Delta St |\ll1$. However, we do not have data for this regime, but it is something that should be investigated in future work. Also, the regime $Fr=O(1)$ should also be explored in future work to see if the clustering enhancement is observed for bidisperse particles in that regime.

We add the cautionary remark that when we used a box length of $2\pi$ for the $Fr=0.052$ case, we observed that the RDF was larger than for the $Fr=0.3$ case, giving the appearance of enhanced clustering due to gravity. However, this apparent enhancement disappeared for the much larger box length of $16\pi$ used in this study. This highlights the need to use large computational domains when $Fr\ll1$ in order to ensure the dynamics of the settling particles are not artificially affected by the periodic boundary conditions used in the DNS, as discussed in detail in \cite{ireland16b}.

\FloatBarrier
\subsection{Collision Kernel}

Finally, we consider the collision kernel for the bidisperse particles. The collision kernel is given by \citep{sundaram4}
\begin{align}
K(d)\equiv4 \pi d^2 g(d) S_{-\parallel}(d),
\end{align}
where $d\equiv (d_1+d_2)/2$ is the collision diameter of two spherical particles with diameters $d_1$ and $d_2$, and $S_{-\parallel}(r)\equiv\langle w^p_\parallel(t)|<0\rangle_r$ is the average of $w^p_\parallel(t)$ conditioned on $w^p_\parallel(t)<0$ for the separation $r$ (i.e. the mean inward relative velocity).

In \cite{ireland16a,ireland16b}, when plotting $K(d)$ we chose a fluid to particle density ratio, for which $d$ is then determined by $St$. Since $d$ so determined is often smaller than the smallest separation for which we have data for $g$ and $S_{-\parallel}$, we had to extrapolate our DNS data down to $d$. For the monodisperse case considered in \cite{ireland16a,ireland16b}, this was justified since the functional forms of $g$ and $S_{-\parallel}$ are known at small separations, i.e. they are power laws \citep{gustavsson11}. However, in the bidisperse case, and with gravity, we do not know how $g$ and $S_{-\parallel}$ depend upon $r$ (they are not power laws), and therefore extrapolation would be unreliable. Therefore, we will simply consider $K(d=0.125\eta)$, corresponding to the smallest separation at which our DNS data is reliable.

\begin{figure}
\vspace{0.1in}
\psfrag{a}[cc][2]{$St_2$}
  \psfrag{b}[cc][2][1][180]{$\widehat{K}(d)$}
    \centering
    \begin{subfigure}[b]{0.5\textwidth}
         \psfrag{L}[cc][2]{$St_1 = 0.4$}
        \centering
        \hspace{-0.0in}
        \includegraphics[width=0.9\textwidth]{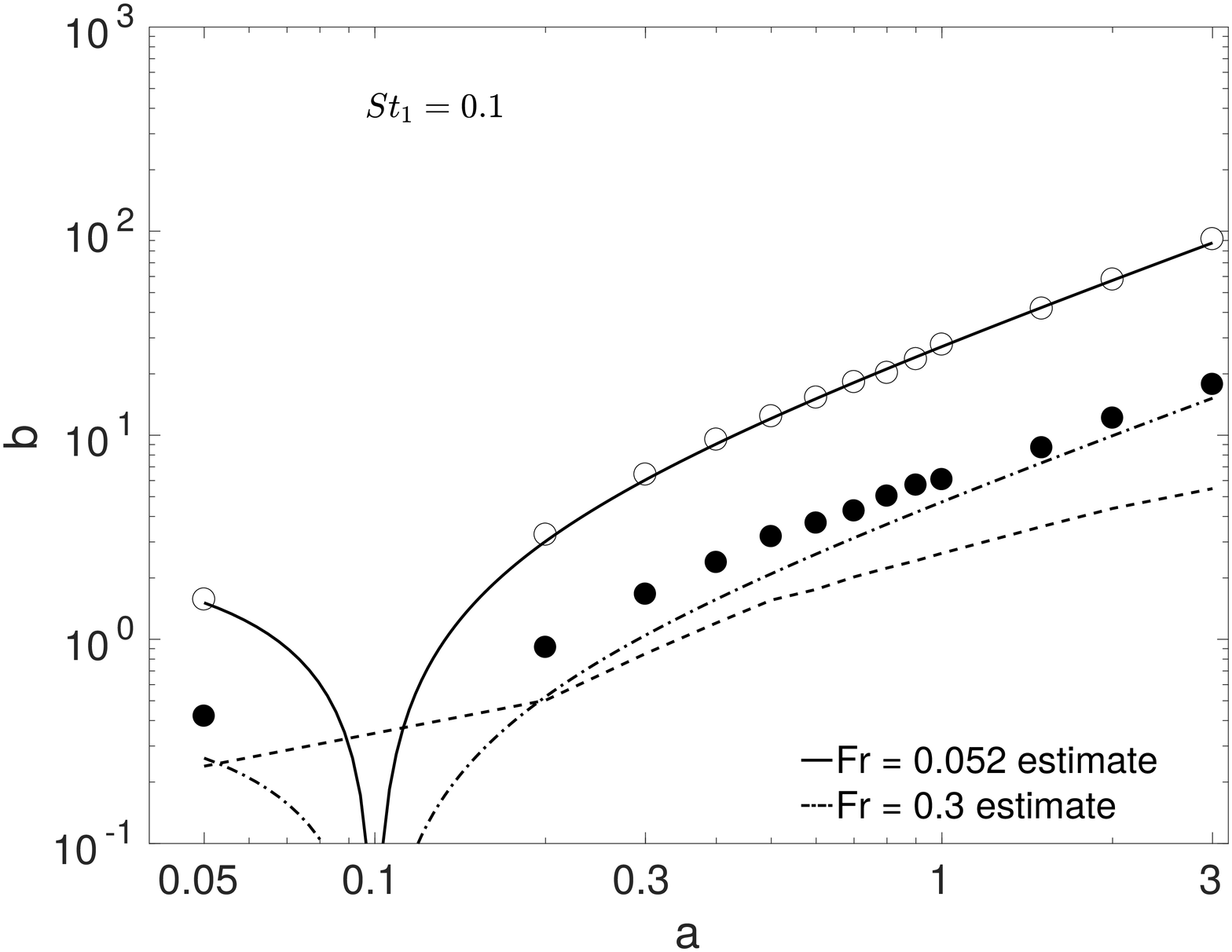}
        \caption{}
    \end{subfigure}%
    ~
    \begin{subfigure}[b]{0.5\textwidth}
     \psfrag{L}[cc][2]{$St_1 = 0.5$}
        \centering
	    \includegraphics[width=0.9\textwidth]{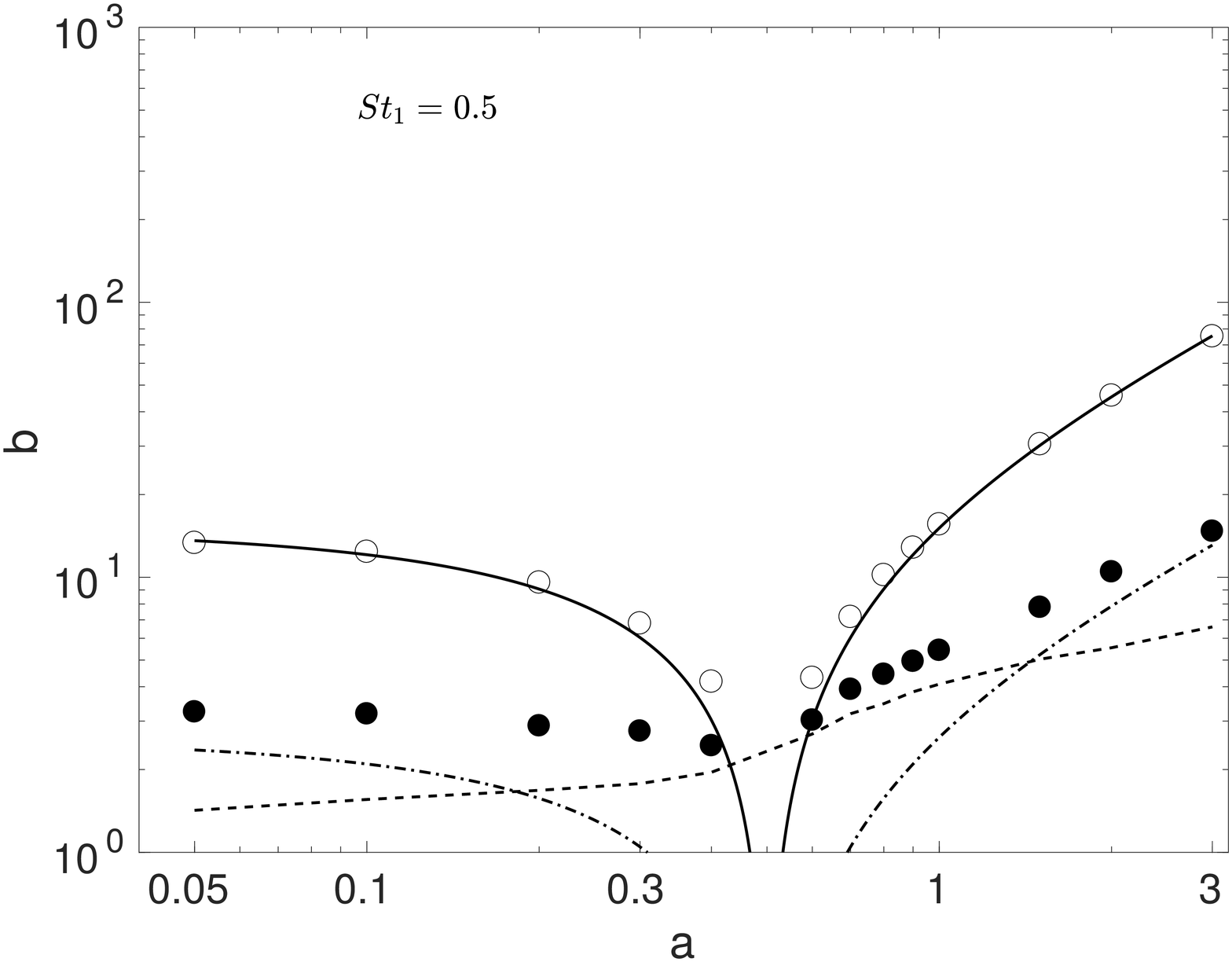}        
	    \caption{}
    \end{subfigure}

\vspace{0.1in}
    \begin{subfigure}[b]{0.5\textwidth}
     \psfrag{L}[cc][2]{$St_1 = 1$}
        \centering
		\includegraphics[width=0.9\textwidth]{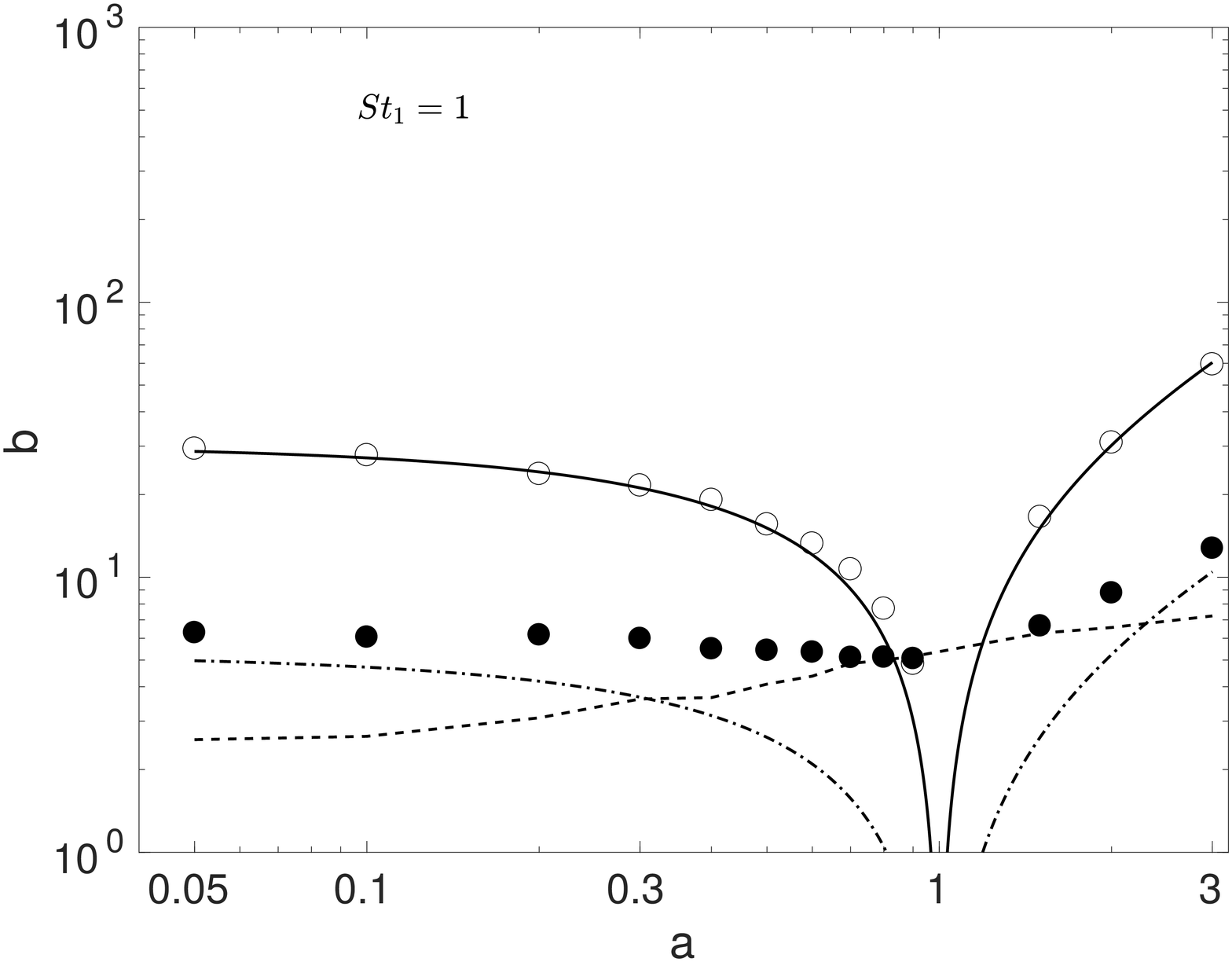}        
		\caption{}
    \end{subfigure}%
    ~
    \begin{subfigure}[b]{0.5\textwidth}
     \psfrag{L}[cc][2]{$St_1 = 3$}
        \centering
         \includegraphics[width=0.9\textwidth]{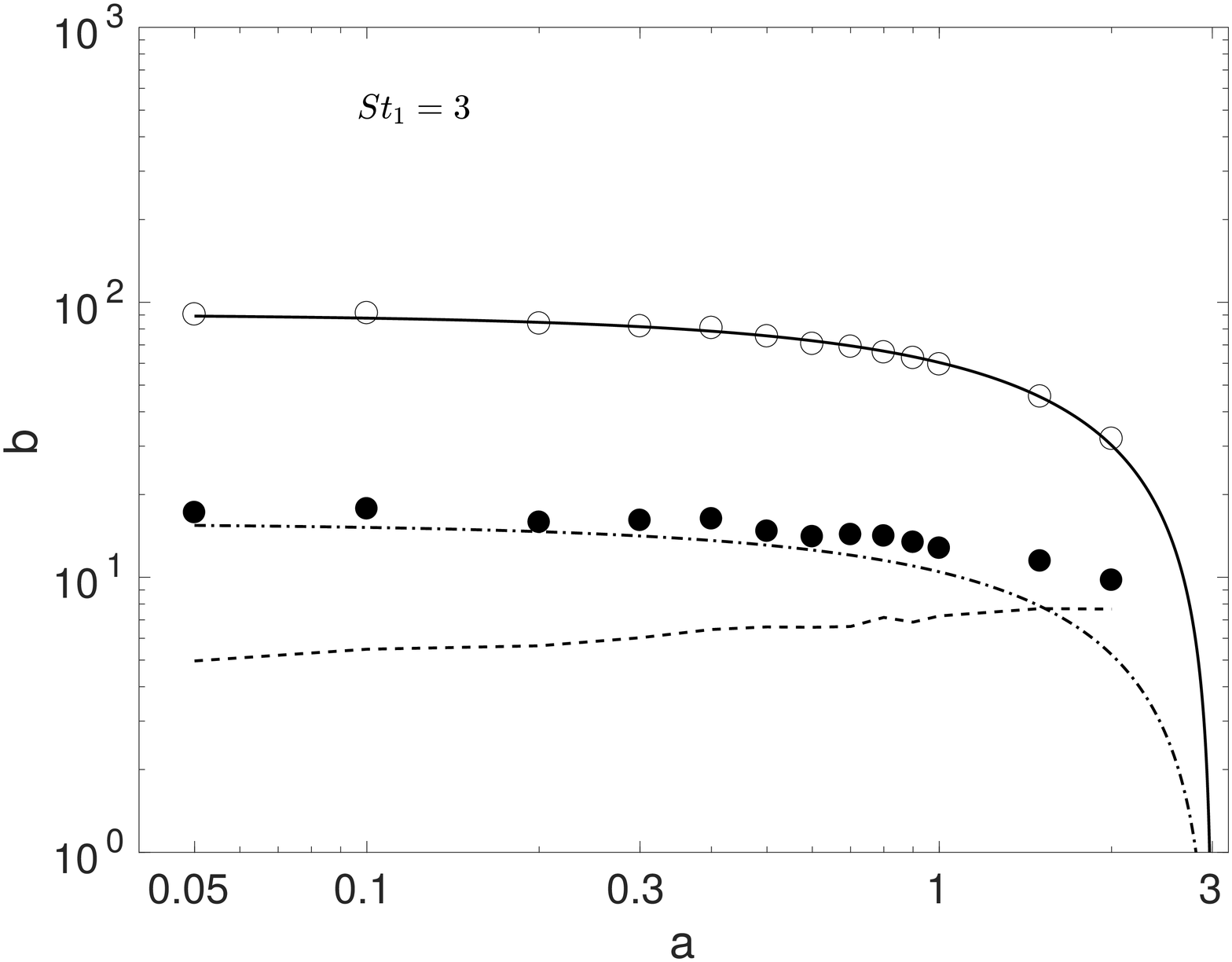}    
          \caption{}
    \end{subfigure}

      \caption{DNS results for $\widehat{K}(d)$ as a function of $St_2$, for $d = 0.125 \eta$ and for different $St_1, Fr$ combinations. Dashed lines correspond to $Fr=\infty$, filled symbols to $Fr=0.3$, open symbols to $Fr=0.052$. The dashed-dot and solid lines correspond to the predictions of \eqref{Kg} for $Fr=0.3$ and $Fr=0.052$, respectively.}
      
  \label{fig:coll_kernel}
\end{figure}

In figure~\ref{fig:coll_kernel} we show results for the normalized collision kernel $\widehat{K}(d)\equiv K(d)/(d^2 u_\eta)$. The results show that decreasing $Fr$ enhances $\widehat{K}(d)$; since clustering reduces as $Fr$ is decreased, this enhancement arises solely because of the enhanced relative velocities due to gravity. Also shown in the plot is the collision kernel for settling bidisperse particles in quiescent flow \citep{grabowski13}
\begin{align}
K_\mathfrak{g}(d)=\pi d^2 |\Delta St | Fr^{-1} u_\eta.\label{Kg}
\end{align}
The results show that for $Fr=0.052$, $K_\mathfrak{g}(d)$ is in excellent agreement with $K(d)$. As we demonstrated earlier, turbulence plays a very significant role on the horizontal relative velocities even when $Fr\ll 1$, however, the relative velocities in the vertical direction are in general much larger, and so the vertical relative velocities dominate the behavior of $S_{-\parallel}(d)$. For $Fr=0.3$, $K_\mathfrak{g}(d)$ is in reasonable agreement with $K(d)$ only if $|\Delta St|\gtrsim 1$, and in general $K_\mathfrak{g}(d)$ significantly underpredicts $K(d)$. 

\begin{figure}\vspace{0.1in}
\psfrag{a}[cc][2]{$St_2$}
  \psfrag{b}[cc][2]{$\widehat{K}(R)$, $\widehat{K}'_\mathfrak{g}(R)$}
    \centering
    \begin{subfigure}[b]{0.5\textwidth}
         \psfrag{L}[cc][2]{$St_1 = 0.4$}
        \centering
        \hspace{-0.0in}
        \includegraphics[width=0.9\textwidth]{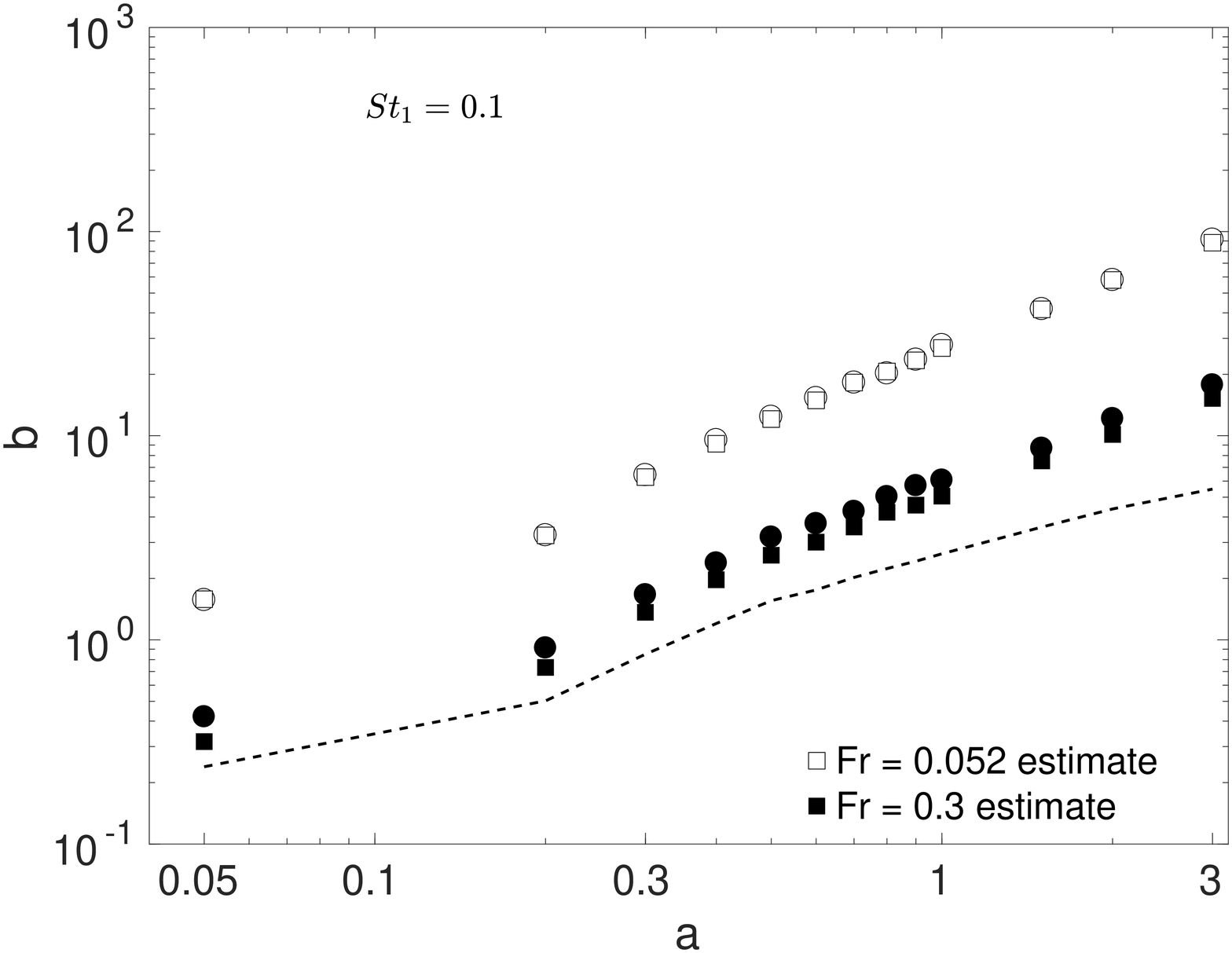}
        \caption{}
    \end{subfigure}%
    ~
    \begin{subfigure}[b]{0.5\textwidth}
     \psfrag{L}[cc][2]{$St_1 = 0.5$}
        \centering
        \includegraphics[width=0.9\textwidth]{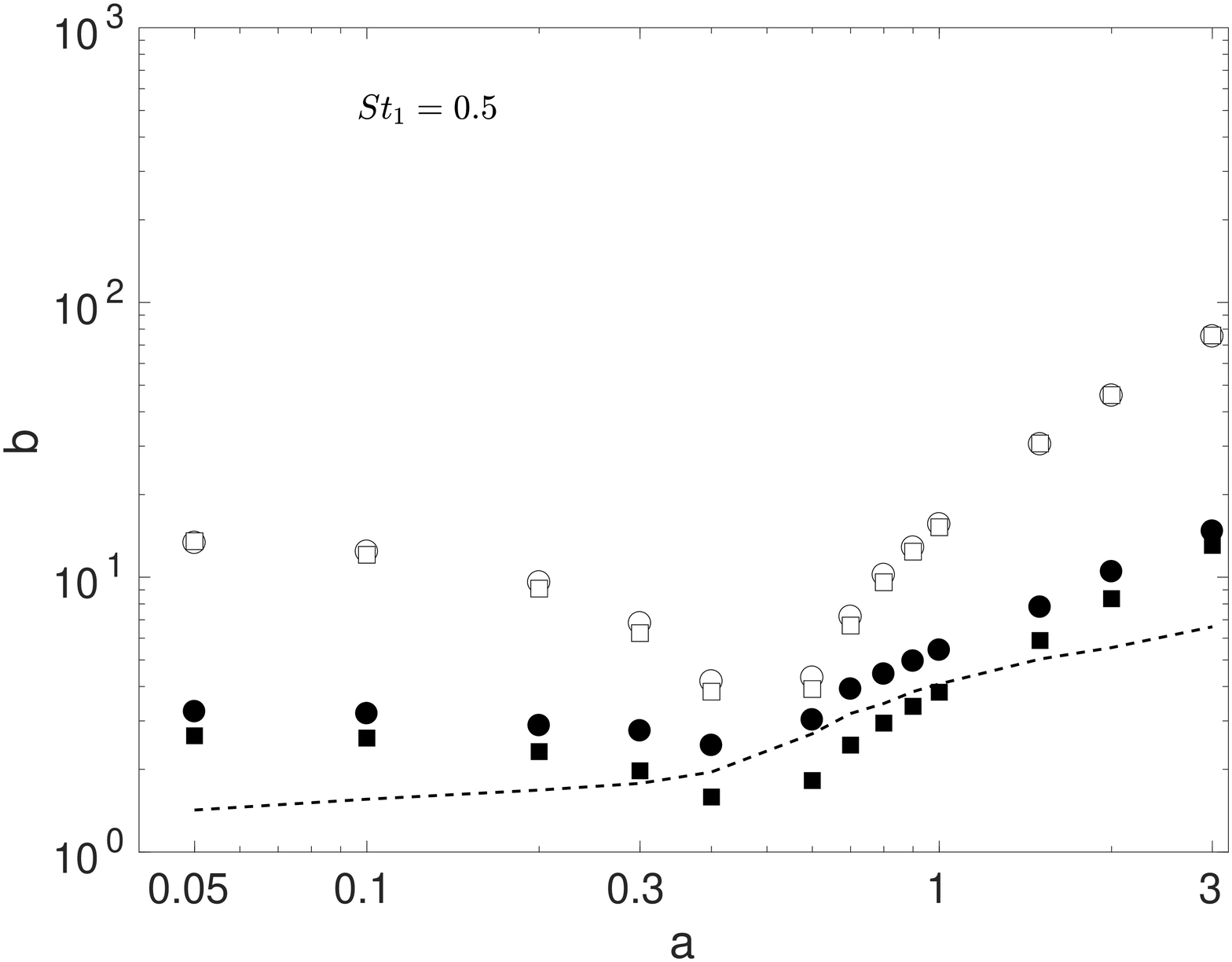}        
        \caption{}
    \end{subfigure}

\vspace{0.1in}
    \begin{subfigure}[b]{0.5\textwidth}
     \psfrag{L}[cc][2]{$St_1 = 1$}
        \centering
        \includegraphics[width=0.9\textwidth]{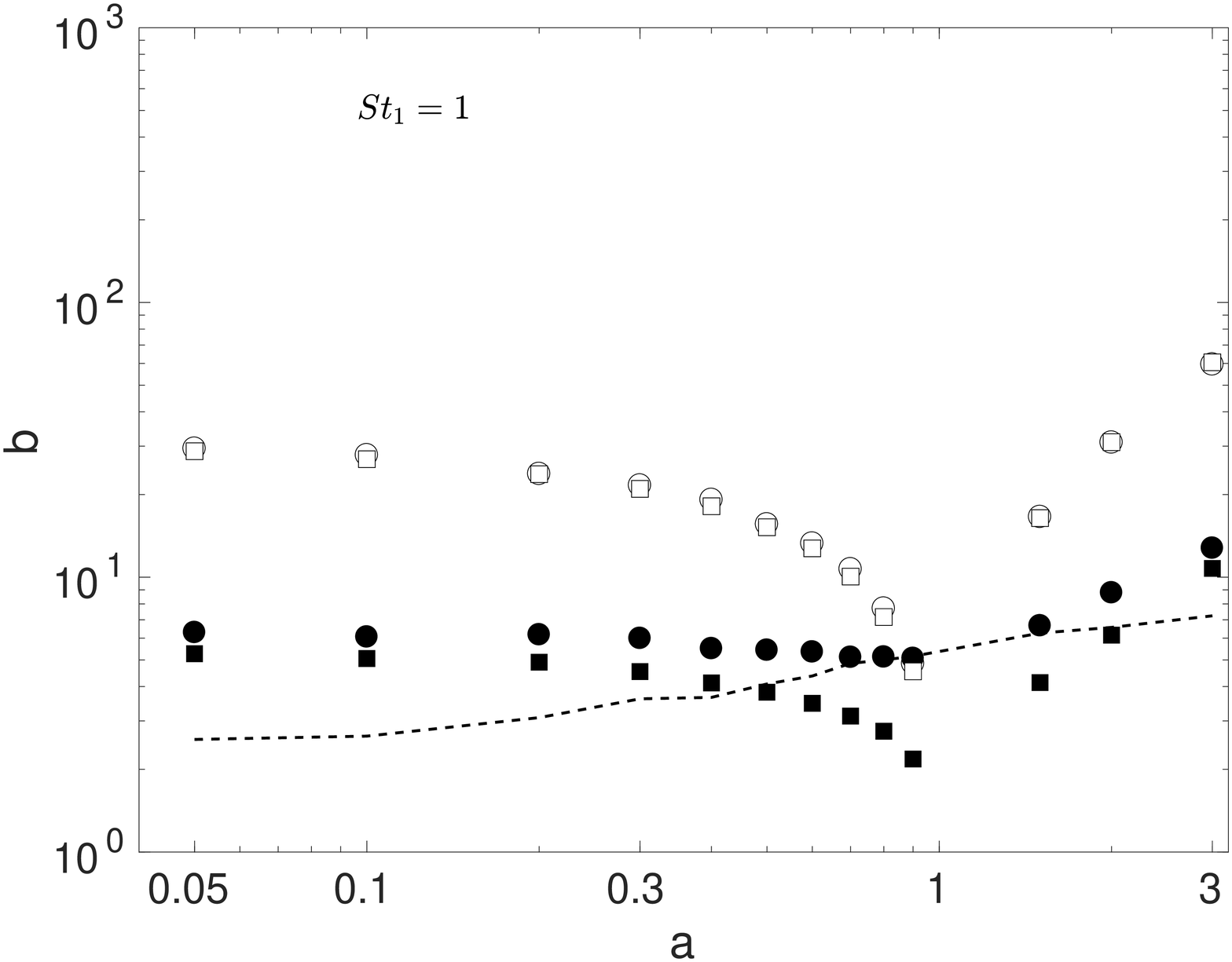}        
        \caption{}
    \end{subfigure}%
    ~
    \begin{subfigure}[b]{0.5\textwidth}
     \psfrag{L}[cc][2]{$St_1 = 3$}
        \centering
         \includegraphics[width=0.9\textwidth]{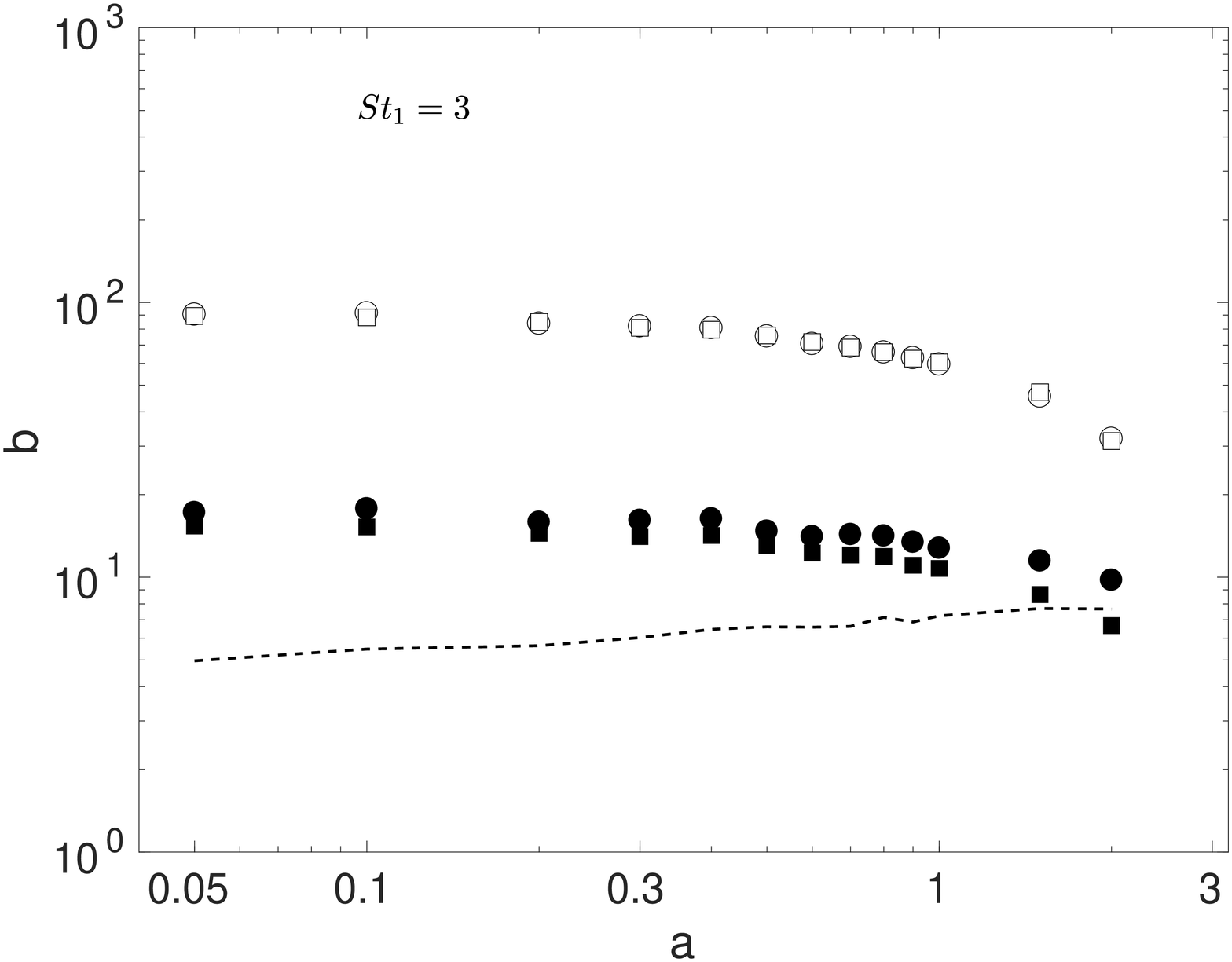}    
          \caption{}
    \end{subfigure}

      \caption{DNS results for $\widehat{K}$ (circles) and $\widehat{K}'_\mathfrak{g}$ (squares)  as a function of $St_2$, for $d = 0.125 \eta$ and for different $St_1, Fr$ combinations. Dashed lines correspond to $Fr=\infty$, filled symbols to $Fr=0.3$, open symbols to $Fr=0.052$.}
      
  \label{fig:coll_kernel2}
\end{figure}

To determine the cause of this underprediction, in figure~\ref{fig:coll_kernel2} we plot $\widehat{K}'_\mathfrak{g}(d)=\pi g(d)|\Delta St | Fr^{-1} $, which uses the DNS data for $g(d)$, but the gravity estimate for $S_{-\parallel}(d)$. The results show that for $Fr=0.052$, $\widehat{K}'_\mathfrak{g}(d)\approx\widehat{K}_\mathfrak{g}(d)\approx \widehat{K}(d)$ since the particle clustering is almost absent. However, for $Fr=0.3$, $\widehat{K}'_\mathfrak{g}(d)$ provides a significantly improved prediction for $\widehat{K}(d)$ compared with $\widehat{K}_\mathfrak{g}(d)$, implying the importance of clustering on the collisions rates in this regime. However, for some combinations of $St_1,St_2$,  $\widehat{K}'_\mathfrak{g}(d)$ still underpredicts $K(d)$ which must be due to errors in the gravity estimate $S_{-\parallel}(r)=(1/4)u_\eta /|\Delta St| Fr^{-1}$.


\FloatBarrier


\begin{figure}\vspace{0.1in}
\psfrag{a}[cc][2]{$r/\eta$}
  \psfrag{b}[cc][2]{$S_{-\parallel}/u_\eta$}
  \psfrag{ St}[cc][2]{{$St_2$}}
  \psfrag{.}[cc][2]{}
      \psfrag{TT}[cc][2]{Increasing $St_2$}
    \centering
    \begin{subfigure}[b]{0.5\textwidth}
         \psfrag{L}[cc][2]{$St_1 = 0.4$}
        \centering
        \hspace{-0.0in}
        \includegraphics[width=0.9\textwidth]{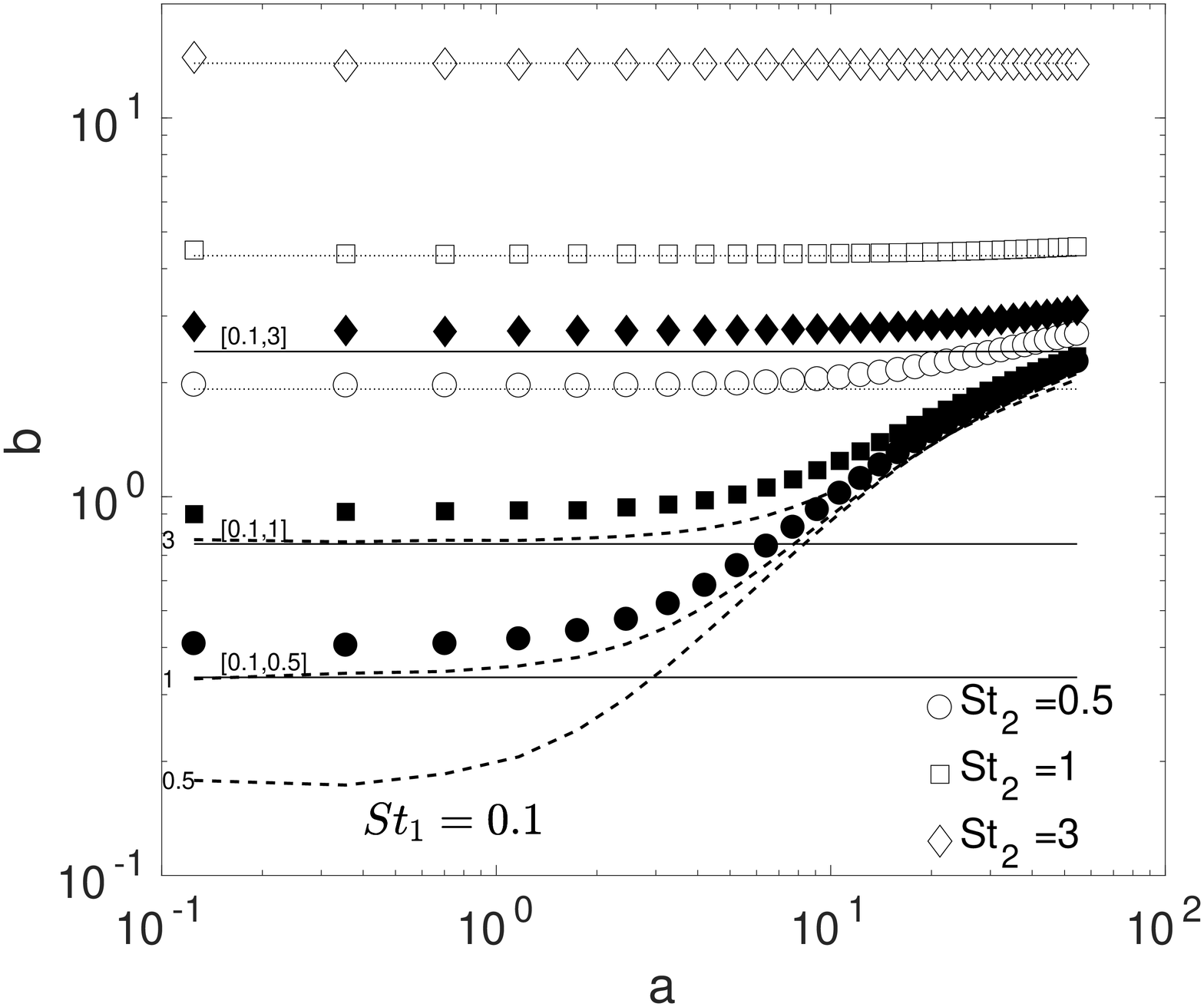}
        \caption{}
    \end{subfigure}%
    ~
    \begin{subfigure}[b]{0.5\textwidth}
     \psfrag{L}[cc][2]{$St_1 = 0.5$}
        \centering
	    \includegraphics[width=0.9\textwidth]{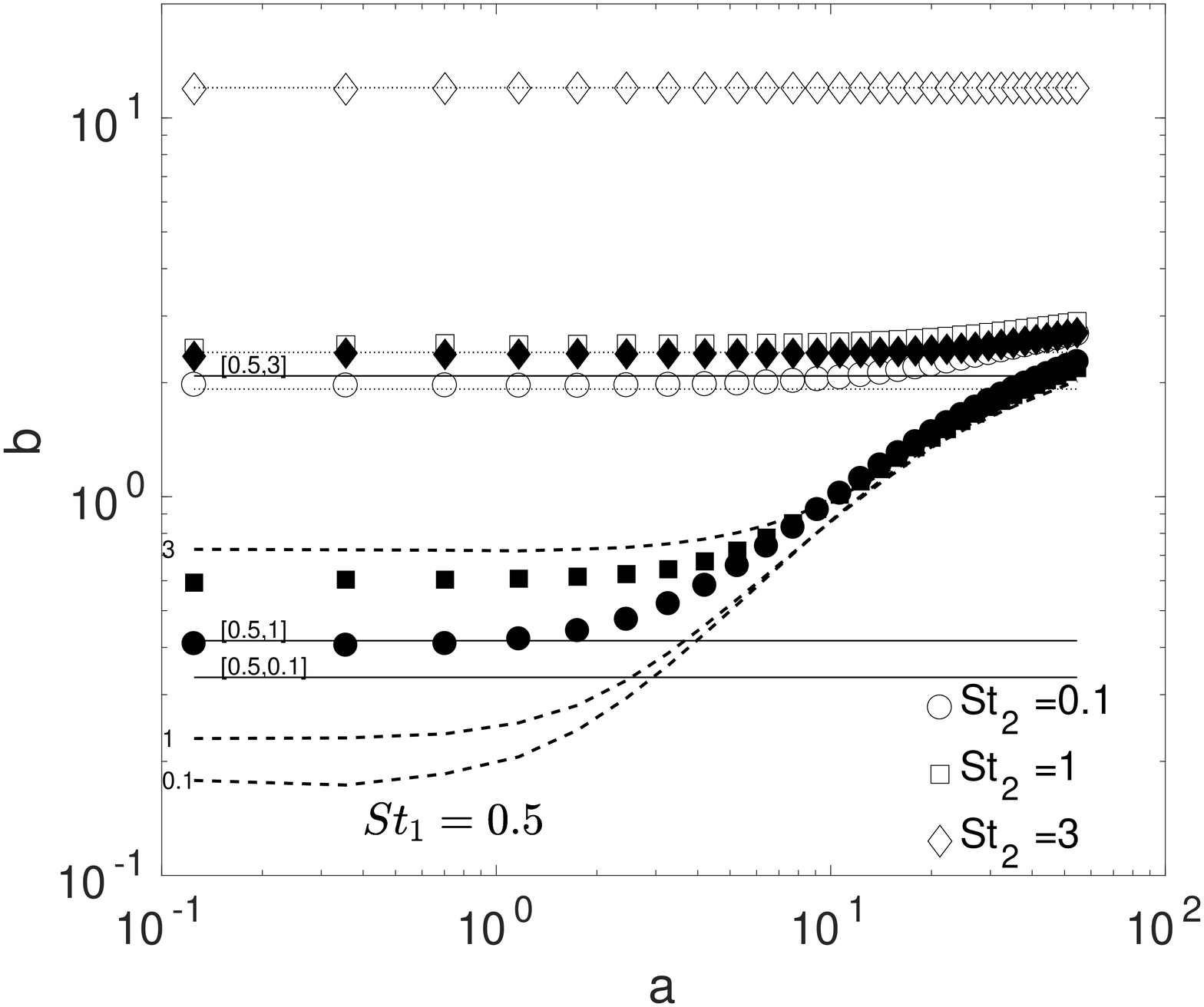}        
	    \caption{}
    \end{subfigure}

\vspace{0.1in}
    \begin{subfigure}[b]{0.5\textwidth}
     \psfrag{L}[cc][2]{$St_1 = 1$}
        \centering
		\includegraphics[width=0.9\textwidth]{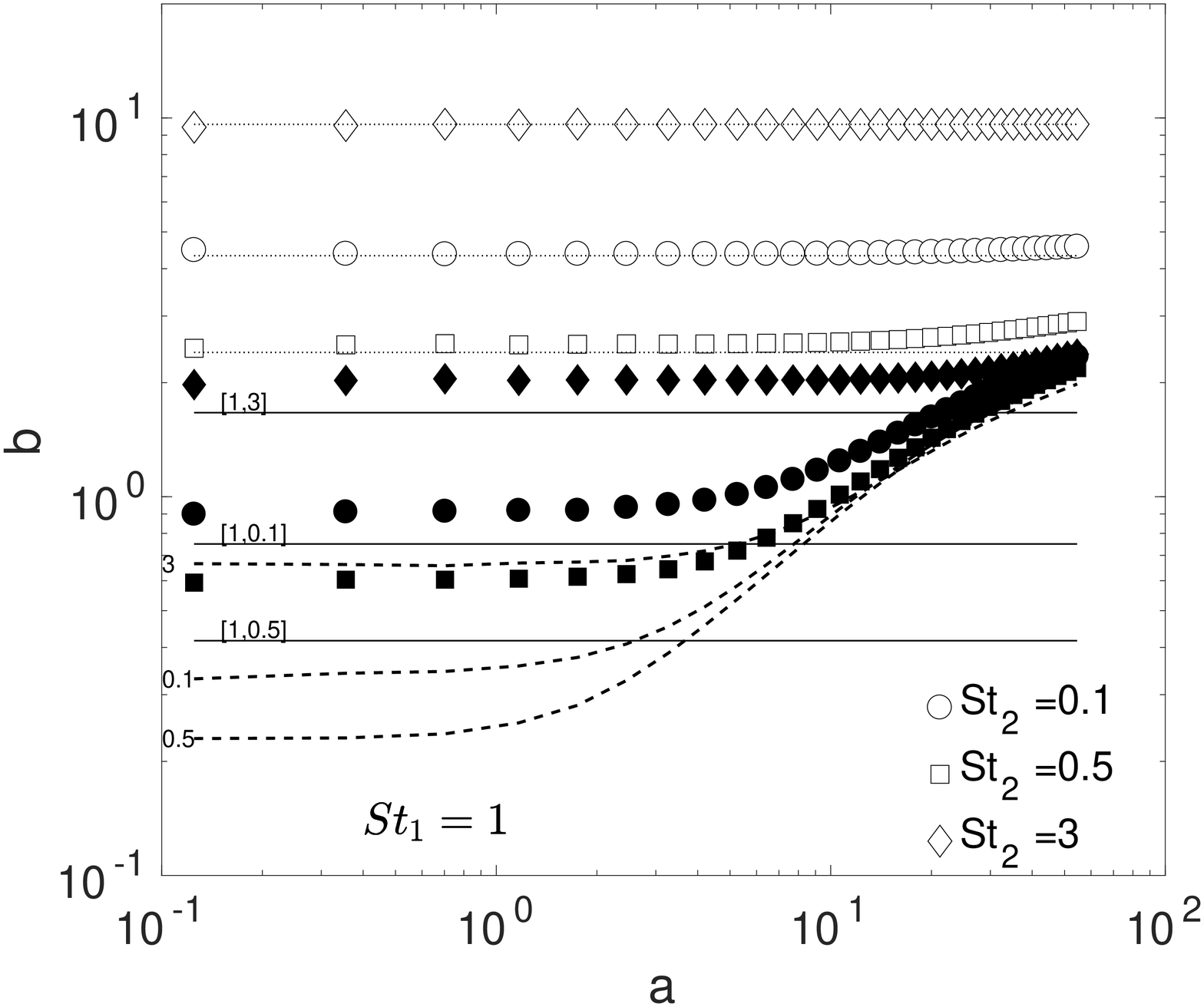}        
		\caption{}
    \end{subfigure}%
    ~
    \begin{subfigure}[b]{0.5\textwidth}
     \psfrag{L}[cc][2]{$St_1 = 3$}
        \centering
         \includegraphics[width=0.9\textwidth]{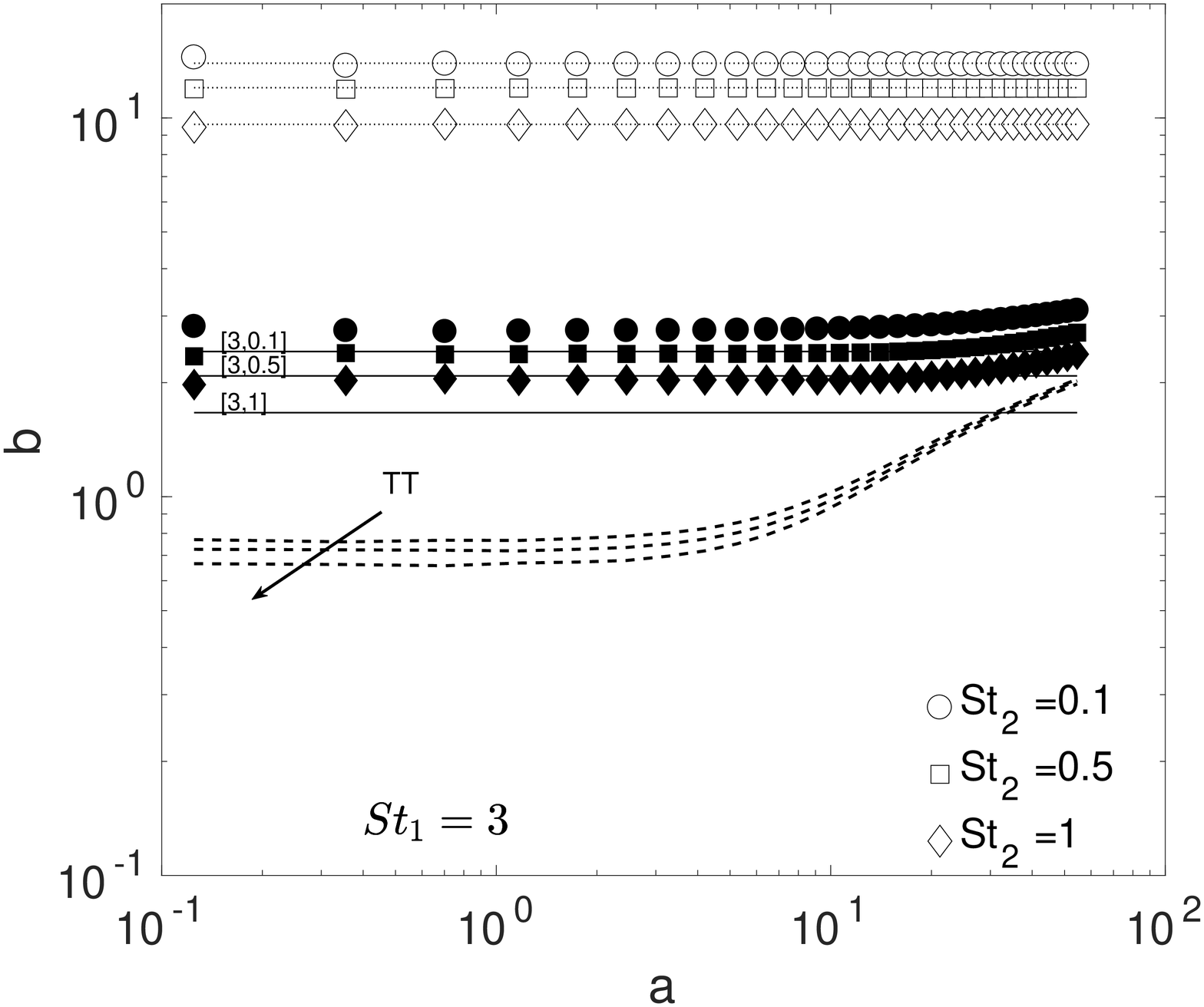}    
          \caption{}
    \end{subfigure}

      \caption{DNS results for $S_{-\parallel}(r)$ as a function of $r/\eta$, for different $St_1, St_2, Fr$ combinations. Dashed lines correspond to $Fr=\infty$, filled symbols to $Fr=0.3$, open symbols to $Fr=0.052$. The numbers next to the dashed lines indicate the $St_2$ value that they correspond to. The dotted lines correspond to the predictions of \eqref{wIn} for $Fr=0.052$, and the solid lines correspond to the predictions of \eqref{wIn} for $Fr=0.3$, where the $St_1, St_2$ combinations that the lines correspond to are shown in square brackets for clarity.}
      
  \label{fig:par_mean_inward}
\end{figure}

 In figure~\ref{fig:par_mean_inward} we plot the DNS data for $S_{-\parallel}(r)$ and compare the results with the gravity estimate. As anticipated, the results show that whereas the gravity estimate
\begin{align}
  S_{-\parallel}(r)=(1/4)u_\eta /|\Delta St| Fr^{-1}, \label{wIn}
\end{align}  
   is in excellent agreement with the DNS data for $Fr=0.052$ and $r<\eta$, it underpredicts the DNS data for $Fr=0.3$ for some combinations of $St_1,St_2$.


%
\section{Conclusions}\label{conc}
We have used DNS to explore the relative motion of settling, bidisperse inertial particles at the small-scales of isotropic turbulence. We found that gravity can strongly enhance the relative velocities of the particles, not only in the direction of gravity, but also in the direction normal to its action. When $Fr\ll1$ the PDF of the relative velocity in the direction of gravity has a mode that is well described by the differential settling velocity, however unless the difference in the Stokes number of the two particles is $\geq O(1)$, the PDF can have a significant spread around this value. These findings are very important for the vertical and horizontal mixing of settling, disperse particles in turbulence since it implies that even for $Fr\ll1$, their vertical mixing rates are not necessarily dominated by gravity, but that turbulence can continue to play a strong role, and further, that their horizontal mixing due to turbulence can be significantly enhanced by gravity.

We find that for the range of $St_2$ and $\Delta St$ considered, gravity always suppresses the clustering of bidisperse particles in turbulence, in contrast to the monodisperse case where it can strongly enhance it. Indeed, for $Fr=0.052$, the particles are almost uniformly distributed throughout the flow for the range of $St$ and $\Delta St$ considered.

Finally, we considered the collision rates of the particles. We found that gravity strongly enhances the collision rates of bidisperse particles, due to their enhanced relative velocities, both in the direction of and normal to gravity. For $Fr=0.3$, the collision kernel prediction based on settling bidisperse particles in quiescent flow under predicts the DNS data except for $|\Delta St|>1$, but is in near perfect agreement for $Fr=0.052$, even for $0<\vert\Delta St\vert\ll1$.

In conclusion, these results have shown how gravity leads to a number of interesting effects on the dynamics of bidisperse particles at the small-scales of turbulence, and that turbulence plays a crucial role in the particle motion even when $Fr\ll1$. A key point is that $Fr$ only characterizes the importance of gravity compared with some typical r.m.s acceleration of the fluid. However, in turbulence the acceleration is extremely intermittent when $R_\lambda$ is large. As a consequence, there is a significant probability for particles to be in regions of the flow where the local, instantaneous Froude number is $>1$, even though the typically defined Froude number is $\ll1$. This could imply, for example, that extreme events in the mixing of settling, bidisperse particles are weakly affected by gravity even when $Fr\ll 1$. To investigate these issues, it is important in future studies to consider higher $R_\lambda$ than we have studied. A significant challenge in this respect is the large computational domains required when $Fr\ll1$. 

\section{Acknowledgements}
We gratefully acknowledge Dr. P. J. Ireland for providing some of the data used in this paper.

\bibliographystyle{jfm}
\bibliography{refs_co12}

\end{document}